\documentclass[12pt,a4paper]{article}

\usepackage[skip=2pt,font=footnotesize]{caption}

\usepackage[margin=20mm]{geometry}
\usepackage[T1]{fontenc}
\usepackage[latin2]{inputenc}
\usepackage{ae}
\usepackage{graphicx}
\usepackage[american]{babel}
\usepackage{amsfonts}

\usepackage{amsthm}
\usepackage{amsmath}
\usepackage{amssymb}
\usepackage{wasysym}
\usepackage{paralist}
\usepackage{multirow}
\usepackage{cite}
\usepackage{pifont}
\usepackage{imakeidx}

\makeindex
\makeatletter
\newcommand{\footnoteref}[1]{%
\ltx@ifpackageloaded{hyperref}{
  \ifHy@hyperfootnotes
    \hbox{\hyperref[#1]{%
            \@textsuperscript {\normalfont \ref*{#1}}}}%
  \else
    \hbox{\@textsuperscript {\normalfont \ref*{#1}}}%
  \fi%
}{
    \hbox{\@textsuperscript {\normalfont \ref{#1}}}%
 }%
}
\makeatother

\newtheorem{Def}{Definition}
\newtheorem{Theorem}{Theorem}
\newtheorem{Lemma}[Theorem]{Lemma}

\newtheorem{Corollary}[Theorem]{Corollary}
\newtheorem{Proposition}[Theorem]{Proposition}

\newtheorem{Conjecture}[Theorem]{Conjecture}

\newtheorem{Note}[Theorem]{Note}

\newcommand{\bs}{\boldsymbol}
\newcommand{\btheta}{\bs{\theta}}
\newcommand{\htheta}{\hat{\theta}}
\newcommand{\bhtheta}{\bs{\hat{\theta}}}

\newcommand{\Alpha}{A}
\newcommand{\A}{\mathcal A}
\newcommand{\Acc}{W}

\newcommand{\Ag}{\mathrm{N^+}}
\newcommand{\B}{\mathcal B}
\newcommand{\Chi}{X}
\newcommand{\Con}{\mathcal C}
\newcommand{\C}{\Con}
\newcommand{\cp}{\tau}

\newcommand{\E}{{{\sf E}}}
\newcommand{\G}{\Gamma}
\newcommand{\I}{\mathcal I}

\renewcommand{\l}{\xi}
\newcommand{\M}{\mathcal M}
\newcommand{\N}{\mathrm N}
\renewcommand{\P}{\mathcal P}
\newcommand{\Pl}{\N}
\renewcommand{\Pr}{{\sf P}}
\newcommand{\R}{\mathbb R}
\newcommand{\X}{\mathcal X}
\newcommand{\Str}{\mathcal S}

\newcommand{\eps}{\varepsilon}
\newcommand{\fii}{\varphi}
\renewcommand{\:}{\colon}
\newcommand{\where}{\ |\ }
\newcommand{\bigwhere}{\ \big|\ }

\newcommand{\argmax}{\operatornamewithlimits{arg\ max}}

\usepackage{color}
\definecolor{red}{rgb}{0.8,0,0}
\definecolor{green}{rgb}{0,0.6,0}

\begin{document}

\title{Efficient Teamwork}
\author{Endre Cs\'oka \\ \normalsize{MTA Alfréd Rényi Institute of Mathematics, Budapest, Hungary}} 
\date{}

\maketitle

\begin{abstract}

Our goal is to solve both problems of adverse selection and moral hazard for multi-agent projects.
In our model, each selected agent can work according to his private ``capability tree''.
This means a process involving hidden actions, hidden chance events and hidden costs in a dynamic manner, and providing contractible consequences which are affecting each other's working process and the outcome of the project.
We will construct a mechanism that induces truthful revelation of the agents' capability trees and chance events and to follow the instructions about their hidden decisions.
This enables the planner to select the optimal subset of agents and obtain the efficient joint execution.
We will construct another mechanism that is collusion-resistant but implements an only approximately efficient outcome.
The latter mechanism is widely applicable, and the major application details will be elaborated.


\medskip


\end{abstract}



\section{Introduction}

Assume that we want to manage a complex project by hiring some agents for different tasks.
The agents have separate but interdependent working processes.
For example, they might have to share common resources (e.g.\ machines or loading areas).
Or a subtask of an agent must precede another subtask of another agent.
Here an efficient cooperation means a stochastic decision plan which chooses each decision considering the current state of the entire project.

However, the current state of the project is not observable, because the capabilities, the chance events and the decisions of each agent are hidden from all other players.
For example, if an agent finishes late (or with any unfavorable outcome), then we cannot ascertain whether he reported better capabilities or higher efforts than the truth or just had unfortunate chance events.
Similarly, if an agent is able to finish a subtask earlier for a little extra cost, or he can adjust his usage of a common resource to the changing demands of others, then he can deny these capabilities or report them more costly.

We are offering a solution to resolve all these cooperation failures, namely, we design a mechanism that incentivizes the agents to truthfully reveal all their detailed capabilities and all chance events, and to follow the instructions of the planner.
In the main part of the paper, we will show it on an idealized model, but in the appendix, we will show how it can be used under more realistic circumstances: if the players are not risk-neutral, they have non-quasi-linear utilities, they have limited liability, they do not precisely know (or are not able to define) their own capabilities, or we do not have unlimited computational capacity.

In our model, there is a principal and some agents. The principal can decide which agents to involve in the game, the others leave the game with utility 0. Each of the remaining players play a dynamic stochastic game according to their capabilities, which exert contractible influences on each other.
The payment to each agent is a function of these contractible events and the communication history.

A capability tree\index{capability tree $\theta_i$}\index{$\theta_i$: capability tree of player $i$} is a dynamic stochastic process describing the entire working capabilities of one agent. (We could have called it a ``capability tree game''.) This contains decision nodes where the agent should choose the branch to continue, and there are chance nodes where the branch is chosen randomly with given probabilities. Each decision node and chance node has an absolute point in time.

\begin{figure}[tb]
\centering
\includegraphics[clip, width=\textwidth]{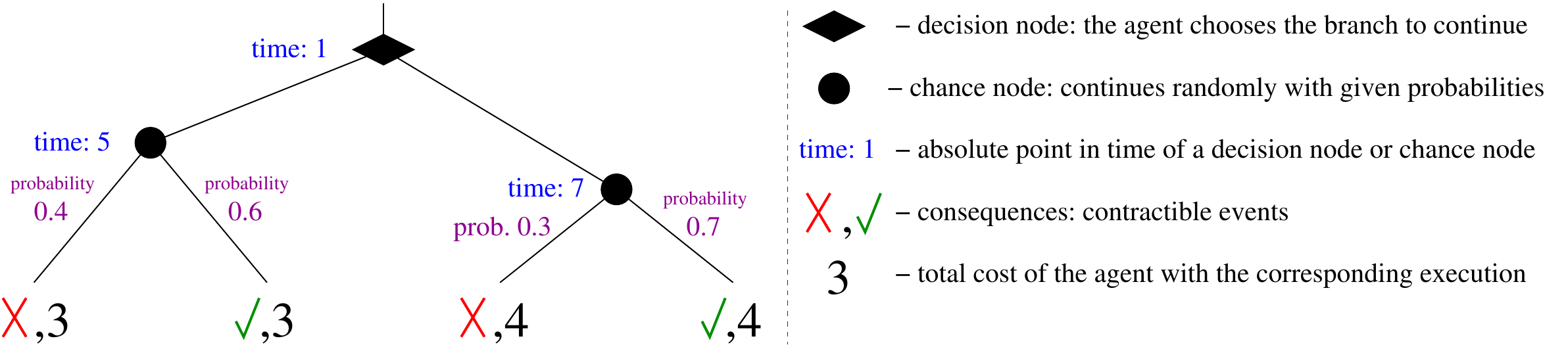}

\smallskip

\caption{The capability tree of an agent describing the potential executions of his working process.}
\label{Fig1}
\end{figure}

A simple example for a capability tree is described by the decision tree in Figure~1, as follows. At time point 1, the agent has to choose between two possible working processes. The left process has a total cost of 3; and with probabilities $0.6$ and $0.4$, the process finishes with a success or failure, respectively. The right process costs 4, but the probabilities are $0.7$ and $0.3$ for the success or failure, respectively. In this simple example, the agent knows only the prior probabilities of the results until time point 5 (left process) or 7 (right process), when he gets to know the result. This capability tree is importantly different from a same tree with different time points, as we will see in Figure~\ref{Fig2}.

The (rules of the) capability tree, the decisions, the chance events and the cost of the work are all private information of the agent. In other words, from the point of view of everybody else, each agent looks like a black box which can communicate and outputs consequences, but about whom nothing else is observable. For example, the agent could easily choose the first, cheaper process instead of the second one (as instructed), or he could even report a completely different capability tree without the risk of being caught.

This was a simplified description of the capability trees. Here, the only consequence provided by the agent was a binary result of his work.
Consequences in general, including direct influences between the capability trees of different players will be introduced in Section~\ref{capability tree}.

\section{Example} \label{example1}

We show a simplified model of the entire project. 
There is a central player called the principal, and some competing agents. Each agent privately knows his capability tree.
Contractible communication between the players is available throughout the game, for example, the agents can send reports to the principal about their capabilities and chance events. The final payments can depend on these reports, but the principal never gets to know whether a report was a lie.
The principal is free to choose which agents to involve in the project, the others get utility 0, and the game ends for them. Then all chosen agents execute their capability trees in parallel, each of them provides a result. At the end, the mechanism determines signed transfers $t_i$ from the principal to each agent, as a function of the results and the entire communication history. The utility of each agent is the transfer $t_i$ he gets minus his costs. The utility of the principal is a joint valuation function of the results of the agents minus all transfers to them.

As a very simple example, consider the following two-task project. The principal must choose one agent per task. Each task $i$ finishes with a binary result: success ({\color{green} \ding{51}$_i$}) or failure ({\color{red}\ding{53}$_i$}). If both tasks succeed, then the principal gets $v(${\color{green} \ding{51}$_1$\ding{51}$_2$}$) = 20$. But if either fails, then the result of the other task is irrelevant, and the principal gets $v(${\color{green} \ding{51}$_1$}{\color{red}\ding{53}$_2$}$) = v(${\color{red} \ding{53}$_1$}{\color{green}\ding{51}$_2$}$) = v(${\color{red} \ding{53}$_1$\ding{53}$_2$}$) = 0$.

\begin{figure}[tb]
\centering
\includegraphics[width=17cm]{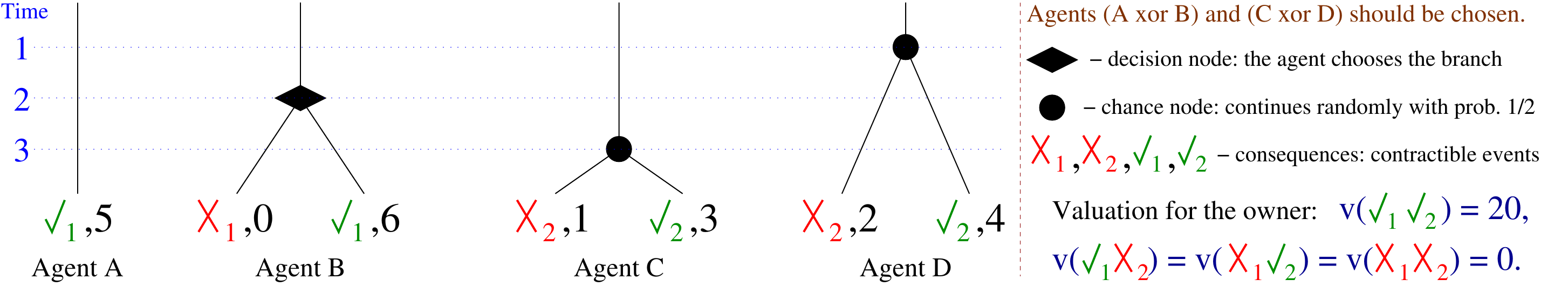}

\bigskip

\includegraphics[width=17cm]{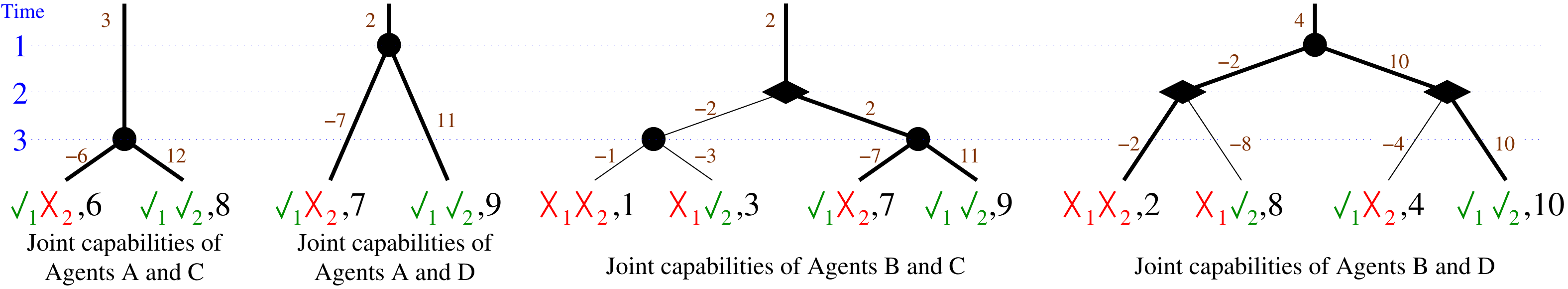}

\medskip

\caption{The capability trees of the agents and the evaluation of the truthfully reported trees.}
 \label{Fig2}
\end{figure}

Agents $A$ and $B$ apply for the first task, Agents $C$ and $D$ apply for the second task. Their capability trees are described in Figure~\ref{Fig2}. In words, Agent~$A$ would complete the first task with a cost of 5. Agent~$B$ would complete the first task with a cost of 6, but he would have an option of quiting at time point 2 (absolute time point, say, the end of the year 2002), with 0 cost. Agent~$C$ would have a cost of 1, until time point 3 (end of 2003), when he either gets to know that he failed to complete the task successfully (result {\color{red}\ding{53}$_2$}), or he will successfully complete it for a further cost of 2 (result {\color{green} \ding{51}$_2$} with a total cost of 3), each option happens with probability $1/2$. Agent~$D$ has the same capability tree but with 1 higher costs, and he observes his result at time point 1 (end of 2001).

In order to understand what the desired outcome is, let us ignore first the incentive constraints, and assume that every player reports his private information truthfully and make the desired actions obediently.

At the beginning, the principal should choose the pair of agents. If she were to choose Agents $A$ and $C$, then for an expected total cost of $5 + \frac{1+3}{2} = 7$, the principal would get $v(${\color{green} \ding{51}$_1$\ding{51}$_2$}$) = 20$ with probability 1/2. This would provide $\frac{20+0}{2} - 7 = 3$ expected total utility. If the principal were to choose Agents $A$ and $D$, then the expected total utility would be $\frac{20+0}{2} - \big(5 + \frac{2+4}{2}\big) = 2$. If the principal were to choose Agents $B$ and $C$, then the efficient strategy would be that Agent~$B$ should choose the right option, and the expected total utility would be $\frac{20+0}{2} - \big(6 + \frac{1+3}{2}\big) = 2$. But if the principal were to choose Agents $B$ and $D$, then the efficient strategy would be for Agent~$B$ to choose the left option if Agent~$D$ succeeds, or the right option if Agent~$D$ fails. This way, the expected total utility would be $\frac{20+0}{2} - \frac{(0+6) + (2+4)}{2}  = 4$. Therefore, implementing efficiency means for the example that the principal should choose Agents $B$ and $D$ (based only on communication, without observing the capabilities of the agents) and the agents should execute the corresponding efficient joint plan.

At the end, the principal pays to the agents for they work according to a fixed rule.
For example, if the principal pays 4 to $B$ and 4 to $D$ in expectation, then the expected utilities are $u(B) = 4 - \frac{0+6}{2} = 1$, $u(D) = 4 - \frac{2+4}{2} = 1$, $u(A) = u(C) = 0$ and the expected utility of the principal is $u(P) = \frac{20+0}{2} - 4 - 4 = 2$. 

In this example, if one of these four agents reports his capability tree truthfully, then he cannot cheat later. We emphasize that this is not the case in general, at all. See for example Figure~\ref{Fig1}, where the agent can choose the (cheaper) left option instead of the (more expensive) right option, without the risk of being caught. This issue did not appear in our example just because we wanted to keep the example as simple as possible.
We also note that these high risks are only the properties of our simple and extreme example. Most real-life projects contain a large number of smaller risks, providing much smaller total risk. We can think about the example so that we have a large number of small independent projects of this kind. For further discussion of this topic, see Appendix~\ref{limited}.

\section{The mechanisms -- via examples} \label{offer-example}\index{mechanism}

The mechanism is the following. First, the principal asks an offer for contract from each agent. Then the principal will accept a subset of them, the other agents will be rejected. After the offers for contracts are received, the principal considers all of her possible strategies about which offers to accept and what messages to send depending on the incoming messages. The principal commits in advance that she will choose a strategy which maximizes his minimum possible payoff.

This mechanism may look counter-intuitive, even for a single-agent project. Therefore, we show first a single-agent example.

\subsection{Example for the mechanisms for a single-agent project}

Assume that we have a project which should be completed by only one agent. There are two possible results: early or late completion, the difference for the principal is equivalent to $d = \$10$. The agents are competing for the job. Consider the dilemma of one competing agent about which proposal\index{proposal $\pi_i$} (offer for contract) to make. A naive proposal would be the following.

\medskip

\noindent
\textbf{Proposal 1.} The agent asks for $\$20$, and he claims that he would finish early with probability $50\%$.

\smallskip
\noindent
\emph{Evaluation.} The worst outcome for the principal would be to have a late completion. Therefore, Proposal 1 is as competitive as a guaranteed late completion for $\$20$. E.g., another proposal by another agent for making it with late completion for $\$19.99$ would be preferred by the principal.

\medskip

\noindent
\textbf{Proposal 2.} The agent asks for $\$25$ if he completes the task early, and $\$15$ if late.

\smallskip
\noindent
\emph{Evaluation.} As good as if the principal believed the probabilities in Proposal 1, because

1. the principal is indifferent about the two outcomes,

2. and the agent would get $\$20$ in expectation.

\medskip

\noindent
\textbf{Proposal 3.} The principal should send a message about an amount of money $x$ in advance. Then the principal should pay $\$20 + 0.5x$ if the agent completes the task early, and $\$20 - 0.5x$ if late.
(Negative payments mean positive payment in the opposite direction.)

\smallskip
\noindent
\emph{Evaluation.} Denote by $v$ the value of an early completion for the principal, and $v - d$ for a late completion with $d = \$10$. If she accepts this proposal, then her maximin utility will be
\begin{equation*}
\max\limits_{x \in \R}\ \min\big( v - (\$20 + 0.5x),\ v - d - (\$20 - 0.5x) \big) = v - \frac{d}{2} - \$20 = v - \$25,
\end{equation*}
and $x = d = \$10 $ should be chosen by the principal. Therefore, this proposal is as competitive as Proposal 2, and provides the same payment for the agent in case of acceptance. This proposal is the same good even if the agent does not know the value of $d$.

\medskip

\noindent
\textbf{Proposal 4.} For an arbitrary amount of money $x$ chosen by the principal in advance, the agent asks for $\$\sqrt{399 + (x/\$ 10)^2} + 0.5x $ if he completes the task early, and $\$\sqrt{399 + (x/\$ 10)^2} - 0.5x$ if late.

\smallskip
\noindent
\emph{Evaluation.} For $d = \$10$, this offers the same as Proposals 2 and 3. But this offer is aware of his risk-aversion about the value of $d$: the agent asks for higher expected payment if he has to take higher risks.

\medskip

The main part of the paper will focus on the risk-neutral case, therefore, we will use the technique in Proposal 3 but not the technique in Proposal 4. However, in Appendix~\ref{obsapp}, we will show how to use the richness of the language of contracts for handling risk-aversion or some other slight relaxations of the assumptions.

Note that if we converted the mechanism to a direct revelation mechanism, then we would lose these robustness features, and Proposal 4 would no longer be a valid proposal. This is the main reason why we do not apply the revelation principle.

\subsection{Truthful proposals with the capability tree in Figure~\ref{Fig1}}

The \textbf{truthful proposal}\index{proposal $\pi_i$!truthful proposal $\pi^*_i$} (or cost price proposal) of an agent with the capability tree in Figure~\ref{Fig1} (page \pageref{Fig1}) is the following.

\medskip

\noindent
{\color{red} \textbf{``}} The principal should send me a message ``Left'' or ``Right'' before time point 1.

\smallskip

\textbf{Case Left.} If she chooses ``Left'', then she should choose an amount of money $x$ before time point 5 (and send a message to me about it). At time point 5, I will send a message ``yes'' or ``no''. If I send ``yes'', then I will provide the result {\color{green}\ding{51}} and the principal should pay me $3 + 0.4 \cdot x$. If I send ``no'', then I will provide the result {\color{red}\ding{53}} and the principal should pay me $3 - 0.6 \cdot x$.

\smallskip

\textbf{Case Right.} If she chooses ``Right'', then she should choose an amount of money $x$ before time point 7. At time point 7, I will send a message ``yes'' or ``no''. If I send ``yes'', then I will provide the result {\color{green}\ding{51}} and the principal should pay me $4 + 0.3 \cdot x$. If I send ``no'', then I will provide the result {\color{red}\ding{53}} and the principal should pay me $4 - 0.7 \cdot x$. {\color{red} \textbf{''}}

\bigskip

We will show two similar mechanisms called first- and second-price mechanisms, in an analogous sense as in auction theory. Under the second-price mechanism, the equilibrium strategy profile will use truthful strategies, and the winners get some extra second-price compensation, defined later. Under the first-price mechanism, the agents should ask for a somewhat higher payment than the (cost-price) truthful proposal. We will explain these issues in Section~\ref{literature}.

\subsection{The mechanisms for the Example}

Consider the Example in (Section~\ref{example1}).
We show an example of proposals of the agents under the \underline{first-price mechanism}\index{mechanism!first-price mechanism}, when all of the agents ask for $1$ more payment in case of acceptance. We will call them \textbf{fair proposal}s\index{proposal $\pi_i$!fair proposal $\pi^*_i + x_i$} with profit $1$.

\medskip

\textbf{Proposal of $A$:} \textbf{``} I will provide the result {\color{green}\ding{51}$_1$} and the principal should pay me $6$. \textbf{''}

\medskip

\textbf{Proposal of $B$:} \textbf{``} If the principal sends me a message ``start'' until time point 2, then I will provide the result {\color{green}\ding{51}$_1$} and she should pay me $7$. Else I will provide the result {\color{red}\ding{53}$_1$} and the principal should pay me $1$. \textbf{''}

\medskip

\textbf{Proposal of $C$:} \textbf{``} The principal should send me a message before time point 3 about an amount of money $x$. At time point 3, I will send a message ``yes'' or ``no''. If I send ``yes'', then I will provide the result {\color{green}\ding{51}$_2$} and the principal should pay me $4 + x$. If I send ``no'', then I will provide the result {\color{red}\ding{53}$_2$} and the principal should pay me $2 - x$. \textbf{''}

\medskip

\textbf{Proposal of $D$:} \textbf{``} The principal should send me a message before time point 1 about an amount of money $x$. At time point 1, I will send a message ``yes'' or ``no''. If I send ``yes'', then I will provide the result {\color{green}\ding{51}$_2$} and the principal should pay me $5 + x$. If I send ``no'', then I will provide the result {\color{red}\ding{53}$_2$} and the principal should pay me $3 - x$. \textbf{''}

\medskip

The execution is the following. The principal accepts $B$ and $D$ and sends ``6'' to $D$. At time point 1, $D$ sends ``yes'' or ``no'' to the principal. If $D$ is truthful, then the report corresponds to his chance event. In the case of ``yes'', the principal sends ``start'' to $B$, otherwise she sends no message (meaning ``do not start'').

There are two possible outcomes, let us see the payments according to the contracts, and the utilities of the players in the two cases. If agent $D$ succeeds, then he gets $5 + 6 = 11$ payment with a cost of $4$, so $u(D) = 11 - 4 = 7$; $B$ gets a payment $7$ with a cost of $6$, so $u(B) = 7 - 6 = 1$. For the principal, $u(P) = 20 - 11 - 7 = 2$. If $D$ fails, then his payment is $3 - 6 = -3$ with a cost of $2$, so $u(D) = (-3) - 2 = -5$; $B$ gets a payment $1$ with a cost of $0$, so $u(B) = 1$. For the principal, $u(P) = 0 - (-3) - 1 = 2$. To sum up, $u(B) = 1$ and $u(P) = 2$ in both cases, while $u(D) = 1 \pm 6$ depending only on his own chance event, and $E\big(u(D)\big) = 1$.

\medskip

The \underline{second-price mechanism}\index{mechanism!second-price mechanism} means the same as the first-price mechanism except that the principal should pay an extra amount of money to each agent, which money is equal to the difference between her maximin utility with all proposals and her maximin utility with all but the agent's proposal.

Under the second-price mechanism, we expect the agents to submit truthful proposals. Namely, they ask for the same payments as in the leaves of the capability trees, which are 1 less payment than in the four proposals above. In this example, this means that the principal chooses agents $B$ and $D$, and after the same execution, the principal pays 1 less (6 or 0 to $B$, and 10 or $-4$ to $D$) plus
the second-price compensations. The utility of the principal is 4 minus the second price compensations.
Without the proposal of $B$, the maximin utility of the principal would have been $3$ (not counting the second-price compensations) by accepting $A$ and $C$. Therefore, $B$ gets a second-price compensation $4 - 3 = 1$. By the same reason, $D$ also gets $4 - 3 = 1$ second-price compensation.

In this example, the players had the same utilities under the first and second price mechanisms, but this was only because of our choice of the profits of the fair proposals. See Section~\ref{literature} for a general comparison.

Under both mechanisms, each agent promises a result in each case meaning that he accepts a huge penalty in the case if he provides a different result. We emphasize that this does not mean at all that the agent is forced to tell the truth. He can cheat about costs and probabilities. But even if he tells the truth in the beginning, he might be able to cheat about his decisions. E.g.\ in the case of Figure~\ref{Fig1}, if the principal asks for the right branch, the agent is free to choose the left branch without the risk of being caught.

\section{Related literature and results} \label{literature}

Let us recall what we know about the following special cases of the problem.

\bigskip \noindent
\textbf{Problem 0.} \emph{Our model for single-agent projects with fixed result: auction problem.}
\smallskip

Consider the case when the principal has to choose only one of the agents to complete the entire project, and the only possible result is the completion of the task.
This problem does nothing about cooperation failures or moral hazard, this is just a single-item auction of a negative-valued good, with a reservation value. This good is the commitment for completing the task.

If the reservation value is fixed, then the second-price single-item auction (or Vickrey-auction) implements the efficient outcome in dominant strategy equilibrium: every agent makes a bid simultaneously, and the agent with the lowest bid wins the task for the payment equal to the second lowest bid. \cite{Vickrey}
However, if the principal can set a higher reservation value, then (in the Bayesian game) this may generate a higher expected revenue for her, and the outcome may no longer be efficient.

If we want to maximize revenue for the principal, namely, we want to get the task completed for the lowest possible expected price, then it is a more difficult problem.
In the Bayesian game, the optimal prior-dependent mechanism was found by Myerson \cite{myerson1981optimal}, but it is used only as a benchmark for more realistic mechanisms.
One of the most commonly used mechanisms is the first-price single-item auction, which is simply that all agents bid with a price simultaneously, and whoever bids with the lowest price wins the task for that price.
The first-price single-item auction does not implement the efficient allocation of the good, and calculating the equilibrium strategy profile and expected revenue is a difficult task except for some simple prior distributions.
The theoretical background why this is one of the most common mechanisms in practice is not entirely complete, but this is believed to be a reasonably good solution.

First- and second-price auctions for Problems 0, A and B are special cases of our first- and second-price mechanisms.\index{mechanism!first-price mechanism}\index{mechanism!second-price mechanism}

\bigskip \noindent
\textbf{Problem A.} \emph{Our model for single-agent projects: selling the project to an agent.}
\smallskip

Consider the case when the principal has to choose a single agent to complete the entire project, but the agent works according to his dynamic stochastic capability tree with different possible results (e.g.\ completion time). Here, there is no risk of cooperation failures, but moral hazard can be a problem. For example, the agent may be free to choose a different effort level than the socially efficient one (or the first branch instead of the second one in Figure~\ref{Fig2}), because none of his actions, costs, luck (chance events) and capabilities are observable, only his result.

The valuation of the principal on the possible results (consequences) $c \in \C$ is denoted by $v\: \C \rightarrow \R$.
The following class of mechanisms $\mathcal{M}$ solves the moral hazard problem.
We use an arbitrary single-item selling mechanism where the good to sell is the contract ``if the agent provides a result $c$, then the principal will pay him $v(c)$''.
For example, using a first-price single-item auction, the mechanism is the following. Each agent submits a bid $b$ and the agent with the highest bid wins the task and will get $v(c) - b$ payment.
These mechanisms guarantee that whoever wins will be incentivized for choosing his hidden decisions (e.g.\ effort levels) in the socially efficient way.

This means that considering mechanisms from $\mathcal{M}$ reduces Problem A to Problem 0.
For example, the second-price auction version implements efficiency in dominant strategies, if the valuation function $v$ is fixed (or if the principal aims to maximize social welfare).
However, it does not mean that $\mathcal{M}$ always includes a best mechanism for maximizing the expected utility of the principal.
It is not hard to construct a (not very natural) prior distribution of the Bayesian game and a mechanism which beats all mechanisms in $\mathcal{M}$ for that prior distribution.
Accordingly, it is very hard to say anything about how good the first-price auction mechanism is compared to other mechanisms, but this is believed to be a reasonably good mechanism for this problem.

In practice, there are other very serious issues to handle, like risk-aversion of the agents.
While there are practically useful results about it, we cannot expect efficiency or any other mathematically clear result without risk-neutrality. In this paper, we will focus first on the idealized case with risk-neutral agents, but in Appendix~\ref{obsapp}, we will show that these well-known risk-sharing techniques are compatible with our mechanisms.

\bigskip \noindent
\textbf{Problem B.} \emph{Our model with trivial capability trees: combinatorial auction.}
\smallskip

Consider the case when the principal has to choose multiple agents and distribute the tasks of the project between them, and completion is the only possible result of each task.
This is a combinatorial auction problem of negative-valued goods with reservation value(s).

The second-price combinatorial auction means that every agent reports his costs of completing each subset of tasks, the principal chooses the efficient allocation of the tasks calculated from these reports, and each agent gets paid his reported costs plus the second-price compensations.
This mechanism implements the socially efficient outcome in dominant strategies if the reservation values are fixed.
However, this mechanism is extremely vulnerable to collusion if the colluding agents can submit bids which are useless alone but useful together, see Appendix~\ref{problem2}.
In contrast, the first-price combinatorial auction is collusion-resistant in the sense that forming a consortium and submitting a joint bid can be beneficial, but bidding separately with a coordinated bid is never better than forming a consortium.
Another issue is that the first-price combinatorial auction is individually rational for all players, but the second-price combinatorial auction is not (always) individually rational for the principal.

The first-price combinatorial auction is commonly used in practice, and this is believed to be a reasonably good mechanism, despite the fact that we can prove almost nothing about the level of its efficiency or the expected revenue in the general case.
The two simple results about special cases worth mentioning are that the first-price combinatorial auction is efficient (1) if the types are common knowledge, or (2) under perfect competition.

\bigskip \noindent
\textbf{Problem AB.} \emph{Our model.}
\smallskip

This problem generalizes both Problems A and B, but it also includes much harder difficulties. As we have already seen, the efficient joint strategy profile means here a high-level of cooperation. Namely, every agent should reveal all of his true abilities and chance events (e.g.\ faster or slower progress) and they should make the efficient decisions according to the entire current reported state of the project (e.g.\ always choosing the currently desired effort level). But they can lie and cheat without the risk of being caught. Moreover, in our full model, we will allow all trees to produce contractible consequences which affects the working processes of each other, and we will allow the principal to also have a capability tree.
We clearly cannot avoid the issues we had in Problems A and B, but we will manage to resolve the much more serious problem of cooperation failures.

We will show that the second-price mechanism implements efficiency in quasi-dominant equilibrium if the valuation function of the principal is fixed, although the mechanism is vulnerable to collusion.\label{collusion-resistance}
The first-price mechanism is collusion-resistant and works reasonably efficiently. However, as we saw with Problems A and B, it has to be very hard to tell that in what extent it is efficient or provide high utility for the principal.
The main message is that this mechanism eliminates the potential losses due to cooperation failures or moral hazard, and it works as well as a first-price combinatorial auction.
We will support this message firstly by proving that it works efficiently under special cases: under perfect competition or if the capability trees are common knowledge. Secondly, we will analyse the general case under practically reasonable assumptions or approximations and a reasonably strong competition between the agents.

For practice, we recommend using the first-price mechanism extended by the practical observations shown in Appendix~\ref{obsapp}.

\bigskip \noindent
\textbf{Problem C.} \emph{Our model but with no option of excluding players from participation: dynamic mechanism design.}
\smallskip

The most important difference between our model and the most relevant earlier papers about dynamic mechanism design is that we consider tendering problems.
Or equivalently, we assume that there is a special player called the principal with the power to allow or disallow each of the other players to participate, and rejected players get utility 0.
In return for this assumption, we got much stronger results.

Earlier results in dynamic mechanism design started with \emph{online mechanisms} by Friedman and Parkes (2003) \cite{FrPa}, and Parkes and Singh (2003) \cite{PaSiSa},
where agents can arrive and depart during the game with hidden utility functions. Cavallo, Parkes, and Singh (2006) \cite{CPS} proposed a Markovian environment for general allocation problems.

The most closely related results to ours are presented by Athey and Segal (2013) \cite{AtSe} and Bergemann and Välimäki (2010) \cite{BeVa}.
Both consider environments in which all players (agents) have the same role, they may have evolving private information and introduce mechanisms which implement the efficient strategy profile in perfect Bayesian equilibrium.
In contrast to our continuous-time model, these models use discrete periods of time in a way that makes the two models importantly different.
We assume a complete chronological ordering of the chance events. Namely, we assume that for any two chance events, the report of the earlier one of them cannot depend on the later event. In contrast, they considered the possibility of same-round chance events, namely, some players mutually observe the chance events (stochastic changes of private states) of each other by the time when they report their own chance events.
This extra assumption in our model makes it possible to find a mechanism that implements efficiency in a stronger equilibrium and has further important features.

Athey and Segal implement socially efficient decision rules by giving a transfer to each agent that equals the sum of the other agents' flow utilities. This works even if the agents' private signals are correlated. But with independent signals across the agents, they also provide a method of converting any incentive-compatible mechanism into a budget-balanced one. The equilibrium they propose have important weaknesses. It is not always unique, but there can also be inefficient equilibria, moreover, this mechanism is very vulnerable to collusion. Furthermore, the model assumes fixed starting states (initial types). Appendix~\ref{comparison} describes a lossy translation of our result to their language, and Appendix~\ref{AtSeComp} describes a detailed comparison.

Bergemann and Välimäki proposed a dynamic pivot mechanism where each agent gets a reward equal to his flow marginal contribution to the total utility.
They assume independent signals across agents. Their mechanism has individual rationality in a generalized sense. (This generalization is not meaningful when the principal can reject agents.) But their environment does not have private decisions, which are essential in our model. Furthermore, their equilibrium is not guaranteed to be unique, and the mechanism is vulnerable to collusion. Detailed comparison can be found in Appendix~\ref{BeVaComp}.

\section{Definitions, the model and the goals} \label{secmodel}

This section formalizes and generalizes the problem described by the Example in Section~\ref{example1}. We expect the reader to \textbf{understand the example} in detail \textbf{before reading this section}. On the other hand, if you do not completely understand something in this section, then you can go on leaning on the understanding of the examples, and come back later to the skipped part.

Section~\ref{notions} is about the mathematical precision of some basic notions. The reader can even skip it and come back to it only if some clarification is required. Section~\ref{capability tree} defines and generalizes the capability tree of an agent, including the possibilities of interdependencies like precedencies between subtasks of different agents or sharing a common resource. This will make us ready to define the model of the entire project in Section~\ref{model}.

The index of the most important terms at the last page of this paper might be useful.

\subsection{General notations and clarification of basic definitions} \label{notions}

For any symbol $x$ and any set $S$ of indices, $\bs{x_S}$ denotes the vector $(x_i)_{i \in S}$, and denotes $\bs{x}_{S^*}$ where $S^*$ is the set of indices $i$ for which $x_i$ is defined. $\bs{x}_S \in \bs{R}_S$ means $\forall i \in S\: x_i \in R_i$ (and therefore, $\bs{x} \in \bs{R}$ means $\bs{x_{S^*}} \in \bs{R_{S^*}}$). The pair $(\bs{x}_S, \bs{y}_S)$ is identified with $\big((x_i, y_i)\big)_{i \in S}$ . If $f: A \rightarrow B$ and $\bs{x}_S \in A^S$, then $\bs{f}(\bs{x}_S)$ denotes $\big(f(x_i)\big)_{i \in S}$ . $\bs{x}_{-j}$ means the vector of all $x_i$ except $x_j$, and $(y, \bs{x}_{-j})$ means the vector $\bs{x}$ by exchanging $x_j$ to $y$. The power set of $X$ is denoted by $2^X = \{Y : Y \subseteq X\}$.

Game means a dynamic game with an arbitrary ordered set of time points.
We will distinguish between information and belief, in order to be able to use belief-independence.
In Appendix~\ref{belief} we show the reason why this difficulty could not be avoided by simpler techniques.

The following definition clarifies how we will use the terms emphasized by bold text. The notation ($N^*, T_<, \Alpha, \alpha, \tilde{\alpha}, \I, \phi, \tilde{\phi}, \tau, \B, \beta, \tilde{\beta}, \xi, \Sigma$) will be used in this subsection only.

\begin{Def}
A \textbf{game form}\index{game form, game} is a tuple $(N^*, T_<, \Alpha, \alpha, \I, \phi, \B, \tilde{\beta}, \bs{\Str}, \Sigma)$ as follows.
\begin{itemize}
\item $N^*$ is the set of strategic and non-strategic \textbf{players}.
\item $T_<$ is the ordered set of \textbf{time points}.
\item $\Alpha$ is a set of possible actions.
\item $\xi^t \in \Alpha^{N^* \times \{t' \in T\: t' < t\}}$ is the \textbf{execution} (or history) until a time point $t$, where $\xi^t(i, t')$ is the action made by player $i$ at time point $t'$.
\item $\alpha: N^* \times \big\{(t, \xi^t)\: t \in T,\ \xi^t \in \Alpha^{N^* \times \{t' \in T\: t' < t\}} \big\} \rightarrow 2^{\Alpha}$ defines the \textbf{action set}, namely, $\alpha(i, t, \xi^t)$ is the set of feasible actions of player $i$ at time point $t$ after the execution $\xi^t$.
\item $\I$ is the set of possible information (or we could have called them ``information identifiers'').
\item $\phi: N^* \times \big\{(t, \xi^t)\: t \in T,\ \xi^t \in \Alpha^{N^* \times \{t' \in T\: t' < t\}} \big\} \rightarrow \I$ defines the \textbf{information} of a player at a time point. This includes the current time and the action set of the agent, namely, there exist mappings $\tilde{\alpha} : \I \rightarrow 2^A$ and $\tau : \I \rightarrow T$ such that
    \begin{align*}
        \forall i \in N^*,\ \forall t \in T,\ \forall \xi^t \in \Alpha^{N^* \times \{t' \in T\: t' < t\}}\: &&&& \tilde{\alpha}\big(\phi(i, t, \xi^t)\big) = \alpha(i, t, \xi^t), &&&& \tau\big(\phi(i, t, \xi^t)\big) = t. &&&&
    \end{align*}
\item $\B$ is the set of possible beliefs (or we could have called them ``belief identifiers'').
\item $\tilde{\beta}: N^* \times \big\{(t, \xi^t)\: t \in T,\ \xi^t \in \Alpha^{N^* \times \{t' \in T\: t' < t\}} \big\} \rightarrow \B$ defines the \textbf{belief} of a player at a time point, which includes his information, namely, there exists a $\tilde{\phi} : \B \rightarrow \I$ satisfying
    \begin{align*}
        \forall i \in N^*,\ \forall t \in T,\ \forall \xi^t \in \Alpha^{N^* \times \{t' \in T\: t' < t\}}\: &&&& \tilde{\phi}\big(\tilde{\beta}(i, t, \xi^t)\big) = \phi(i, t, \xi^t). &&&&&&&&&&&&&&&&&&&&&&
    \end{align*}
\item $\bar{\Str}_i \subset \Alpha^{\B}$ is the set of feasible 
     \textbf{pure strategies}, with a sigma-algebra $\Sigma_i$ on $\bar{\Str}_i$. A pure strategy always chooses an action from the feasible action set, namely
    \begin{align*} \index{$s_i$: strategy of player $i$}
        \forall i \in N^*,\ \forall s_i \in \Str_i,\ \forall t \in T,\ \forall \xi^t \in \Alpha^{N^* \times \{t' \in T\: t' < t\}}\: &&&& s_i\big(\tilde{\beta}(i, t, \xi^t)\big) \in \alpha(i, t, \xi^t). &&&&&&&&&&&&&&
    \end{align*}
    Furthermore, each pure strategy profile uniquely determines the \textbf{execution}\index{execution $\l_i$ of a capability tree}\index{$\l_i$: execution of $\theta_i$} $\xi$, namely, \smallskip
    \\ $\forall \bs{s} = \bs{s}_{N^*} \in \bs{\Str}_{N^*},\ \exists !\ \xi : N^* \times T \rightarrow \Alpha,\ \forall i \in N^*,\ \forall t \in T\:$
    \begin{equation} \label{consistency}
        s_i\Big(\tilde{\beta}\big(i, t, \xi|_{N^* \times \{t' \in T\: t' < t\}}\big)\Big) = \xi(i, t).
    \end{equation}
\item $\Sigma_i$ is a sigma-algebra on the set of pure strategies of each player $i$. $\Str_i$ denotes the probability distributions on $(\Str_i, \Sigma_i)$, called (mixed) \textbf{strategies}.\index{$\Str_i$: strategy set of player $i$} 
\end{itemize}
\end{Def}

A \textbf{game} is a game form with a set $N \subseteq N^*$ of \textbf{strategic players} and a measurable utility function $u : N \times \Alpha^{N^* \times T} \rightarrow \R$. This assigns a real utility $u_i(\xi)$ to each strategic player $i \in N \subseteq N^*$ and each execution $\xi \in \Alpha^{N^* \times T}$. The utility function should satisfy that for each $\bs{\sigma} \in \bs{\Delta\Str}$, if we choose each pure strategy $s_i$ from the (mixed) strategy $\sigma_i$ independently, then the expected utilities are finite (and exist), namely, $E_{\bs{s} \in \bs{\sigma}}\big(\bs{u}(\bs{s})\big) \in \R^N$.

Note that we defined game (form) in the sense of complete-information game (form).

An \textbf{incomplete-information game} can be identified with a set $\Gamma$ of games with the same set of players, and each player $i$ has the same strategy set $\Str_i$ in all games $G \in \Gamma$. Playing an incomplete-information game means that first each player chooses a strategy $s_i \in \Str_i$, then nature chooses one of the games $G \in \Gamma$ and $G$ will be played with $\bs{s}$. By default, the choice of $G$ is nondeterministic, therefore, the utility function is $u : \Gamma \times N \times \Alpha^{N^* \times T} \rightarrow \R$ is a function of $G$.

The \textbf{Bayesian game} is an incomplete-information game along with a probability distribution on $\Gamma$ satisfying that for each $\bs{\sigma} \in \bs{\Delta\Str}$ and $i \in N$ and an initial information $\theta_i \in \I$, the expected utilities are finite, namely, $E\big(u_i(\bf{s}) \bigwhere \theta_i \big) \in \R$.

\textbf{Subgame} form \index{game form, game!subgame} means the game form after some periods of the game.
We define it only for game forms with perfect information, namely, when $\phi$ is the identity function.
Formally, for a game form with perfect information $(N^*, T_<, \Alpha, \alpha, \bs{\Str})$, an upper-closed set of time points $T' \subseteq T$ (i.e. with $t' \in T',\ t \in T,\ t' < t \Rightarrow t \in T'$) and a partial execution $\xi^{T \setminus T'} : N^* \times (T \setminus T') \rightarrow A$, subgame is defined as a tuple $\big(N^*, T'_<, \Alpha, \alpha|_{N \times T'}, \bs{\Str'}\big)$ where $\bs{\Str'}$ is the non-empty set of strategy profiles which are consistent with the partial execution $\xi^{T \setminus T'}$, namely,
\begin{align*}
    \bs{s} \in \bs{\Str'} && \Leftrightarrow && \bigg( \forall i \in N,\ \forall t \in T \setminus T'\: &&
    s_i\Big(\tilde{\beta}\big(i, t, \xi|_{N \times \{t' \in T\: t' < t\}}\big)\Big) = \xi(i, t) \bigg).
\end{align*}

In subgame form structures (e.g.\ subgames), further but analogous consistency conditions should apply (e.g.\ for the utility function $\bs{u'}\big(\xi'(\bs{s})\big) = \bs{u}\big(\xi(\bs{s})\big)$ ). A subgame ``preceding'' or ``after'' a time point $t$ (or an event at $t$) means the subgame with $T' = \{t' \in T: t' \ge t\}$ or $T' = \{t' \in T: t' > t\}$, respectively.
A \textbf{state}\index{state: identified with subgame} of a game is a synonym of the subgame of the game with perfect information (from a specific time point, or after an event, etc.).
The dependence of any function or relation $f$ on the game $G$ is denoted by the form $f^{G}$, but we may omit this superindex when it causes no confusion.

We say that $i$ does not use his beliefs, or $s_i$ is belief-independent if
$\exists s_i^{\prime} \in \Alpha^{\I},\ \forall \beta \in \B\:\ s_i(\beta) = s_i^{\prime}\big(\tilde{\phi}(\beta)\big)$.
The term of common knowledge refers to the information (not just the beliefs) of the players.

\subsubsection{Ignored mathematical details}

We have some implicit technical assumptions. All of them will be true for finite games, and most of the readers should simply ignore them. We list these assumptions here because we will not reduce the readability of the paper with mentioning these issues later.

The first technical assumption was \eqref{consistency}. We note that it is not automatical, this could fail e.g.\ in continuous-time games. This is the reason why we allowed restrictions on the strategy sets (i.e. $\Str_i \subset {\Alpha}^{\B}$ and not $\Str_i = {\Alpha}^{\B}$). There are different ways to add such restrictions, but we have no universal one. In this paper, any restriction could work as long as this implies \eqref{consistency} and allows some specific simple strategies, e.g.\ the truthful strategy.

We will assume that all expected values exist and finite.
We will also assume that all suprema are maxima: there always exists a strategy profile maximizing the total expected utility.
Finally, we assume that the conspiracy-fearing strategy of the principal is always well-defined: the max-min theorem applies for all possibly occurring two-player zero-sum games.
These will give implicit restrictions on the set of capability trees and proposals, just like \eqref{consistency} gives a restriction on the possible strategy sets.

\subsection{Capability tree} \label{capability tree}

The (initial) type of each player including the principal will be a rich structure, describing his entire working capabilities and preferences.
We call them capability trees rather than types.
Loosely speaking, the capability tree of a player is a dynamic description of
\begin{itemize}
\item what decisions he can make during his work, e.g.\ his options about which technology he uses, how much effort he makes, or how many people he employs;
\item what stochastic feedbacks from his working process he expects to observe, e.g.\ faster progresses or failures;
\item how his task is affected by the consequences (externalities) of other tasks, e.g.\ completion time of a preceding subtasks, reservation of a common resource;
\item what consequences the task provides, e.g.\ the examples above, or the result (main output) of his entire working process;
\item the player's hidden total costs or benefits.
\end{itemize}

\medskip

The working process of $i$ can be influenced through contractible consequences of others. But we want to define the working capabilities independently from the other players. This is the reason why we include an abstract player called the world. The effects of the consequences by others on the working process of $i$ will be expressed by the moves of the world in $i$'s capability tree.


\begin{Def} \label{capability treedef}
A \textbf{capability tree} $\theta_i$\index{capability tree $\theta_i$}\index{$\theta_i$: capability tree of player $i$} is defined as a 3-player game form (or 2-player stochastic game form) with a consequence function $c$ and a valuation function $v$, which satisfies the following properties.
\begin{itemize}
\item The players are: the worker, the world and nature. (They personalize the working efforts of player $i$, the externalities of others to this capability tree, and the chance events, respectively.)
\item The set of time points is $I$, the same set for all capability trees.
\item  At each time point in $I$, the action set of the world is a globally fixed set $\C$, where $\emptyset \in \C$ represents doing nothing, and if $X, Y \in \C$, then $X \cup Y \in \C$. \\ (The action sets of the worker and nature have no restrictions. These can be arbitrary functions of the earlier moves of all three players, at each time point.)
\item Nature makes its moves with a probability distribution determined by the earlier moves of the three players of the capability tree. These are called the \textbf{chance events}\index{chance event}, at a \textbf{chance node}\index{chance node} of the capability tree.
\item At each time point $t \in I$, the capability tree outputs a consequence $c(\theta')$, where $c\: \Theta' \rightarrow C$ and $\Theta'$ denotes the set of possible states of the capability tree.
\item The valuation function\index{valuation $v(\l_i)$} $v\: \A^{\{\text{worker, nature, world}\} \times I} \rightarrow \R$ assigns a real number (money) $v(\l_i)$ to the \textbf{execution} \index{execution $\l_i$ of a capability tree}\index{$\l_i$: execution of $\theta_i$} $\l_i$.
\end{itemize}
We say that a capability tree is \textbf{omissible}\index{capability tree $\theta_i$!omissible capability tree} (or idle-allowing) if it has the following properties.
\begin{itemize}
\item The worker has a distinguished strategy, called the do-nothing strategy, that provides no consequence (provides consequence $\emptyset$ at all points in time), and ends with valuation 0, whatever nature and the world do.
\end{itemize}
\end{Def}

\bigskip

For example, the capability trees of the players in the Example (Section~\ref{example1}) can be defined in the following way.
\begin{center}
\begin{tabular}[center]{c||c|c|c|}
\textbf{Example 1}& capability tree: game form & consequences & valuation \\
\hline \hline
\multirow{3}{*}{principal}
& \multirow{2}{*}{only the world} & \multirow{2}{*}{$\emptyset$} & if the union of the moves \\
& \multirow{2}{*}{makes actions} & \multirow{2}{*}{in all time points} & of the world was \{{\color{green} \ding{51}$_1$},{\color{green} \ding{51}$_2$}\}, \\
& & & then 100, otherwise 0 \\
\hline
\multirow{2}{*}{Agents $A$, $B$} & \multirow{2}{*}{Figure~3 without results} & before the end: $\emptyset$, & \multirow{3}{*}{the negative of} \\
& \multirow{2}{*}{and costs at the leaves;} & at the end: {\color{green} \ding{51}$_1$} or {\color{red}\ding{53}$_1$} & \multirow{3}{*}{the cost at the leaf} \\
\cline{1-1}
\cline{3-3}
\multirow{2}{*}{Agents $C$, $D$} & \multirow{2}{*}{the world does nothing} & before the end: $\emptyset$, & \\
& & at the end: {\color{green} \ding{51}$_2$} or {\color{red}\ding{53}$_2$} & \\
\hline
\end{tabular}
\end{center}

To be more precise, to the capability tree of each agent in Example 1, we should include an option of doing nothing, always providing consequence $\emptyset$ and having valuation 0. Hereby the capability tree of each agent will be omissible. When the principal rejects an agent, this means that she enforces the agent to choose the do-nothing strategy.

We define the \emph{empty capability tree}\index{capability tree $\theta_i$!empty capability tree} to be the capability tree which always provides consequence $\emptyset$, and the valuation is constantly 0. Having an empty capability tree means having no capability to do any work.

A capability tree of an agent should be interpreted so as it describes the dynamics of the \emph{knowledge of the agent about his working process}.
Therefore, the assumption that each agent has perfect information about his own capability tree should be interpreted as his knowledge about his own capabilities weakly dominates the joint knowledge of everybody else about it.

\subsection{The model} \label{model}


The primary motivation of the following definition is to describe a general environment where we want to get a particular project completed by some agents.
We divide the owner of the project to a planner and a principal. The planner personalizes the behavior of the owner of the project that she can commit to, and the principal personalizes the strategic owner.

\bigskip

We define the Project Management Model,\index{Project Management Model $\G$} or in short, the \textbf{\underline{project}}\index{project $\G$}, denoted by $\G$\index{$\G$: project}, as an incomplete-information game consisting of (complete-information) games satisfying the following conditions.

There is a \underline{player} $0$ called the \textbf{principal}\index{principal (player $0$)} and players $1, 2, ..., n$ called \textbf{agents}. The set of all strategic players is $\N = \{0, 1, 2, ..., n\}$, and the set of agents is $\N^+ = \{1, 2, ..., n\}$\index{$\N^+ = \{1, 2, ..., n\}$: set of all agents}\index{$\N = \{0, 1, 2, ..., n\}$: set of all players}. $N^* = N \cup \{\text{nature}\}$.
There are two non-strategic players: the \textbf{planner}\index{planner} and \textbf{nature}.
The actions of nature are chosen with given probabilities. The planner commits to a strategy in advance.
From now on, the word \emph{``player''} will refer to the strategic players, and $\Str_i$\index{$\Str_i$: strategy set of player $i$} will denote the set of mixed strategies of the players, $\bs{\Str} = \bs{\Str}_N$.
In each game $G \in \G$, the principal has a capability tree $\theta_0$ and each agent $i$ has an omissible capability tree $\theta_i$.\footnote{Or some of the agents may get non-omissible but verifiable capability trees, see Note~\ref{verifiable-capability-tree}.}
The set of time points $T \supseteq \{t_0, t_1\} \cup I$ includes $I$ and at least two earlier time points $t_0$ and $t_1$, where $t_0$ is the very first time point in $T$.

The \underline{actions}\index{action} of the players and of nature include the following.
\begin{itemize}
\item The players can send arbitrary contractible time-stamped instant messages to the planner at each real time point and vice versa.
\item The capability trees of all players are concurrently executed during $I$, where each player controls the worker of his capability tree, except that, at the time point $t_1$, the planner can choose to enforce him following the do-nothing strategy. The actions of the world in each capability tree are defined as the union of the same-time consequences of other capability trees. Each move of nature in a capability tree is made with the given probabilities, independently of everything else up to the current time.\footnote{If a chance event is contractible (e.g.\ currency exchange rates are of this kind), then we can allow dependence of this chance event with the same-time messages of nature to other players, see Appendix~\ref{infchance}.}
\item At the end (at a time point after $I$), the planner determines the \textbf{payment} (or transfer) vector\index{payment $p_i$} $\bs{p} \in \R^{\N}$\index{$p_i$: payment to player $i$} as a function of the contractible events. This must be balanced (including the principal):
\begin{equation} \label{balanced}
\sum\limits_{i \in \N} p_i = 0.
\end{equation}
\end{itemize}

The \underline{information} of each player $i$ consists of his perfect information in his capability tree and all messages he receives. Namely:
\begin{itemize}
\item He has perfect information in his own capability tree $\theta_i$. In detail, he knows $\theta_i$ and all strictly earlier moves of the three players in it, and his chance event at the current time.\footnote{This is just an unimportant condition for mathematical convenience that we identify the time point of the move of nature with the first time point when the player can react to it. With the formalism introduced in Section~\ref{notions}, this can be expressed so as the set of time points is (a subset of) $\R \times \{0, 1\}$ with the lexicographical ordering, where nature makes actions only at $\big\{(t, 0): t \in \R\big\}$, and the players make actions only at $\big\{(t, 1): t \in \R\big\}$.}
\item His information contains all messages he received from the planner strictly previously.
\end{itemize}

The \underline{belief} of each player at each time point is assumed to be independent from the concurrent chance events of other players, conditional on the entire past.\footnote{Conditional independence between beliefs and \emph{later} chance events is already implied by the independence assumption on the moves of nature.} This expresses that each player can report the outcome of each of his chance events before any correlated event happens outside of the capability tree. This is our only assumption about beliefs.

By default, the information of the planner is the consequences by all players and the messages she receives until the current time. As a non-default version of the model, the planner may be informed also about the capability tree of the principal. (The planner will not use beliefs. The information and belief of nature is irrelevant as long as it can choose the outcome of the chance events with the desired probabilities.)

The \underline{utility}\index{utility $u_i$} $\bs{u}$\index{$u_i$: utility of player $i$} of the players are defined as follows.
\begin{align} \label{utility}
\forall G \in \G,\ \forall \xi \in \A^{N^* \times T},\ \forall i \in \N \:
&&&& u_i(G, \xi) &= v(\theta_i, \xi_i) + p_i(G, \xi),
&&&&&&&&&&&&&&
\end{align}
where $\xi \in \A^{N^* \times T}$ is the execution of $G$ and $\xi_i = \xi_i(\xi) \in \A^{\{\text{worker, nature, world}\} \times I}$ is the execution of $\theta_i$, the capability tree of $i$.
We will use the alternative notation $u_i(G, \bs{s}) = u_i\big(G, \xi(G, \bs{s})\big)$.



\bigskip

\underline{Interpretation.} The owner of the project should agree with the agents about the payment rules, namely, how the payments depend on the achievements (consequences) of the players and on the messages they send to each other during the work. Agreement is optional: on one hand, individual rationality of the agents will mean that no agent can be punished if he says no and uses his do-nothing strategy. On the other hand, the planner will be free to reject each agent. Then all agents who agreed with the owner, as well as the owner concurrently execute their capability trees. At the end, the agents are paid according to the agreements. These payments must depend only on the consequences of the working processes and the communication. Hidden efforts, chance events and other privately observable things of the capability tree cannot directly affect the payment.

Using no prior about the capability trees of the players expresses that we are looking for ex post equilibria with respect to these capability trees.
Similarly, we did not specify the beliefs of the players because our solution concept will be robust to them.
Therefore, the Project Management Model includes environments when different forms of communication like signaling and cheap talk are also possible.

\subsection{Goal} \label{Goalsection}

For any $S \subset \Pl$, let $u_S = \sum\limits_{i \in S} u_i$\index{$u_i$S@$u_S$: total utility of all players in $S \subseteq \Pl$}, and we call $u_{\Pl}$ the \textbf{total utility}\index{utility $u_i$!total utility $u_{\Pl}$}. Clearly,
\begin{equation}
u_{\Pl}(G, \xi) \mathop{=}^{\eqref{balanced}\eqref{utility}} \sum_{i \in \N} v(\theta_i, \xi_i). \label{psum}
\end{equation}

For a finite set of capability trees $\btheta_S$, we define an optimization problem as follows. The capability trees are executed simultaneously, one player controls the worker in all trees, and the goal is to maximize $\E\big(\sum\limits_{i \in S} v(\theta_i, \xi_i) \big)$. We denote this maximum by $V(\btheta_S)$. Clearly,
\begin{align} \label{V-opt}
  \forall G \in \G,\ \forall \bs{s} \in \bs{\Str}\: && \E\big(u_N(G, \bs{s})\big) &\le V(\btheta) && \phantom{\forall G \in \G,\ \forall \bs{s} \in \bs{\Str}\:}
\end{align}

We want to design a mechanism under which a strategy profile $\bs{s^*}$ satisfies the following goals.
\begin{itemize}
\item \emph{Efficiency (if all players are risk-neutral):} $\forall G \in \G\:\ \E\big(u_{\Pl}(G, \bs{s^*})\big) = V(\btheta)$.
\item \emph{Incentive compatibility:} $\bs{s^*}$ is a ``convincing'' equilibrium (see Section~\ref{justification}).
\item \emph{Collusion-resistance} (see Section~\ref{results1})\index{collusion-resistance}.
\item \emph{Individual rationality (or optional participation) for each agent.}
\item \emph{Avoids free-riding:} $\forall G \in \G,\ \forall i \in \Ag\: \text{ if } \theta_i = 0 \text{, then } \E\big(u_i(\bs{s^*})\big) = 0$\index{free-riding}.
\end{itemize}

We will achieve these goals except that either we have to give up collusion-resistance, or we will achieve weaker goals about efficiency.

\section{The mechanisms}

A \textbf{mechanism}\index{mechanism} is identified with the strategy of the planner.
We define two versions of the mechanism, because there is a tradeoff between them. Their relation will be analogous to the relation between the first and second-price auctions (see Section~\ref{literature}).

\subsection{The first-price mechanism}\index{mechanism!first-price mechanism}

We define the first-price mechanism in the following way.
Denote the space of messages by $\M$.\index{$\M$: message space}
We define contract (referring to a contractible payment rule) as a function $\C^I \times \M^{\{0, i\} \times I} \rightarrow \R$ that determines the payment between the principal and an agent $i$, depending on the consequences provided by all players and the communication between them (at all time points $t \in T$ of the game $G$). At the beginning, each agent sends a contract to the principal, we call it $i$'s proposal $\pi_i$. Then the principal accepts some of them, and forces the other agents to use the do-nothing strategy (they are removed from the rest of the game with utility 0). The only role of the planner is that she observes the communication including the proposals, and she determines the payments at the end, according to the accepted contracts. It will not matter whether these messages are observable by other players.

\subsection{The conspiracy-fearing strategy of the principal}

We define the ``conspiracy-fearing'' or \textbf{truthful strategy} of the principal $\cp_0 = \cp_0(\theta_0)$. \index{strategy $s_i$!truthful strategy $\cp_i$}
After the principal receives the proposals, she uses the strategy out of all her possible strategies by which \emph{her minimum possible expected utility is the largest}, in the following sense.
After the proposals are submitted, we define the following two-player game, called \textbf{principal-devil game} $D(\theta_0, \bs{\pi})$.\index{principal-devil game}
One player is the principal with the same capability tree. The other player is called the devil who controls the moves of all agents, and we replace their capability trees with universal capability trees. Namely, the devil has the capability to provide arbitrary consequences and to send messages in the name of each agent. The devil observes everything in the past including the capability tree of the principal and its execution. The payments at the end are determined according to the contracts. The aim of the devil is to minimize the expected utility of the principal, while the principal aims to maximize it. This is a two-player zero-sum game with perfect information, therefore, the principal has a maximin strategy which is belief-independent. The conspiracy fearing strategy of the principal in our original game is defined so as she makes the very same moves as what she would do by the maximin strategy in this game against the devil. We call the value of the principal-devil game the \textbf{maximin utility}\index{utility $u_i$!maximin utility} of the principal. Correspondingly, we define the \textbf{joint value} $V(\bs{\pi}_S) = V(\theta_0, \bs{\pi}_S)$ \textbf{of a set} $\bs{\pi}_S$ \textbf{of proposals} as the maximin utility of the principal if she accepts this set of proposals. Therefore,
\begin{align}\label{V(pi)-ineq}
  \forall G \in \G,\ \forall \bs{s}_{-0} \in \Str_{\Ag}\: &&&&
  \E\Big(u_0\big(G, (\cp_0, \bs{s}_{-0}) \big) \Big) &\ge V\big(\theta_0, \bs{\pi}(G, \bs{s}_{-0})\big), &&&&&&&&&&&&&&
\end{align}
(and $V$ is the largest value such that there exists a strategy denoted by $\cp_0$ with this property, provided that $\G$ is rich enough).

\subsection{The second-price mechanism}\index{mechanism!second-price mechanism}


The \textbf{value of a proposal} $\pi_i$\index{value of a proposal $v^+_i = v^+_i(\theta_0, \bs{\pi})$} is the marginal contribution of the proposal to the maximin utility:
\begin{equation} \label{vpdef}
v^+_i = v^+_i(\bs{\pi}) = v^+_i(\theta_0, \bs{\pi}) = V(\theta_0, \bs{\pi}) - V(\theta_0, \bs{\pi}_{-i}).
\end{equation}\index{$v^+_i = v^+_i(\theta_0, \bs{\pi})$: value of a proposal $\pi_i$}

The second-price mechanism is the same as the first-price mechanism except that the principal pays $v^+_i$ more to each agent $i \in \N^+$, namely,
\begin{equation*}
p_i(G^2, \xi) = p_i(G^1, \xi) + v^+_i(\theta_0, \bs{\pi}),
\end{equation*}
where $G^1$ and $G^2$ are defined precisely in the next subsection.


We note that this definition requires that the planner knows $\theta_0$. This is not a more serious assumption than the same for single-item auctions, see Section~\ref{literature}.

\subsection{Definitions and notations}

For a game $G$ let $G/m$ denote the game given the mechanism $m$, namely, the planner is no longer a non-strategic player, but his strategy becomes a part of the rules of the game.
Let
$G/m$ denote the game $G$ given the mechanism $m$, and $\G/m = \big\{ G/m \where G \in \G \big\}$ denote the incomplete-information game $\G$ given the mechanism $m$. Let $m_1$ and $m_2$ denote the first and the second-price mechanisms, respectively, and $G^i = G/m_i$ and $\G^i = \G/m_i$\index{$\G^i$: project $\G$ under the first- or second-price mechanism}\index{$G^i$: game $G$ under the first- or second-price mechanism}.
All notions defined in $G$ and $\G$ are used correspondingly in $G^i$ and $\G^i$.
From now on, we will omit the quantifier $\forall G \in \G$.

We identify the game forms and the strategy sets of $G^1 = G/m_1$ and $G^2 = G/m_2$;\footnote{Here we treat payments not as actions but a part of the definition of the utility functions.} the only difference between the two games is the utility function.
\begin{align}
\forall i \in \Ag\:&& u_i(G^2, \xi) &= u_i(G^1, \xi) + v^+_i(\bs{\pi}), &&\phantom{\forall i \in \Ag\:} \label{p2i}
\\ && u_0(G^2, \xi) &= u_0(G^1, \xi) - \sum_{i \in \Ag} v^+_i(\bs{\pi}). &&  \label{p2C}
\end{align}

We say that the principal rejects an agent $i$ when she forces him to follow the do-nothing strategy. Otherwise she accepts $i$.
The set of accepted agents is denoted by $\Acc$ and $\Acc_0 = \Acc \cup \{0\}$.\index{$\Acc_0 = \Acc \cup \{0\} =$ set of accepted agents and the principal}

\section{Results}

The Reader should refer to Problems 0, A, B and AB in Section~\ref{literature} where the results were already summarized, focusing on the interpretation. Here, we summarize the results in more detail but with less interpretation.

We focus only on the goals listed in Section~\ref{Goalsection}. In Appendix~\ref{obsapp}, we will see some practically important extensions of these goals.


The \textbf{truthful proposal}\index{proposal $\pi_i$!truthful proposal $\pi^*_i$} (or cost price proposal) $\pi^*(\theta)$ of a capability tree $\theta$ were informally described in Section~\ref{offer-example}. The precise mathematical formulation of it is the following.
To each capability tree $\htheta$, the proposal $\pi^*(\htheta)$ of an agent $i \in N^+$ is defined as the following function $\C^I \times \M^{\{0, i\} \times I} \rightarrow \R$.
Let $\A$ and $\X$ denote the action set of the worker and nature in $\htheta$, respectively.\footnote{Without loss of generality, we can assume that the action sets are the same at each point in time.}
We fix surjective dictionary functions $(a, w): \M \rightarrow \A \times \R^{\X}$ and $x: \M \rightarrow \X$.
We use the notations $a^t = a(m_{0 \rightarrow i}^t)$, $w^t = w(m_{0 \rightarrow i}^t)$, $x^t = a(m_i^t)$, where $m_{0 \rightarrow i}^t$ and $m_i^t$ are the messages of the principal and the agent at time point $t \in T$, respectively.
Let $\xi(\htheta, \bs{a}, \bs{x}, \bs{c})$ denote the state of the game $\htheta$ after the (complete or partial) execution $\bs{a}$, $\bs{x}$ and $\bs{c}$ denoting the history of actions of the worker, nature and the world, respectively.
A vector $w_t$ is legal if
\begin{equation}\label{Ew=0}
  \E_{x_t}\big(w_t(x_t) \bigwhere \xi(\htheta, \bs{a}^{<t}, \bs{x}^{<t}, \bs{c}^{<t}) \big) = 0.
\end{equation}
If the principal sends an illegal vector, then we replace it to the legal vector $w \equiv 0$.
Now for a $\bs{c} \in \C^I$, $\bs{m} \in \M^{\{0, i\} \times I}$, we define
\begin{equation} \label{pi-star}
\pi^*(\htheta)(\bs{c}, \bs{m}) = - v\big(\xi(\htheta, \bs{a}, \bs{x}, \bs{c})\big) + \sum_{t \in T} w_t(x_t)
\end{equation}
The truthful proposal is defined as $\pi^*_i = \pi^*(\theta_i)$.
\index{$\pi^*_i$: truthful proposal of agent $i$}

We will assume an infinitesimal flexibility on the proposals in the following sense. In the degenerate case when two chance nodes $X$, $Y$ of the capability trees of two different agents have the same time points, then the proposals should let the principal to consider one of the chance nodes, say, $X$ earlier than $Y$, and to choose the vector $w_Y$ for $Y$ dependent on the chance event of $X$. Alternatively, this can be handled by defining $w_Y$ as a function of the chance event of $X$. (And analogously for multiple coincides.)

We define the \textbf{truthful strategy} $\cp_i = \cp(\theta_i)$\index{strategy $s_i$!truthful strategy $\cp_i$}\index{$\cp_i$: truthful strategy of player $i$} (or cost price strategy) of a player $i \in N$ as follows. He sends the proposal $\pi^*_i$, and then, at each time point $t \in I$, he always makes the move $d(m_0^t)$ according to the message $m_0^t$ of the principal, and he reports the true outcome $\chi$ of the chance event (sends a message from $d^{-1}(\chi)$) at each chance node.

Despite the fact that no dominant strategy equilibrium exists in such a general model (see Appendix~\ref{nodominant}), the second-price mechanism implements the efficient strategy profile in a slightly weaker equilibrium called quasi-dominant equilibrium.

\begin{Theorem} \label{equilibrium2}
If the planner knows the capability tree of the principal, then the truthful strategy profile $\mathcal{\bs{\cp}}$ maximizes the expected total utility of the players, and $\mathcal{\bs{\cp}}$ is a quasi-dominant equilibrium under the second-price mechanism.
The mechanism is individually rational and avoids free-riding.
\end{Theorem}

This means that the second-price mechanism satisfies all of our goals except individual rationality for the principal and collusion resistance.
This mechanism does not achieve these two goals because of the second-price compensations, and this issues are already present in the special cases of the second-price single-item and combinatorial auctions (Problems 0 and B), respectively.
We note that Conjecture~\ref{conjecture} can be interpreted so as the mechanism "works well" without the assumption that the planner knows the capability tree of the principal.

\bigskip

Similarly to the auction problems mentioned in Section~\ref{literature}, under the first-price mechanism, the agents should follow almost the truthful strategies but with asking for higher payments in their proposals.

Compared to the second-price mechanism, the first-price mechanism is
\begin{itemize}
  \item collusion-resistant in a particular sense, see Section~\ref{collusion},
  \item individually rational also for the principal (but avoids free-riding in a slightly weaker sense), see Section~\ref{indiv-rat-0},
  \item but in exchange, this implements only an approximately efficient outcome, see Section~\ref{leveleff}.
\end{itemize}

In Section~\ref{results2}, we will prove Theorem~\ref{equilibrium2} with Nash-equilirium.
In Section~\ref{results1}, we will explain the three bullet points about the first-price mechanism.

\begin{Note}[Non-rejectable agents with verifiable capability trees] \label{verifiable-capability-tree}
{\normalfont Consider the extension of the Project Management Model so that some agents cannot be rejected but their capability trees are observable by the planner. We emphasize that the executions of their capability trees are still hidden. All of our results extend to this case with essentially the same proof (except individual rationality for these agents, which is not a reasonable goal for them): we only need to assume that each of those agents $i$ is forced to submit his truthful proposal $\cp_i$ and we define $v^+_i = 0$, namely, he does not get any second-price compensation. Or $v^+_i$ could also be defined as an arbitrary function of the proposals $\bs{\pi}$.
}
\end{Note}

\subsection{Proofs about the second price mechanism} \label{results2}

First, we state Lemma~\ref{u(cp)=0-thm} and Proposition~\ref{Veq-thm}, and we will prove them after showing how and why these imply the equilibrium.

\begin{Lemma} \label{u(cp)=0-thm}
If an agent uses the truthful strategy, then his expected utility will be 0, no matter what the other players do. Formally,
\begin{align}\label{u(cp)=0}
  \forall i \in \Ag,\ \forall \bs{s}_{-i} \in \bs{S}_{-i}\:&&&&
  \E \Big( u_i\big(G^1, (\cp_i, \bs{s}_{-i})\big) \Big) &= 0.
  &&&&&&&&&&&&&&
\end{align}
\end{Lemma}

\begin{Proposition} \label{Veq-thm}
  \begin{equation}\label{Vineq}
    V\big(\theta_0, \pi^*(\bhtheta_{-0})\big) \ge V\big((\theta_0, \bhtheta_{-0})\big)
  \end{equation}
\end{Proposition}

\begin{Theorem} \label{efficient2}
  \eqref{Vineq} is true with equality, and therefore, the truthful strategy profile $\bs{\cp}$ is efficient. Namely,
  \begin{equation}\label{Veq}
    V\big(\theta_0, \pi^*(\bhtheta_{-0})\big) = V\big((\theta_0, \bhtheta_{-0})\big),
  \end{equation}
  \begin{equation} \label{eq:efficient}
    \E\big(u_N(G^2, \bs{\cp})\big) = V(\btheta).
  \end{equation}
\end{Theorem}

\begin{proof}
  \begin{equation*}
    \E\big(u_N(G^2, \bs{\cp})\big)
    \mathop{=}^{\eqref{p2i}\eqref{p2C}}
    \E\big(u_N(G^1, \bs{\cp})\big)
    \mathop{=}^{\eqref{u(cp)=0}}
    \E\big(u_0(G^1, \bs{\cp})\big)
    \mathop{\ge}^{\eqref{V(pi)-ineq}}
    V\big(\theta_0, \bs{\cp^*}(\btheta_{-0})\big)
    \mathop{\ge}^{\eqref{Vineq}}
    V(\btheta)
    \mathop{\ge}^{\eqref{V-opt}}
    \E\big(u_N(G^2, \bs{\cp})\big) \qedhere
  \end{equation*}
  This proves \eqref{eq:efficient}, and if we apply this for a game with the capability trees $(\theta_0, \bhtheta_{-0})$, then we can conclude \eqref{Veq}, as well.
\end{proof}

\begin{Theorem} \label{Nash-proof}
  The truthful strategy profile $\bs{\cp}$ is a Nash-equilibrium under the second-price mechanism, provided that the planner knows the type of the principal.
\end{Theorem}

\begin{proof}
  We prove that if only one player $i$ deviates, then it does not decrease $u_{\Pl \setminus \{i\}}$, the total expected utility of all other players. Namely,
  \begin{align}\label{deviate}
    \forall i \in \Pl,\ \forall s_i \in \Str_i\: &&&&
    \E\Big(u_{N \setminus \{i\}}\big(G^2, (s_i, \bs{\cp}_{-i})\big)\Big) &\ge
    \E\big(u_{N \setminus \{i\}}(G^2, \bs{\cp})\big) &&&&&&&&&&&&
    \end{align}
  This will imply that $\bs{\cp}$ is a Nash-equilibrium, because
  \begin{equation*}
    \E\Big(u_i\big(G^2,(s_i, \bs{\cp}_{-i})\big)\Big)
    = \E\Big(u_N\big(G^2,(s_i, \bs{\cp}_{-i})\big)\Big) - \E\Big(u_{N \setminus \{i\}}\big(G^2,(s_i, \bs{\cp}_{-i})\big)\Big)
  \end{equation*}
  \begin{equation*}
    \mathop{\le}^{\eqref{V-opt}\eqref{deviate}} V(\btheta) - \E\big(u_{N \setminus \{i\}}(G^2, \bs{\cp})\big)
    \mathop{=}^{\eqref{u(cp)=0}\eqref{Veq}} \E\big(u_N(G^2, \bs{\cp})\big) - \E\big(u_{N \setminus \{i\}}(G^2, \bs{\cp})\big)
    = \E\big(u_i(G^2, \bs{\cp})\big).
  \end{equation*}
  If only the principal deviates, then for all agent $i \in \Ag$,
  \begin{equation*}
    \E\Big(u_0\big(G^2,(s_0, \bs{\cp}_{-0})\big)\Big)
    \mathop{=}^{\eqref{p2i}}
    \E\Big(u_i\big(G^1,(s_0, \bs{\cp}_{-0})\big)\Big) + \sum_{i \in \Ag} v^+_i\big(\theta_0, \bs{\pi^*}(\btheta_{-0})\big)
    \mathop{=}^{\eqref{u(cp)=0}} \sum_{i \in \Ag} v^+_i\big(\theta_0, \bs{\pi^*}(\btheta_{-0})\big),
  \end{equation*}
  which is invariant of $s_0$. This implies \eqref{deviate} for the principal, because both sides are the same.

  For an agent $i \in \Ag$, the proof of \eqref{deviate} is the following.
  \begin{equation*}
    \E\Big(u_{N \setminus \{i\}}\big(G^2, (s_i, \bs{\cp}_{-i})\big)\Big)
    \mathop{=}^{\eqref{p2i}\eqref{p2C}}
    \E\Big(u_{N \setminus \{i\}}\big(G^1, (s_i, \bs{\cp}_{-i})\big)\Big) - v^+_i\Big(\theta_0, \big(\pi(G, s_i), \bs{\pi^*}(\btheta_{-0-i})\big)\Big)
  \end{equation*}
  \begin{equation*}
    \mathop{=}^{\eqref{u(cp)=0}\eqref{vpdef}}
    \E\Big(u_0\big(G^1, (s_i, \bs{\cp}_{-i})\big)\Big) - V\big(\theta_0, \bs{\pi}(s_i, \bs{\cp}_{-i})\big) + V\big(\theta_0, \bs{\pi}_{-i}(\bs{\cp}_{-i})\big)
    \mathop{\ge}^{\eqref{V(pi)-ineq}} V\big(\theta_0, \bs{\pi}_{-i}(\bs{\cp}_{-i})\big). \qedhere
  \end{equation*}
\end{proof}

We will show that Lemma~\ref{u(cp)=0-thm} and Propostion~\ref{Veq-thm} imply that $\mathcal{\bs{\cp}}$ is an equilibrium in a very strong sense.
We will introduce this concept called quasi-dominant equilibrium in Section~\ref{justification}.
Then in Section~\ref{eq-comparison}, we will show that it is a stronger concept than perfect Bayesian equilibrium, as well as essentially every refinement of Nash-equilibrium provided that this equilibrium exists in all (finite) games. Finally, the proof of quasi-dominant equilibrium will be presented in Section~\ref{quasi-proof}.

\textbf{Proof of individual rationality and avoiding free-riding.}
If an agent $i \in \N^+$ submits the \emph{empty proposal}, namely, promises no consequence and asks for 0 payment, and then he chooses the do-nothing strategy, then his payment will be $p_i = 0$. This proves individual rationality for the agents. Also, the mechanism avoids free-riding, because if agent $i$ has an empty capability tree and he is truthful, then he gets payment $p_i = 0$, and incentive-compatibility will imply that this is the best he can do.

\begin{proof}[Proof of Lemma~\ref{u(cp)=0-thm}]
  If $i \notin \Acc$, then $u_i = 0$. Assume that $i \in \Acc$.
  With the truthful strategy $\cp_i$, the execution $\xi_i\big(G, (\cp_i, \bs{s}_{-i})\big)$ of the capability tree of $i$ is the same as $\xi(\htheta_i, \bs{a}, \bs{x}, \bs{c})$ in \eqref{Ew=0} and \eqref{pi-star}. Therefore,
  \begin{equation*}
    \E \Big( u_i\big(G^1, (\cp_i, \bs{s}_{-i})\big)\Big)
    \mathop{=}^{\eqref{utility}}
    \E \Big( v\big(\theta_i, \xi_i(\cp_i, \bs{s}_{-i})\big) + p_i\big(G^1, \xi(\cp_i, \bs{s}_{-i})\big) \Big)
  \end{equation*}
  \begin{equation*}
    = \E \Big( v\big(\theta_i, \xi_i(\cp_i, \bs{s}_{-i})\big) + \pi^*(\theta_i)\big( \bs{c}(\cp_i, \bs{s}_{-i}), \bs{m}(\cp_i, \bs{s}_{-i}) \big) \Big)
    \mathop{=}^{\eqref{pi-star}}
    \E \Big( \sum_{t \in I} w_t(x_t) \Big) \mathop{=}^{\eqref{Ew=0}} 0. \qedhere
  \end{equation*}
\end{proof}

\begin{proof}[Proof of Proposition~\ref{Veq-thm}]
  We prove it by constructing a strategy $s_0$ of the principal such that if all agents submit the truthful proposals, then the principal gets $V(\btheta)$ utility, in expectation. Formally,
  \begin{align}\label{s_0-prop}
    \big(\forall \bs{s} \in \Str_{\Ag} \bigwhere \bs{\pi}(G, \bs{s}_{-0}) = \bs{\pi^*}(\btheta) \big)\:
    &&&& \E\big(u_0(G, \bs{s})\big) &= V(\btheta).
    &&&&&&&&&&&&&&&&&&&&
  \end{align}
  By the definition of $V\big(\theta_0, \pi^*(\bhtheta_{-0})\big)$, this will imply \eqref{Veq}.

  We can assume that each proposal is of the form $\pi(G, s_i) = \pi^*(\htheta_i)$, otherwise \eqref{s_0-prop} have no restriction on $s_0$.
  The principal can reconstruct all these $\htheta_i$ from the proposals, up to equivalence.
  The principal considers the optimization problem $(\btheta_0, \bhtheta_{-0})$, and sends the messages (action suggestions) $\bs{a}$, and makes the actions in her own tree according to the optimal strategy.
  If all agents are truthful, then it leads to the efficient expected outcome, formally,
  \begin{equation}\label{prop-eff}
    \E\Big( u_N\big(G, (s_0, \bs{\cp}_{-0})\big) \Big) = V(\btheta),
  \end{equation}
   no matter which weights $\bs{w}$ the principal is sending.

   Now we are specifying the strategy $s_0$ about choosing $\bs{w}$ so as to make the principal indifferent about what the agents do.
   Let us identify each chance event $\chi$ during the execution of $(\theta_0, \bhtheta_{-0})$ with the subgame after the chance event.
   Now $\chi$ is a stochastic variable, and let $\bar{\chi}$ denote the subgame before the chance event. Let
   \begin{equation}\label{delta-def}
     \delta(\chi) = V(\chi) - V(\bar{\chi}).
   \end{equation}
   Clearly, $V(\chi) = \E\big( V(\chi) \big)$, or equivalently,
   \begin{equation}\label{E-delta}
     \E\big(\delta(\chi)\big) = 0.
   \end{equation}
   Let us define $s_0$ so that it chooses $w(\chi) = \delta(\chi)$. We are going to show that it satisfies \eqref{s_0-prop}.

   Let $\hat{\xi}$ denote the virtual execution of $\bhtheta = (\theta_0, \bhtheta_{-0})$ according to the messages, and $\bhtheta(\hat{\xi}^{<t})$ the state before time point $t$, and $\Chi(\hat{\xi}^{<t})$ the set of chance events before $t$. Let $\hat{\xi}_0 = \xi_0$.
   According to the definition of $\delta$ in \eqref{delta-def},
   \begin{equation}\label{invariant}
     V\big( \bhtheta(\hat{\xi}^{<t}) \big) - \sum_{\hat{\chi} \in \Chi(\hat{\xi}^{<t})} \delta(\hat{\chi})
   \end{equation}
   is invariant of $t$, or in other words, it does not change during the virtual execution of $\bhtheta$.
   Therefore,
   \begin{equation*}
     \E\big( u_0(G^1, \bs{s}) \big)
     \mathop{=}^{\eqref{utility}\eqref{balanced}}
     \E\Big( v(\theta_0, \xi_0) - \sum_{i \in N^+} p_i(G^1, \bs{s}) \Big)
     = \E\Big( v(\theta_0, \xi_0) - \sum_{i \in N^+} \pi^*(\theta_i)(G, \bs{s}) \Big)
   \end{equation*}
   \begin{equation*}
     \mathop{=}^{\eqref{pi-star}}
     \E\bigg( \sum_{i \in N} v(\theta_i, \hat{\xi}_i) - \sum_{i \in \Ag} \sum_{\hat{\chi} \in \Chi(\hat{\xi}_i)} \delta(\hat{\chi}) \bigg)
     = \E\bigg( \sum_{i \in N} v(\theta_i, \hat{\xi}_i) \bigg) - \sum_{\hat{\chi} \in \Chi(\hat{\xi})} \delta(\hat{\chi}) \bigg) + \E \bigg( \sum_{\hat{\chi} \in \Chi(\hat{\xi}_0)} \delta(\hat{\chi}) \bigg)
   \end{equation*}
   \begin{equation} \label{s_0-prop-eq}
     \mathop{=}^{\eqref{E-delta}}
     \E\bigg( \sum_{i \in N} \Big( v(\theta_i, \hat{\xi}_i) - \sum_{\hat{\chi} \in \Chi(\hat{\xi}_i)} \delta(\hat{\chi}) \Big) \bigg)
     = \E\Big( V\big(\bhtheta(\hat{\xi}_i)\big) - \sum_{\hat{\chi} \in \Chi(\hat{\xi})} \delta(\hat{\chi}) \Big).
   \end{equation}
   Using the invariance of \eqref{invariant} for the virtual starting state and the endstate, we get that
   \begin{equation*}
     \eqref{s_0-prop-eq} = \E\big( V(\bhtheta) \big) = \E\big( V(\btheta) \big). \qedhere
   \end{equation*}
\end{proof}

\subsection{Proofs about the first-price mechanism} \label{results1}

\subsubsection{Collusion resistance} \label{collusion}

Forming a consortium may be beneficial, because it allows a joint offer with a joint pricing, but we are going to show that colluding in a secret way is never better than that. We suggest that it should be interpreted so as this situation is analogous to the collusion resistance of the first-price combinatorial auctions, see Problem B in Section~\ref{literature}.

We model the case when a set of agents form a consortium, namely, we replace them to a new agent $x$ with multiple capability trees.
Having multiple capability trees is equivalent to having one joint capability tree (like in Example 1), which is the capability tree with the product of the action sets, the union of the consequences and the sum of the corresponding utilities.

Now we compare this game $G_x$ with $(N^+)^{G_x} = N^+ \setminus X \cup \{x\}$ with the original game $G$ with the colluding set of players $X \subset N^+$.
The sketch of this comparison is the following. Any joint strategy $\bs{s}^G_X$ naturally corresponds to a strategy $s_x$ for the game $G_x$, and $u_x(G_x, s_x) = \sum_{i \in X} u(G_i, \bs{s}_X)$. This provides a reduction of the problem with colluding agents to the original problem where the new agents are consortiums formed by the original agents. Applying the results for this reduced game shows the collusion resistance of the first-price mechanism.

We show this argument more formally. We define the combination (or the product) of some capability trees $\theta_x = \prod\limits_{i \in X} \theta_i$ as the following capability tree.
For all $i \in X$, the games $\theta_i$ are played simultaneously by the three players, making the moves in chronological order.
The worker of the joint capability tree $\theta_x$ plays as the worker in each $\theta_i$.
Nature uses the strategies specified in the capability trees $\theta_i$, independently. At each time point, if the action of the world in the joint capability tree $\theta_x$ is $c \in \C$, and the consequence provided by each capability tree $\theta_j$ is $c_j$, then the action of the world in $\theta_i$ is defined as $c \cup \bigcup\limits_{j \in X \setminus \{i\}} c_j$. The consequence of $\theta_x$ is defined as $\bigcup\limits_{i \in X} c_i$. The valuation of an execution of $\theta_x$ is the sum of the valuations of the individual capability trees, namely for an execution $\xi_X = (\xi_i)_{i \in X}$, we define $v(\xi_x) = \sum\limits_{i \in X} v(\xi_i)$.

We define the product of proposals $\sum\limits_{i \in X} \pi_i$ as the following proposal.
For all $X' \subseteq X$, we define a simulated $|X'|$-channel communication as follows. Each message between $x$ and the principal should be an element of $X' \times \M$ with the only exception that each time $x$ outputs a consequence $c_x \in \C$, he immediately sends to the principal a vector $\bs{c}_{X'} \in \C^{X'}$ with $\bigcup\limits_{i \in X'} c_i = c_x$.
At $t_1$, the principal sends a set $X' \subseteq X$ to the agent.
We assume that the communication during $I$ follows the protocol according to $X'$.\footnote{E.g. messages (or lack of messages) which are incompatible with the protocol are ignored or replaced by a default message.} Let $\bs{m}_i$ and $\bs{m}_{0\rightarrow i}$ denote the history of the second arguments of messages $(i, .)$ sent from $x$ to the principal, and from the principal to $x$, respectively.
Now the payment is defined as
\begin{equation*}
  \Big(\sum\limits_{i \in X} \pi_i\Big)(\bs{c}, \bs{m}) = \sum\limits_{i \in X'} \pi_i(\bs{c}_i, \bs{m}_{0\rightarrow i}, \bs{m}_i).
\end{equation*}

For the principal, this proposal $\sum\limits_{i \in X} \pi_i$ is equivalent to receiving all proposals $\pi_i$. The principal-devil games in the two cases are also equivalent, and therefore, the principal makes the corresponding decision about the acceptance (including the choice of $X' \subseteq X$).

We define the product strategy $\prod\limits_{i \in X} s_i \in \Str_x\big(\prod\limits_{i \in X} \theta_i\big)$ of agent $x$ as follows. He sends the proposal $\sum\limits_{i \in X} \pi(G, s_i)$, and simulates the case as if he would follow the strategy $s_i$ in each capability tree $\theta_i$, namely, he makes the corresponding actions in the capability trees, he always reports all states what he would report by each $s_i$.

Assume that each of the other players $i \in N \setminus X$ does not observe whether $X$ play as one agent in a game $G_x$ or as different agents in $G$. Or at least, this change in the belief of $i$ does not affect his actions. Then the executions of $G_x$ correspond to the executions of $G$ by a natural transformation which preserves probabilities, therefore,
\begin{equation*}
u_x(G_x, \xi_x) = \sum\limits_{i \in X} u_i(G, \xi).
\end{equation*}

\subsubsection{Individual rationality for the principal and avoiding free-riding} \label{indiv-rat-0}

Individual rationality for the principal is implied by the fact that she can reject all agents, and this option is equivalent to the outside option.

Free-riding is avoided because of the incentive-compatibility of the mechanism and the fact that the truthful strategy provides utility 0. This argument is weaker here than under the second-price mechanism because we have a weaker equilibrium here. However, we have another argument. Namely, each agent can also play any free-riding strategy (like a consortium of himself and an imaginary freerider), and if it is beneficial, then it means that his original strategy can be improved.

\subsubsection{Level of efficiency and revenue maximization} \label{leveleff}

We want to prove the same results that are true for the first-price auction mechanisms (Problems 0 and B). In particular, we prove efficiency in two extreme cases: $(1)$ under perfect competition and $(2)$ with publicly known capability trees (but still with private decisions and hidden chance events). Then we show a justification that, in a typical case, the players are approximately incentivized to follow essentially truthful strategies, and we prove an inequality which also suggests that it is very exceptional having incentives to an essential deviation from it.

\bigskip

We say that there is \textbf{perfect competition}\index{perfect competition} if $\forall G \in \G,\ \forall i \in \Ag\: V(\btheta_{-i}) = V(\btheta)$.
It can be interpreted so as it is common knowledge between the agents that the marginal contribution of each agent to the maximum possible total utility is approximately 0.

\begin{Theorem} \label{price1perfect} \index{quasi-dominant equilibrium}
The truthful strategy profile $\bs{\cp}$ under the first-price mechanism is a quasi-dominant equilibrium under perfect competition.
\end{Theorem}

\noindent The proof is presented in Section~\ref{price1perfectsection}.

\bigskip

In the general case (under imperfect competition), analogously to the case of first-price auctions (Problems 0 and B), we expect that the agents ask higher payments than their real costs.
The requested amount of this extra payment is prior-dependent even in Problem 0, therefore, we need a probability distribution on $\G$ which extends $\G$ to a Bayesian game. We call this game a \emph{Bayesian Project Management Model}.

For a proposal $\pi_i$ and a number $x$ (which depends on the belief of agent $i$ when he submits his proposal), let $\pi_i + x$ be the same proposal except that the payment is increased by a constant $x$.
For a strategy $s_i$ using proposal $\pi_i$, let $s_i + x$ mean the strategy $s_i$ except that $i$ sends the proposal $\pi_i + x$ instead of $\pi_i$. We call a strategy $\cp_i + x$ a \textbf{fair strategy}\index{strategy $s_i$!fair strategy $\cp_i + x$} and a proposal $\pi^*_i + x$ a \textbf{fair proposal}\index{proposal $\pi_i$!fair proposal $\pi^*_i + x_i$} with \textbf{profit} $x$.

Consider the extreme case when the players can observe the capability trees of each other. We emphasize that this assumption does not mean that the players can observe anything about the executions of the capability trees of others.

\begin{Theorem} \label{price1common}
If the capability trees of the players are common knowledge between the agents, then there exists a weak quasi-dominant equilibrium\label{quasi-dominant equilibrium!weak quasi-dominant eq.} where the agents use fair strategies, the principal uses truthful strategy, they maximize the expected total utility, and the principal gets at least as much utility as under the second-price mechanism (with the quasi-dominant strategy profile).
\end{Theorem}

\noindent The precise definitions, the theorems and proofs are presented in Sections \ref{weakquasi} and \ref{commonknowsection}.

\bigskip

In Section~\ref{Bayesian1}, we will elaborate more about the general case, including an inequality suggesting that the agents have very weak incentives in deviating from their best fair strategies. In Appendix~\ref{conjecturesection}, we introduce an environment for analyzing different versions of general auction mechanisms for problems including the Project Management Model. For example, this could help analyzing the version with ``the core compensation closest to the second-price compensation''. This version might be a good tradeoff between efficiency and collusion-resistance, but we will not elaborate on it.

\section{The quasi-dominant equilibrium} \label{justification}\index{quasi-dominant equilibrium|(}

Theorem~\ref{Nash-proof} was about Nash-implementation, but we show that the proof implies a strong kind of incentive compatibility, which works even in dynamic
environments. In this section, we define the quasi-dominant equilibrium, which is designed to catch the incentive compatibility the proof provides. Then, in
Section~\ref{quasi-proof}, we will see that the proof indeed works with this equilibrium.

The strength of an equilibrium concept is an interpretational question. However, we try to provide a justification as close to a mathematical proof as possible.
Starting with a very special case, we arrive at the quasi-dominant equilibrium and its justification in several steps.

In Section~\ref{eq-comparison}, we will show that quasi-dominant equilibrium is stronger than perfect Bayesian equilibrum, in a reasonable sense. These justifications also give a good intuition for that proof.

At all steps, $\Str_i$ denotes the strategy set and $u_i$ denotes the utility function of player $i$.

\bigskip

\textbf{Step 1.} Players $A$ and $B$ play an arbitrary deterministic \emph{dynamic} game with \emph{imperfect information}. Suppose that there is a strategy profile
$\bs{s^*} \in \bs{\Str}$ satisfying the following.
\begin{align}
\forall s_B \in \Str_B\:&&&&&&&&
u_A(s^*_A, s_B) &\ge 1 &&&&&&&&\phantom{\forall s_B \in \Str_B\:}\label{pA1}
\\ \forall s_A \in \Str_A\:&&&&&&&&
u_B(s_A, s^*_B) &\ge 1 &&&&&&&&\phantom{s_A \in \Str_A\:} \label{pB1}
\\ \forall \bs{s} \in \bs{\Str}\:&&&&&&&&
u_A(\bs{s}) + u_B(\bs{s}) &\le 2 &&&&&&&&\phantom{\forall \bs{s} \in \bs{\Str}\:} \label{pAB1}
\end{align}
Then $\bs{s^*}$ is an equilibrium.

\begin{proof}[Justification.]

In an arbitrary two-player game, if either player can guarantee himself utility 1, and even with collusion, they cannot get more utility in total, then both players will use the strategy guaranteeing utility 1.

More formally, \eqref{pA1} shows that $A$ can get utility at least $1$, therefore, if she is selfish and rational, then she will get expected utility at least $1$.
Comparing this with \eqref{pAB1}, this shows that $B$ has no hope of getting expected utility more than $2 - 1 = 1$. But \eqref{pB1} shows that $s^*_B$ guarantees him $1$. Therefore, $B$ has no incentive to deviate from $s^*_B$. And the same argument holds for $A$, as well. Therefore, we can rightfully say that $\bs{s^*}$ is an equilibrium.
\end{proof}

Notice that this was already a new reasoning of being an equilibrium in a dynamic game. According to the knowledge of the author, $\bs{s^*}$ does not satisfy the conditions of any other equilibrium concepts for dynamic games. For example, $\bs{s^*}$ is not a perfect Bayesian equilibrium, not even a subgame-perfect equilibrium for games with perfect information, because each player might miss the opportunity to completely utilize when the other player makes a bad move. As a simple counterexample, consider the following game. Player $A$ has to choose his utility $u_A$ from $[0, 1]$. Then player $B$, after observing $u_A$, has to choose his desired utility $u'_B$ from $[0, 2]$. If $u_A + u'_B \le 2$, then $u_B = u'_B$, otherwise $u_B = 0$. Then the strategy profile of choosing 1 by both players satisfies all \eqref{pA1}, \eqref{pB1} and \eqref{pAB1}. But this is not subgame-perfect: if agent $A$ chose e.g.\ $u_A = 0.7$, then the best choice for agent $B$ would be $u'_B = 1.3$, but he chooses 1 instead. The only subgame-perfect equilibrium is that $A$ chooses $1$ and $B$ chooses $2 - u_A$. But the utilities of the players are the same with both equilibria, and in this sense, we will show that quasi-dominant equilibrium is a stronger concept than perfect Bayesian equilibrium.

\bigskip

\textbf{Step 2.} There is a set of players $N$ playing a deterministic dynamic game with imperfect information. Suppose that there is a strategy profile $\bs{s^*} \in \bs{\Str}$ and constants $C_i$ satisfying the following.
\begin{align}
\forall i \in N,\  \forall \bs{s}_{-i} \in \bs{\Str}_{-i}\:&&
u_i(s^*_i, \bs{s}_{-i}) &\ge C_i \label{pi2}
&&\phantom{\forall i \in N,\  \forall \bs{s}_{-i} \in \bs{\Str}_{-i}\:}
\\ \forall \bs{s} \in \bs{\Str}\:&&
\sum_{i \in N} u_i(\bs{s}) &\le \sum_{i \in N} C_i \label{pN2}
&&\phantom{\forall \bs{s} \in \bs{\Str}\:}
\end{align}
Then $\bs{s^*}$ is an equilibrium.

\begin{proof}[Justification.]

\eqref{pi2} implies that each player $i \in N$ can get utility at least $C_i$. (We say that $C_i$ is \emph{guaranteed} for $i$.) Each player $i \in N$ has no hope of getting more expected utility than the maximum possible total utility of all players minus the sum of the guaranteed utilities of the other players. Therefore, $i$ has no hope of getting more expected utility than
\begin{equation*}
\sup_{\bs{s} \in \bs{\Str}} \sum_{j \in N} u_{j}(\bs{s}) - \sum_{j \in N \setminus \{i\}} u_j(\bs{s^*}) \mathop{=}^{\eqref{pN2}} \sum_{j \in N}
C_j - \sum_{j \in N \setminus \{i\}} C_j = C_i.
\end{equation*}
Consequently, each player $i \in N$ has no incentive to deviate from $s^*_i$, and therefore, we can rightfully say that $\bs{s^*}$ is an equilibrium.
\end{proof}

\bigskip

\textbf{Step 3.} There is a set of players $N$ playing a deterministic dynamic game with imperfect information. At the same initial time point, the players do simultaneous actions denoted by $a_i = \alpha(s_i) \in \A_i$ for each player $i \in N$. Suppose that there are functions $f_i: \bs{\A}_{-i} \rightarrow \R$ for all $i \in N$, and a strategy profile $\bs{s^*} \in \bs{\Str}$ with $a^*_i = \alpha(s^*_i)$ satisfying the following.

\begin{align}
\forall i \in N,\ \forall \bs{s}_{-i} \in \bs{\Str}_{-i}\:&&
u_i(s^*_i, \bs{s}_{-i}) &\ge f_i\big(\bs{\alpha}(\bs{s}_{-i})\big)
&&&&&&&&&& \label{fi3}
\\ \forall \bs{s} \in \bs{\Str},\ \forall i \in N,\  \forall a_i \in \A_i\:&&
\sum_{j \in N} u_j(\bs{s}) &\le f_i(\bs{a}^{\bs{*}}_{-i}) + \sum_{j \in N \setminus \{i\}} f_j(a_i, \bs{a}^{\bs{*}}_{-i-j})
&&&&&&&&&& \label{f-i3}
\end{align}
Then $\bs{s^*}$ is an equilibrium.

\begin{proof}[Justification.]

Each player $i \in N$ has no influence on $\bs{a}_{-i}$. Therefore, $f_i(\bs{a}_{-i})$ is independent of $s_i$, and \eqref{fi3} implies that $i$ can get at least $f_i(\bs{a}_{-i})$ utility. Each player $i \in N$ has no hope of getting more expected utility than the maximum possible total utility of all players minus the sum of the guaranteed utilities of the other players. Therefore, if the players other than $i$ take $\bs{a}^{\bs{*}}_{-i}$, then $i$ has no hope of getting more expected utility than
\begin{equation*}
\sup_{\bs{s} \in \bs{\Str}} \sum_{j \in N} u_{j}(\bs{s}) - \inf_{a_i \in \A_i} \sum_{j \in N \setminus \{i\}} f_j(a_i, \bs{a}^{\bs{*}}_{-i-j})
\mathop{\le}^{\eqref{f-i3}} f_i(\bs{a}^{\bs{*}}_{-i}).
\end{equation*}
Consequently, each player $i \in N$ has no incentive to deviate from $s^*_i$, and therefore, we can rightfully say that $\bs{s^*}$ is an equilibrium.
\end{proof}

\bigskip

\textbf{Step 4.} There is a set of players $N$ playing a dynamic \emph{stochastic} game with imperfect information. At the same initial time point, the players do simultaneous actions denoted by $a_i = \alpha(s_i) \in \A_i$ for each player $i \in N$. Suppose that there are functions $f_i: \bs{\A}_{-i} \rightarrow \R$ for all $i \in N$, and a strategy profile $\bs{s^*} \in \bs{\Str}$ with $a^*_i = \alpha(s^*_i)$ satisfying the following.

\begin{align}
\forall i \in N,\  \forall \bs{s}_{-i} \in \bs{\Str}_{-i}\:&&
\E \big( u_i(s^*_i, \bs{s}_{-i}) \big) &\ge f_i\big(\bs{\alpha}(\bs{s}_{-i})\big)
&& \label{fi4}
\\ \forall \bs{s} \in \bs{\Str},\ \forall i \in N,\  \forall a_i \in \A_i\:&&
\sum_{j \in N} \E\big(u_j(\bs{s})\big) &\le f_i(\bs{a}^{\bs{*}}_{-i}) + \sum_{j \in N \setminus \{i\}} f_j(a_i,
\bs{a}^{\bs{*}}_{-i-j})
&& \label{f-i4}
\end{align}
Then $\bs{s^*}$ is an equilibrium.

\begin{proof}[Justification.]

Each player $i \in N$ has no influence on $\bs{a}_{-i}$. Therefore, $f_i(\bs{a}_{-i})$ is independent of $s_i$, and \eqref{fi4} implies that $i$ can get at least $f_i(\bs{a}_{-i})$ expected utility. Each player $i \in N$ has no hope of getting more expected utility than the maximum possible expected total utility of all players minus the sum of the guaranteed utilities of the other players. Therefore, if the players other than $i$ take $\bs{a}^{\bs{*}}_{-i}$, then $i$ has no hope of getting more expected utility than
\begin{equation*}
\sup_{\bs{s} \in \bs{\Str}} \sum_{j \in N} \E \big( u_{j}(\bs{s}) \big) - \inf_{a_i \in \A_i} \sum_{j \in N \setminus \{i\}} f_j(a_i,
\bs{a}^{\bs{*}}_{-i-j}) \mathop{\le}^{\eqref{f-i4}} f_i(\bs{a}^{\bs{*}}_{-i}).
\end{equation*}
Consequently, each player $i \in N$ has no incentive to deviate from $s^*_i$, and therefore, we can rightfully say that $\bs{s^*}$ is an equilibrium.
\end{proof}

\bigskip

\textbf{Step 5.} We define the quasi-dominant equilibrium as follows.
\begin{Def}
There is a set of players $N$ playing an incomplete-information game $\G$, where each $G \in \G$ is a stochastic dynamic game with imperfect information.\footnote{Remember that $G$ is chosen in a nondeterministic way with no prior probabilities. (See Section~\ref{notions}.) Therefore, $E\big(u^G_i(s)\big)$ is a function of $G$, taking expectation applies only to the random events in $G$.}
$\Str_i$ denotes the strategy set and $u_i: \G \times \bs{\Str} \rightarrow \R$ denotes the utility function of player $i \in N$.
The initial information of each player $i$ in each game $G \in \G$ includes a \emph{type} $\theta_i = \theta_i^G \in \Theta$.
The players do their first actions simultaneously at the same initial time point, denoted by $a_i = \alpha(G, s_i) \in \A_i$ for each player $i \in N$.
Suppose that there is a belief-independent strategy profile $\bs{s^*} \in \bs{\Str}$ where the first action of each $i \in N$ depends only on $\theta_i$, denoted by $\alpha^*(\theta_i) = \alpha(G, s^*_i)$, and there are functions $f_i: \G \times \bs{\A}_{-i} \rightarrow \R$ for all $i \in N$ satisfying the following.\footnote{Remember that with our notation,
$\bs{\alpha}(\btheta_{-i}, \bs{s}_{-i}) = \big(\alpha(\theta_j, s_j)\big)_{j \in N \setminus \{i\}}$ and $\bs{\alpha^*}(\btheta_{-i-j}) =
\big(\alpha^*(\theta_k)\big)_{k \in N \setminus \{i, j\}}$ and $\bs{\alpha^*}(\btheta_{-i}) = \big(\alpha^*(\theta_j)\big)_{j \in N \setminus i}$ .}

\begin{align} \label{fi5}
\forall G \in \G,\  \forall i \in N,\  \forall \bs{s}_{-i} \in \bs{\Str}_{-i}\:&&&&
\E\Big(u_i\big(G, (s^*_i, \bs{s}_{-i})\big)\Big) &\ge f_i\big(G, \bs{\alpha}(G, \bs{s}_{-i})\big)
&&&&&&&&&&
\end{align}
\\ $\forall G \in \G,\ \forall \bs{s} \in \bs{\Str},\ \forall i \in N,\ \forall a_i \in \A_i\:$
\begin{equation} \label{f-i5}
\sum_{j \in N} \E\big(u_j(G, \bs{s})\big) \le f_i\big(G, \bs{\alpha^*}(\btheta_{-i})\big) + \sum_{j \in N \setminus \{i\}}
f_j \Big( G, \big(a_i, \bs{\alpha^*}(\btheta_{-i-j})\big) \Big)
\end{equation}

Then $\bs{s^*}$ is a \textbf{quasi-dominant equilibrium}.
\end{Def}

\begin{proof}[Justification.]
In short, it is publicly known that whatever the game $G$ is, if it was publicly revealed, then $\bs{s}^{\bs{*}}$
\footnote{more precisely, the modification of each $s^*_i$ which ignores this extra revealed information}
would be an equilibrium as defined in Step~4.
Therefore, we can rightfully say that $\bs{s}^{\bs{*}}$ is an equilibrium.

A direct justification is the following. Each player $i \in N$ has no influence on $G$ and $\bs{a}_{-i}$, therefore, $f_i(G, \bs{a}_{-i})$ is independent of $s_i$. \eqref{fi5} implies that $i$ can get at least $f_i(G, \bs{a}_{-i})$ expected utility. Each player $i \in N$ has no hope of getting more expected utility than the maximum possible total expected utility of all players minus the sum of the guaranteed expected utilities of the other players. Therefore, if the players other than $i$ start with $\bs{\alpha^*}(\btheta_{-i})$, then $i$ has no hope of getting more expected utility than
\begin{equation*}
\sup_{\bs{s} \in \bs{\Str}} \sum_{j \in N} \E\big(u_{j}(\btheta, \bs{s})\big) - \inf_{a_i \in \A_i} \sum_{j \in N \setminus \{i\}}
f_j\Big(G, \big(a_i, \bs{\alpha^*}(\btheta_{-i-j})\big)\Big)
\mathop{\le}^{\eqref{f-i5}} f_i\big(G, \bs{\alpha^*}(\btheta_{-i})\big).
\end{equation*}
Consequently, each risk-neutral player $i \in N$ has no incentive to deviate from $s^*_i$, and therefore, we can rightfully say that $\bs{s^*}$ is an equilibrium.
\end{proof}
\index{quasi-dominant equilibrium|)}

\subsection{The weak quasi-dominant equilibrium} \label{weakquasi} \index{quasi-dominant equilibrium!weak quasi-dominant eq.|(}

We also introduce a slightly weaker version of the quasi-dominant equilibrium. We show the corresponding versions of Steps 4 and 5.
The difference is that the function $f$ may depend on $a_i$ here. (Therefore, every quasi-dominant equilibrium is a weak quasi-dominant equilibrium.)

\bigskip

\textbf{Step 4W.} There is a set of players $N$ playing a dynamic stochastic game with imperfect information. At the same initial time point, the players do simultaneous actions denoted by $a_i = \alpha(s_i) \in \A_i$ for each player $i \in N$.
Suppose that there are functions $f_i: \bs{\A}_{-i} \rightarrow \R$ for all $i \in N$, and a strategy profile $\bs{s^*} \in \bs{\Str}$ with $a^*_i = \alpha(s^*_i)$ satisfying the following.

\begin{align}
\forall i \in N,\  \forall \bs{s}_{-i} \in \bs{\Str}_{-i}\:&&
\E \big( u_i(s^*_i, \bs{s}_{-i}) \big) &\ge f_i\big(\bs{\alpha}(\bs{s}_{-i})\big)
&&&&&&&&&& \label{fi3W}
\end{align}
\\ $\forall i \in N,\  \forall a_i \in \A_i,\ \big( \forall \bs{s} \in \bs{\Str} : \bs{a}(\bs{s}) = (a_i,
\bs{a}^{\bs{*}}_{-i}) \big)\:$
\begin{equation} \label{f-i3W}
\sum_{j \in N} \E\big(u_j(\bs{s})\big) \le f_i(\bs{a}^{\bs{*}}_{-i}) + \sum_{j \in N \setminus \{i\}} f_j(a_i,
\bs{a}^{\bs{*}}_{-i-j})
\end{equation}
Then $\bs{s^*}$ is an equilibrium.

\begin{proof}[Justification.]

Each player $i \in N$ has no influence on $\bs{a}_{-i}$. Therefore, if the others use $\bs{a}^{\bs{*}}_{-i}$, then \eqref{fi3W} implies that $i$ can get at least $f_i(\bs{a}^{\bs{*}}_{-i})$ expected utility. Each player $i \in N$ has no hope of getting more expected utility than the maximum possible expected total utility of all players minus the sum of the guaranteed utilities of the other players. Therefore, if the players other than $i$ take
$\bs{a}^{\bs{*}}_{-i}$, then $i$ has no hope of getting more expected utility than
\begin{equation*}
\sup_{\bs{s} \in \bs{\Str} \: \bs{\alpha}(\bs{s}_{-i}) = \bs{a}^{\bs{*}}_{-i}}
\Big( \sum_{j \in N} u_{j}(\bs{s}) - \sum_{j \in N \setminus \{i\}} f_j\big(\alpha(s_i), \bs{a}^{\bs{*}}_{-i-j}\big) \Big)
\mathop{\le}^{\eqref{f-i3W}} f_i(\bs{a}^{\bs{*}}_{-i}).
\end{equation*}
Consequently, each player $i \in N$ has no incentive to deviate from $s^*_i$, and therefore, we can rightfully say that $\bs{s^*}$ is an equilibrium.
\end{proof}

\bigskip

\textbf{Step 5W.}
There is a set of players $N$ playing an incomplete-information game $\G$, where each $G \in \G$ is a stochastic dynamic game with imperfect information.
$\Str_i$ denotes the strategy set and $u_i: \G \times \bs{\Str} \rightarrow \R$ denotes the utility function of player $i \in N$.
The initial information of each player $i$ in each game $G \in \G$ includes a \emph{type} $\theta_i = \theta_i^G \in \Theta$.
The players do their first actions simultaneously at the same initial time point, denoted by $a_i = \alpha(G, s_i) \in \A_i$ for each player $i \in N$.
Suppose that there is a belief-independent strategy profile $\bs{s^*} \in \bs{\Str}$ where the first action of each $i \in N$ depends only on $\theta_i$, denoted by $\alpha^*(\theta_i) = \alpha(G, s^*_i)$, and there are functions $f_i: \G \times \bs{\A}_{-i} \rightarrow \R$ for all $i \in N$ satisfying the following.
\begin{align}
\forall G \in \G,\  \forall i \in N,\  \forall \bs{s}_{-i} \in \bs{\Str}_{-i}\:&&&&
\E\Big(u_i\big(G, (s^*_i, \bs{s}_{-i})\big)\Big) &\ge f_i\big(G, \bs{\alpha}(G, \bs{s}_{-i})\big) \label{fi5W}
&&&&&&&&
\end{align}
\\ $\forall G \in \G,\ \forall i \in N,\ \big( \forall \bs{s} \in \bs{\Str} \bigwhere
\bs{\alpha}(G, \bs{s}_{-i}) = \bs{\alpha^*}(\btheta_{-i}) \big)\:$
\begin{equation} \label{f-i5W}
\sum_{j \in N} \E\big(u_j(G, \bs{s})\big) \le f_i\big(G, \bs{\alpha^*}(\btheta_{-i})\big) + \sum_{j \in N \setminus \{i\}} f_j\Big(G, \big(\alpha(G, s_i), \bs{\alpha^*}(\btheta_{-i-j})\big)\Big)
\end{equation}

Then $\bs{s^*}$ is an equilibrium.

\begin{proof}[Justification.]

Each player $i \in N$ has no influence on $\btheta_i$ and $\bs{a}_{-i}$. Therefore, if the other players use $\bs{\alpha^*}(\btheta_{-i})$, then \eqref{fi5W} implies that $i$ can get at least $f_i\big(G, \bs{\alpha^*}(\btheta_{-i})\big)$ expected utility. Each player $i \in N$ has no hope of getting more expected utility than the maximum possible total expected utility of all players minus the sum of the guaranteed expected utilities of the other players. Therefore, if the players other than $i$ start with $\bs{\alpha^*}(\btheta_{-i})$, then $i$ has no hope of getting more expected utility than
\begin{equation*}
\sup_{\bs{s} \in \bs{\Str} \: \bs{\alpha}(G, \bs{s}_{-i}) = \bs{\alpha^*}(\btheta_{-i})}
\bigg( \sum_{j \in N} \E\big(u_{j}(G, \bs{s})\big) - \sum_{j \in N \setminus \{i\}} f_j\Big(G, \big(\alpha(\theta_i, s_i),
\bs{\alpha^*}(\btheta_{-i-j})\big)\Big) \bigg) \mathop{\le}^{\eqref{f-i5W}} f_i\big(G, \bs{\alpha^*}(\btheta_{-i})\big).
\end{equation*}
Consequently, each risk-neutral player $i \in N$ has no incentive to deviate from $s^*_i$, and therefore, we can rightfully say that $\bs{s^*}$ is an equilibrium.
\end{proof}

We call the equilibrium defined in Step 5W a \textbf{weak quasi-dominant equilibrium}.\index{quasi-dominant equilibrium!weak quasi-dominant eq.}
Every quasi-dominant equilibrium is a weak quasi-dominant equilibrium because \eqref{f-i5} implies \eqref{f-i5W}.

\subsection{The (weak) quasi-dominant equilibrium is a PBE-refinement} \label{eq-comparison}\index{quasi-dominant equilibrium}

The following proposition is the key to show that the weak quasi-dominant equilibrium is stronger than any reasonable equilibrium concept which $(i)$ refines (Bayesian) Nash equilibrium and $(ii)$ which exists in all games (at least, in finite games).
Of those, the most important concept is the perfect Bayesian equilibrium.

\begin{Lemma} \label{PBE-prop}
Assume that a strategy profile $\bs{s^*}$ is a weak quasi-dominant equilibrium in an incomplete-information game $\G$.
Denote by $\G(i, \eps)$ the game $\G$ with the following modifications about the first parallel actions in each $G \in \G$.
Every player $j \in N \setminus \{i\}$ is forced to start with $\alpha^*(\theta_j)$.
Player $i$ is free to choose his first move except that deviating from $\alpha^*(\theta_i)$ has an extra cost of $\eps > 0$ for him.
Let us add a probability distribution on $\G$, and denote the resulting Bayesian games of $\G$ and $\G(i, \eps)$ by $\tilde{G}$ and $\tilde{G}(i, \eps)$, respectively.
We treat $\tilde{G}(i, \eps)$ as the same game as $\tilde{G}$ but with a restricted strategy set $\Str^{\tilde{G}(i, \eps)} \subset \Str^{\tilde{G}}$ and a different utility function.
Then all Nash equilibria $\bs{s}$ of $\tilde{G}(i, \eps)$ satisfy that $\alpha(G, s_i) = \alpha^*(\theta_i)$ with probability 1, and
$\E\big(u_j(\tilde{G}, \bs{s})\big) = \E\big(u_j(\tilde{G}, \bs{s^*})\big)$ for all players $j \in N$ (where expectation also applies to the random variable $G \in \tilde{G}$).
\end{Lemma}

\begin{proof}

\begin{equation*}
\E\big(u_i(\tilde{G}, \bs{s})\big)
\mathop{\le}^{\eqref{fi5W}} \E\big(u_i(\tilde{G}, \bs{s})\big)
+ \sum_{j \in \N \setminus \{i\}} \E\Big(u_j\big(\tilde{G}, (s^*_j, \bs{s}_{-j})\big) - f_j\big(G, \bs{\alpha}(G, \bs{s}_{-j})\big)\Big)
\end{equation*}
\begin{equation*}
\le \E\big(u_i(\tilde{G}, \bs{s})\big) + \sum_{j \in \N \setminus \{i\}} \E\Big(u_j(\tilde{G}, \bs{s}) -
f_j\big(G, \bs{\alpha}(G, \bs{s}_{-j})\big)\Big)
\end{equation*}
\begin{equation*}
= \sum_{j \in N} \E\big(u_j(\tilde{G}, \bs{s})\big)
- \sum_{j \in N \setminus \{i\}} \E\Big(f_j\big(G, \bs{\alpha}(G, \bs{s}_{-j})\big)\Big)
\end{equation*}
\begin{equation*}
= \sum_{j \in N} \E\big(u_j(\tilde{G}, \bs{s})\big)
- \sum_{j \in N \setminus \{i\}} \E\bigg(f_j\Big(G, \big(\alpha(G, s_i), \bs{\alpha^*}(\bs{\theta}_{-i-j})\big)\Big)\bigg)
\mathop{\le}^{\eqref{f-i5W}} \E\Big(f_i\big(G, \bs{\alpha^*}(\bs{\theta}_{-i})\big)\Big)
\end{equation*}
\begin{equation*}
= \E\Big(f_i\big(G, \bs{\alpha}(G, \bs{s}_{-i})\big)\Big)
\mathop{\le}^{\eqref{fi5W}} \E\Big(u_i\big(\tilde{G}, (s^*_i, \bs{s}_{-i})\big)\Big)
= \E\big(u^{G(i, \eps)}_i(s^*_i, \bs{s}_{-i})\big)
\end{equation*}
\begin{equation*}
\le \E\Big(u_i\big(G(i, \eps), \bs{s}\big)\Big)
= \E\big(u_i(\tilde{G}, \bs{s})\big) - \eps \cdot \Pr\big(\alpha(G, s_i) \ne \alpha^*(\theta_i)\big)
\le \E\big(u_i(\tilde{G}, \bs{s})\big).
\end{equation*}
Therefore, all inequalities must hold with equality.
For the last inequality, this means that
\begin{equation*}
\Pr\Big(\alpha(G, s_i) \ne \alpha^*(\theta_i)\Big) = 0.
\end{equation*}
Notice that the entire calculation remains valid if we replace $\bs{s}$ with $\bs{s^*}$. Consider now the first inequality of the calculation. Using that equality in the two cases,
\begin{equation*}
\E\big(u_j(\tilde{G}, \bs{s})\big)
= \E\Big(f_j\big(G, \bs{\alpha}(G, \bs{s}_{-j})\big)\Big)
= \E\Big(f_j\big(G, \bs{\alpha^*}(\bs{\theta}_{-j})\big)\Big)
= \E\big(u_j(\tilde{G}, \bs{s^*})\big). \qedhere
\end{equation*}
\end{proof}

\begin{Corollary}
If $G$ is a finite game\footnote{or a game with a first time point and with compact strategy spaces and with an upper semicontinuous utility function with respect to the topology on the executions induced by the product topology of the strategy spaces}, then there exists a perfect Bayesian equilibrium $s$ in each $G(i, 0)$ satisfying that $\alpha(G, s_i) = \alpha^*(\theta_i)$ with probability 1, and
$\E\big(u_j(G, \bs{s})\big) = \E\big(u_j(G, \bs{s^*})\big)$ for all players $j \in N$.
\end{Corollary}

\begin{proof}
$G$ is finite, therefore, the set of mixed strategy profiles including beliefs is compact, and it contains at least one perfect Bayesian equilibrium of each $G(i, \eps)$ if $\eps > 0$. Choose one for each $G(i, \frac{1}{k})$, where $k \in \mathbb{Z}$. These must have an accumulation point, and this point must be a perfect Bayesian equilibrium of $G(i, 0)$.
\end{proof}

We only need a very technical finishing step constructing a perfect Bayesian equilibrium of $G$. We do not present it here, partially because it is very technical and not interesting and partially because there are multiple slightly different definitions of the perfect Bayesian equilibrium in the literature which differ about the off-equilibrium paths, and these would require different constructions. But we should conclude the following.

\begin{Theorem} \label{QD-PBE}
Assume that a strategy profile $\bs{s^*}$ is a weak quasi-dominant equilibrium in a finite incomplete-information game $\G$.
Then for a Bayesian game $\tilde{G}$ of $\G$, there exists a perfect Bayesian equilibrium $\bs{s}$ satisfying that, for all players $i \in N$,
$\alpha(\tilde{G}, s_i) = \alpha^*(\theta_i)$ with probability 1, and $\E\big(u_i(\tilde{G}, \bs{s})\big) = \E\big(u_i(\tilde{G}, \bs{s^*})\big)$.
\end{Theorem}
\index{quasi-dominant equilibrium!weak quasi-dominant eq.|)}

\section{Proofs of quasi-dominant equilibria}\index{quasi-dominant equilibrium}

\subsection{The second-price mechanism} \label{quasi-proof}

\begin{Theorem} \label{quasi2}
If the planner knows the capability tree of the principal, then the truthful strategy profile $\mathcal{\bs{\cp}}$ is a quasi-dominant equilibrium under the second-price mechanism.
\end{Theorem}

\begin{proof}
We apply the conditions of the equilibrium to our case. The first same-time actions of the agents are identified with the proposals $\bs{a}_{\Ag} = \bs{\pi}_{\Ag}$.
The principal makes no action (her action set is a singleton) at this time point.
Furthermore, $\bs{s^*} = \bs{\cp}$ with $\bs{\alpha^*}(\btheta_{-0}) = \bs{\pi^*}(\btheta_{-0}) = \bs{\pi^*}$.
Let $\P = \P_i = \big\{ \C^I \times \M^{\{0, i\} \times I} \rightarrow \R \big\}$ denote the set of proposals.
We need to prove that there exist functions $f_i: \G \times \P^{\Ag \setminus \{i\}} \rightarrow \R$ for all player $i \in \Pl$ (for $i = 0$ we use $\{0, i\} = \{0\}$), so that the following inequalities hold.

\bigskip

\noindent
\begin{align}
\forall G \in \G,\ \forall i \in \N,\ \forall \bs{s}_{-i} \in \bs{\Str}_{-i}\:&&
\E\Big(u_i\big(G^2, (\cp_i, \bs{s}_{-i})\big)\Big)
&\ge f_i\big(G, \bs{\pi}_{-i}(G, \bs{s}_{-0-i})\big) && \label{fy}
\end{align}
\begin{align}
\forall G \in \G,\ \forall i \in \Ag,\ \forall \pi^{\prime}_i \in \P\:&&
V(\btheta) &\le f_i(G, \bs{\pi}^{\bs{*}}_{-i}) + \sum_{j \in N \setminus \{i\}} f_j\big(G, (\pi^{\prime}_i, \bs{\pi}^{\bs{*}}_{-i-j})\big)&& \label{f-i}
\end{align}
\begin{align}
\forall G \in \G\:&&
V(\btheta) &\le f_0(G, \bs{\pi}^{\bs{*}}) + \sum_{i \in N^+} f_i(G, \bs{\pi}^{\bs{*}}_{-i})&& \label{f0}
\end{align}

We show that the following functions $f_i$ satisfy \eqref{fy}, \eqref{f-i} and \eqref{f0}.
\begin{align}
\forall i \in \Ag\: && f_i(G, \bs{\pi}_{-i}) &= V\big(\theta_0, (\pi^*_i, \bs{\pi}_{-i})\big) - V(\theta_0, \bs{\pi}_{-i}) && \phantom{\forall i \in \Ag\:} \label{fidef}
\\ && f_0(G, \bs{\pi}) &= \sum_{i \in \Ag} V(\theta_0, \bs{\pi}_{-i})
- (n - 1) \cdot V\big(\theta_0, (\pi^*_i, \bs{\pi}_{-i})\big) && \label{fCdef}
\end{align}

Proof of \eqref{fy} for agents. $\forall G \in \G,\  \forall i \in \Ag,\  \forall \bs{s}_{-i} \in \bs{\Str}_{-i}\:$
\begin{equation*}
\E\Big(u_i\big(G^2, (\cp_i, \bs{s}_{-i})\big)\Big) \mathop{=}^{\eqref{p2i}}
\E\Big(u_i\big(G^1, (\cp_i, \bs{s}_{-i})\big)\Big) + v^+_i\big(G, (\cp_i, \bs{s}_{-i})\big)
\mathop{=}^{\eqref{u(cp)=0}} v^+_i\big(G, (\cp_i, \bs{s}_{-i})\big)
\end{equation*}
\begin{equation*}
\mathop{=}^{\eqref{vpdef}}
V\Big(\theta_0, \big(\pi^*_i, \bs{\pi}_{-i}(G, \bs{s}_{-0-i})\big)\Big) -
V\big(\theta_0, \bs{\pi}_{-i}(G, \bs{s}_{-0-i})\big)
\mathop{=}^{\eqref{fidef}}
f_i\big(G, \bs{\pi}_{-i}(G, \bs{s}_{-0-i})\big)
\end{equation*}

Proof of \eqref{fy} for the principal. $\forall G \in \G,\  \forall \bs{s}_{-0} \in \bs{\Str}_{-0}\:$
\begin{equation*}
\E\Big(u_0 \big(G^2, (\cp_0, \bs{s}_{-0})\big)\Big) \mathop{=}^{\eqref{p2C}}
\E\Big(u_0 \big(G^1, (\cp_0, \bs{s}_{-0})\big)\Big)
- \sum_{i \in \Ag} v^+_i\big( \theta_0, \bs{\pi}(G, \bs{s}_{-0}) \big)
\end{equation*}
\begin{equation*}
\mathop{\ge}^{\eqref{V(pi)-ineq}\eqref{vpdef}}
V\big(\theta_0, \bs{\pi}(G, \bs{s}_{-0})\big) - \sum_{i \in \Ag} \Big( V\big(\theta_0, \bs{\pi}(G, \bs{s}_{-0})\big) - V\big(\theta_0, \bs{\pi}_{-i}(G, \bs{s}_{-0-i})\big) \Big)
\end{equation*}
\begin{equation*}
= \sum_{i \in \Ag} V\big(\theta_0, \bs{\pi}_{-i}(G, \bs{s}_{-0-i})\big) - (n-1) \cdot V\big(\theta_0, \bs{\pi}(G, \bs{s}_{-0})\big) = f_0\big( G, \bs{\pi}(G, \bs{s}_{-0}) \big)
\end{equation*}

Proof of \eqref{f-i}. $\forall i \in \Ag,\ \forall G \in \G,\ \forall \pi^{\prime}_i \in \P\:$
\begin{equation*}
f_i\big( G, \bs{\pi}^{\bs{*}}_{-i} \big)
+ \sum_{j \in \Pl \setminus \{i\}} f_j \big(G, (\pi^{\prime}_i, \bs{\pi}^{\bs{*}}_{-i-j}) \big)
\end{equation*}
\begin{equation*}
= f_0 \big(G, (\pi^{\prime}_i, \bs{\pi}^{\bs{*}}_{-i})\big)
+ f_i \big( G, \bs{\pi}^{\bs{*}}_{-i}) \big)
+ \sum_{j \in \Ag \setminus \{i\}} f_j \Big(G, \big(\pi^{\prime}_i, \bs{\pi}^{\bs{*}}_{-i-j})\big) \Big)
\end{equation*}
\begin{equation*}
\mathop{=}^{\eqref{fCdef}\eqref{fidef}}
\Big( \sum_{j \in \Ag \setminus \{i\}} V\big(\theta_0, (\pi^{\prime}_i, \bs{\pi}^{\bs{*}}_{-i-j}) \big) - (n-1) \cdot V\big(\theta_0, (\pi^{\prime}_i, \bs{\pi}^{\bs{*}}_{-i})\big) \Big)
+ \Big( V(\theta_0, \bs{\pi^*}) - V(\theta_0, \bs{\pi}^{\bs{*}}_{-i}) \Big)
\end{equation*}
\begin{equation*}
+ \sum_{j \in \Ag \setminus \{i\}} \Big( V\big(\theta_0, (\pi^{\prime}_i, \bs{\pi}^{\bs{*}}_{-i})\big) - V\big(\theta_0, (\pi^{\prime}_i, \bs{\pi}^{\bs{*}}_{-i-j})\big) \Big)
= V(\theta_0, \bs{\pi^*}) \mathop{\ge}^{\eqref{V(pi)-ineq}} V(\btheta) \qedhere
\end{equation*}

Proof of \eqref{f0}. $\forall G \in \G\:$
\begin{equation*}
  f_0(G, \bs{\pi^*}) + \sum_{i \in \Ag} f_i(G, \bs{\pi}^{\bs{*}}_{-i})
  \mathop{=}^{\eqref{fidef}\eqref{fCdef}} V(\theta_0, \bs{\pi^*}) \mathop{\ge}^{\eqref{V(pi)-ineq}} V(\btheta) \qedhere
\end{equation*}
\end{proof}

\subsection{The first-price mechanism under perfect competition} \label{price1perfectsection}

Assume that
\begin{align}\label{perfect}
\forall G \in \G,\ \forall i \in \Ag\: &&
V(\btheta_{-i}) &= V(\btheta). &&
\phantom{\forall G \in \G,\ \forall i \in \Ag\:}
\end{align}
We call it a Project Management Model under \textbf{perfect competition}. For such a $\G$, let $\G^{\text{perf}(N)} = \G / m_1$ denote a Project Management Model under perfect competition and under the first frice mechanism.

\newtheorem*{price1perfect}{Theorem \ref{price1perfect}}
\begin{price1perfect}
The truthful strategy profile $\bs{\cp}$ under the first-price mechanism is a quasi-dominant equilibrium under perfect competition.
\end{price1perfect}

\begin{proof}
The required inequalities applied in our case are the following.

\begin{align}
\forall G \in \G^{\text{perf}(N)},\ \forall i \in \N,\ \forall \bs{s}_{-i} \in \bs{\Str}_{-i}\:&&
\E\Big(u_i\big(G^1, (\cp_i, \bs{s}_{-i})\big)\Big)
&\ge f_i\big(G, \bs{\pi}_{-i}(G, \bs{s}_{-0-i})\big) &&\label{fy-perf}
\end{align}
\begin{align}
\forall G \in \G^{\text{perf}(N)},\ \forall i \in \Ag,\ \forall \pi^{\prime}_i \in \P\:&&
V(\btheta) &\le f_i(G, \bs{\pi}^{\bs{*}}_{-i}) + \sum_{j \in N \setminus \{i\}} f_j\big(G, (\pi^{\prime}_i, \bs{\pi}^{\bs{*}}_{-i-j})\big)&& \label{f-i-perf}
\\ \forall G \in \G^{\text{perf}(N)}\:&&
V(\btheta) &\le f_0(G, \bs{\pi^*}) + \sum_{i \in \Ag} f_i(G, \bs{\pi}^{\bs{*}}_{-i})&& \label{f-0-perf}
\end{align}

We show that the following functions $f_i$ satisfy \eqref{fy-perf}, \eqref{f-i-perf} and \eqref{f-0-perf}.
\begin{align}
\forall i \in \Ag\: && f_i(G, \bs{\pi}_{-i}) &= 0 && \phantom{\forall i \in \Ag\:} \label{fi1def}
\\ && f_0(G, \bs{\pi}) &= V(\theta_0, \bs{\pi}) && \label{fC1def}
\end{align}

Proof of \eqref{fy-perf} for agents. $\forall G \in \G^{\text{perf}(N)},\  \forall i \in \Ag,\  \forall \bs{s}_{-i} \in
\bs{\Str}_{-i}\:$
\begin{equation*}
\E\Big(u_i\big(G^1, (\cp_i, \bs{s}_{-i})\big)\Big) \mathop{=}^{\eqref{u(cp)=0}} 0 \mathop{=}^{\eqref{fi1def}} f_i(G, \bs{\pi}_{-i})
\end{equation*}

Proof of \eqref{fy-perf} for the principal. $\forall G \in \G^{\text{perf}(N)},\  \forall \bs{s}_{-0} \in \bs{\Str}_{-0}\:$
\begin{equation*}
\E\Big(u_0\big(G^1, (\cp_0, \bs{s}_{-0})\big)\Big)
\mathop{=}^{\eqref{V(pi)-ineq}} V\big(\theta_0, \bs{\pi}(G, \bs{s}_{-0})\big)
\mathop{=}^{\eqref{fC1def}} f_0\big(G, \bs{\pi}(G, \bs{s}_{-0})\big)
\end{equation*}

Proof of \eqref{f-i-perf}.
Notice that
\begin{align}\label{V-monotone}
  \forall \theta_0 \in \Theta,\ \forall S' \subset S\: &&
  V(\theta_0, \bs{\cp}_{S'}) &\le V(\theta_0, \bs{\cp}_S), &&
  \phantom{\forall \theta_0 \in \Theta,\ \forall S' \subset S\:}
\end{align}
because in the principal-devil game $D(\theta_0, \bs{\cp}_S)$, the principal can apply her optimal strategy in $D(\theta_0, \bs{\cp}_{S'})$ with rejecting all agents in $S \setminus S'$. Therefore,
$\forall G \in \G^{\text{perf}(N)},\ \forall i \in \Ag,\ \forall \pi^{\prime}_i \in \P\:$
\begin{equation*}
f_i(G, \bs{\pi}^{\bs{*}}_{-i}) + \sum_{j \in N \setminus \{i\}} f_j\big(G, (\pi^{\prime}_i, \bs{\pi}^{\bs{*}}_{-i-j})\big)
\mathop{=}^{\eqref{fi1def}\eqref{fC1def}}
V\big(\theta_0, (\pi^{\prime}_i, \bs{\pi}^{\bs{*}}_{-i})\big)
\mathop{\ge}^{\eqref{V-monotone}}
V(\theta_0, \bs{\pi}^{\bs{*}}_{-i})
\mathop{\le}^{\eqref{V(pi)-ineq}}
V(\btheta_{-i})
\mathop{=}^{\eqref{perfect}}
V(\btheta),
\end{equation*}
where $(*)$ holds because the principal can reject $i$, therefore, his proposal $\pi^{\prime}_i$ cannot decrease the principal's maximin utility.

\medskip
Proof of \eqref{f-0-perf}. $\forall G \in \G^{\text{perf}(N)}\:$
\begin{equation*}
  f_0(G, \bs{\pi^*}) + \sum_{i \in \Ag} f_i\big(G, \bs{\pi}^{\bs{*}}_{-i})\big) \mathop{=}^{\eqref{fi1def}\eqref{fC1def}} V(\theta_0, \bs{\pi^*}) \mathop{=}^{\eqref{Veq}} V(\btheta) \qedhere
\end{equation*}
\end{proof}

\section{The first-price mechanism in general cases} \label{Bayesian1}

For calculation purposes, we introduce the following notations. For any game $G \in \G$ and any agent $i \in \N^+$, let $G_{+i}$ denote the game we can obtain from $G$ by excluding the option of the do-nothing execution from the capability tree $\theta_i$ of agent $i$. Similarly, $G_{-i}$ is the game when $i$ is forced to do nothing, or equivalently, he must be rejected. Finally, for a set of agents $S \subseteq \N^+$, $\G_S$ is the modified game when the agents in $S$ do not have the option to do nothing, and all other agents are forced to do nothing.
For any notation $* \in \{``+i", ``-i", ``S"\}$, we define $\G_* = \{G_* \where G \in \G\}$.
We modify the first-price mechanism accordingly, namely, the corresponding agents must be accepted or rejected, and the conspiracy-fearing strategy $\cp_0$ is defined by the maximin utility of the principal under these restrictions.
Let $\G^1_* = \G_* / m_1 = \{G^1_* \where G \in \G\} = \{G_*/m_1 \where G \in \G\}$ and $V_* = V_*(\theta_0, \bs{\pi})$ denote the principal's maximin utility in $\G^1_*$.
Notice that the proofs of Proposition~\ref{u(cp)=0} and Lemma~\ref{Veq} are valid in these games, as well.
Namely,
\begin{equation} \label{pcvgnew}
\E\Big(u_0\big(G^1_*, (\theta_0, \cp_0)\big)\Big) = V_*(\theta_0, \bs{\pi}),
\end{equation}
\begin{align} \label{cppnew}
\forall i \in \Ag\: &&
\E\big(u_i(G^1_*, \cp_i)\big) &= 0.&&
\phantom{\forall i \in \Ag\:}
\end{align}

Furthermore, it does not matter whether $i$ does not play at all or he takes part in the game but the planner rejects him for sure.
Therefore,
\begin{equation}\label{M-i}
  V_{-i}(\theta_0, \bs{\pi}) = V(\theta_0, \bs{\pi}_{-i}).
\end{equation}
The fact that the principal either accepts or rejects $i$ implies that
\begin{equation} \label{pmi}
V(\theta_0, \bs{\pi})
= \max\big(V_i(\theta_0, \bs{\pi}), V_{-i}(\theta_0, \bs{\pi})\big)
\mathop{=}^{\eqref{M-i}} \max\big(V_i(\theta_0, \bs{\pi}), V(\theta_0, \bs{\pi}_{-i})\big),
\end{equation}
and the planner
accepts $i$ if $V_i(\theta_0, \bs{\pi}) > V(\theta_0, \bs{\pi}_{-i})$,
and rejects if $V_i(\theta_0, \bs{\pi}) < V(\theta_0, \bs{\pi}_{-i})$.

\medskip
Let the \textbf{signed value of a proposal} $\pi_i$ be\index{signed value of a proposal $v^\pm_i$}\index{$v^\pm_i$: signed value of a proposal $\pi_i$}
\begin{equation} \label{sgnval}
v^\pm_i = v^\pm_i(\pi_i) = v^\pm_i(\theta_0, \bs{\pi})
= V_i(\theta_0, \bs{\pi}) - V_{-i}(\theta_0, \bs{\pi})
\mathop{=}^{\eqref{M-i}} V_i(\theta_0, \bs{\pi}) - V(\theta_0, \bs{\pi}_{-i}).
\end{equation}
We can easily see that $v^+$ is the positive part of $v^{\pm}$, namely,
\begin{equation}
v^+_i
\mathop{=}^{\eqref{vpdef}} V(\theta_0, \bs{\pi}) - V(\theta_0, \bs{\pi}_{-i})
\mathop{=}^{\eqref{pmi}}
\max\big(V_i(\theta_0, \bs{\pi}), V(\theta_0, \bs{\pi}_{-i})\big) - V(\theta_0, \bs{\pi}_{-i})
\mathop{=}^{\eqref{sgnval}} \max(0, v^\pm_i),  \label{vpvpm}
\end{equation}
and $v^\pm_i > 0$ if and only if $i \in \Acc$.

\begin{equation} \label{VMi+}
  V_i\big(\theta_0, (\pi_i + x, \bs{\pi}_{-i})\big) = V_i(\theta_0, \bs{\pi}) - x,
\end{equation}
because the two principal-devil games are the same except that the principal should pay $x$ more to agent $i$ when she accepts $\pi_i + x$ instead of $\pi_i$.
Therefore,
\begin{equation*}
v^\pm_i(\pi_i + x)
\mathop{=}^{\eqref{sgnval}}
V_i\big(\theta_0, (\pi_i + x, \bs{\pi}_{-i})\big) - V(\theta_0, \bs{\pi}_{-i})
\end{equation*}
\begin{equation} \label{vpm+x}
\mathop{=}^{\eqref{VMi+}}
V_i(\theta_0, \bs{\pi}) - x - V(\theta_0, \bs{\pi}_{-i})
\mathop{=}^{\eqref{sgnval}}
v^\pm_i(\pi_i) - x.
\end{equation}
This implies that for each agent, there is exactly one fair proposal with a given signed value.

Let the \textbf{strategy form}\index{strategy form $\fii$} of a strategy $s_i$ mean $\fii(s_i) = \{s_i + x \where x \in \R \}$.\index{$\fii$: strategy form} Let $\fii(\cp_i) = \{\cp_i + x \where x \in \R\}$ be called the \textbf{fair strategy form}.\index{strategy form $\fii$!fair strategy form $\fii(\cp_i)$}



Given $G$ and $\bs{s}_{-i}$, we have that $\E\big(u_i(G^1_i, s_i + x)\big) = \E\big(u_i(G^1_i, s_i)\big) + x$. With \eqref{vpm+x}, these imply that $\E\big(u_i(G^1_i, s_i)\big) + v^\pm_i(\pi_i)$ depends only on $\fii(s_i)$. We call this sum the \textbf{value of the strategy form}\index{value of a strategy form $V_i\big(\fii(s_i)\big)$}
\begin{equation*}
  V_i\big(\fii(s_i)\big) = V_i\big(\fii(s_i)\big)(G, \bs{s}_{-i}) = \E\big(u_i(G^1_i, s_i)\big) + v^\pm_i(\pi_i) = \E\big(u_i(G^1_i, \bs{s})\big) + v^\pm_i\big(\bs{\pi}(\bs{s})\big).
\end{equation*}

In each strategy form, there exists a strategy $s_i$, using a proposal $\pi_i$, for which $\pi_i + x$ would be accepted if $x > 0$ and rejected if $x < 0$. Then
\begin{equation*}
\E\big(u_i(G^1, s_i + x)\big) = \{ 0\text{ if }x < 0;\text{ and }V_i\big(\fii(s_i)\big) - x\text{ if }x > 0 \}.
\end{equation*}
For each capability tree $\theta \in \Theta$, let $\theta - x$ denote the same tree but with valuation decreased (or costs increased) by $x$. Clearly, $\pi^*(\theta - x) = \pi^*(\theta) + x$ and $\cp_i(\theta_i - x) = \cp_i(\theta_i) + x$. Therefore, applying Lemma~\ref{u(cp)=0}, we get the following. Each agent $i$ with fair strategy $\cp_i + x$ has a guaranteed expected utility $x$ in the case of acceptance, or formally,
\begin{equation} \label{fairpieq}
\E\big(u_i(G^1, s_i = \cp_i + x) \bigwhere i \in \Acc \big) = x.
\end{equation}

The following theorem shows that if all agents use a fair strategy, then after the choice of $\Acc$, the mechanism works efficiently. This implies that the planner chooses the set $\Acc$ for which the expected total utility \emph{minus the sum of the demanded profits of the agents in $\Acc$} is the largest possible; and this choice of $\Acc$ is the only inefficient decision throughout the game.
\begin{Theorem} \label{optimality1}
If in a game $G^1 \in \G^1$, all accepted agents use fair strategies, and the principal follows the conspiracy-fearing strategy, then they play optimally in $\btheta_{\Acc_0}$. Formally,
\begin{equation}
\E\big(u_{\Pl}(G^1, \bs{s})\big) = V(\btheta_{\Acc_0}), \label{psfair}
\end{equation}
provided that $s_0 = \cp_0$ and $\forall i \in \Acc\: s_i = \cp_i + x_i$.
\end{Theorem}

\begin{proof}
Using the notation $\Acc = \Acc(\cp_0, \theta_0, \bs{\pi})$,
\begin{equation*}
\E(u_{\Pl})
\mathop{\ge}^{\eqref{V(pi)-ineq}}
V(\theta_0, \bs{\pi})
= V_{\Acc}(\theta_0, \bs{\pi}_{\Acc})
\mathop{=}^{\eqref{Veq}} V_{\Acc}\big(\theta_0, (\bs{\theta - x})_{\Acc}\big)
= V_{\Acc}(\btheta_{\Acc_0}) - \sum_{i \in \Acc} x_i.
\end{equation*}
$\eqref{fairpieq}$ shows that if $i \in \Acc$, then $E(u_i) = x_i$.
By definition, if $i \in \Ag \setminus \Acc$, then $E(u_i) = 0$.
To sum up,
\begin{equation*}
\E(u_{\Pl})
= \E(u_0) + \sum_{i \in \Acc} \E(u_i)
 + \sum_{i \in \Ag \setminus \Acc} \E(u_i)
\end{equation*}
\begin{equation*}
= \Big( V_{\Acc}(\btheta_{\Acc_0}) - \sum_{i \in \Acc} x_i \Big) + \sum_{i \in \Acc} x_i + \sum_{i \in \Ag \setminus \Acc} 0
= V_{\Acc}(\btheta_{\Acc_0}) = V(\btheta_{\Acc_0}).
\qedhere
\end{equation*}
\end{proof}

Let a \textbf{fair-looking strategy}\index{strategy $s_i$!fair-looking strategy} of an agent $i \in \Ag$ mean a strategy which is statistically equivalent to a truthful strategy $\cp(\htheta_i)$ of an agent with a capability tree $\htheta_i$.
Namely, if $i$ was replaced with an agent with a capability tree $\htheta_i$ and the truthful strategy, then it would induce the very same probability distribution on the execution of the game $G$ excluding the execution of $\theta_i$ or $\htheta_i$.
Clearly, the truthful strategy of an agent is the fair strategy with profit 0, and all fair strategies are fair-looking strategies of the form $\cp(\theta_i + x)$. The most important example of fair-looking strategies is when the agent sends the proposal of his real capability tree but with costs increased by different amounts at different executions, and then he plays truthfully. For the sake of generality, we apply the definition of fair-looking strategies $\cp(\htheta_i)$ also for the principal.

\begin{Theorem} \label{interest1}
Consider an arbitrary game $G \in \G$ and an agent $i \in \Ag$. Suppose that every other player $j \in \Pl \setminus \{i\}$ uses a fair-looking strategy $s_j$ corresponding to $\cp^*(\hat{\theta}_j)$. Then the fair strategies $s_i = \cp_i + x$ maximize the value of the strategy form
\begin{equation} \label{interest1eq}
V\big(\phi(\cp_i)\big) = v^{\pm}_i\big(\hat{\theta}_0, \bs{\pi}(\bs{s})\big) + \E\big(u_i(G^1_i, \bs{s})\big).
\end{equation}
\end{Theorem}

\begin{proof}
There exists a game $\hat{G}$ in which each player $j \in \Pl \setminus \{i\}$ has the capability tree $\hat{\theta}_j$, and from the point of view of $i$, the game $G$ with the strategy subprofile $\bs{s}_{-i}$ is equivalent to $\hat{G}$ with $\bs{\cp}^{\hat{G}}_{-i} = \bs{\cp}(\bs{\hat{\theta}}_{-i})$. Therefore, for any $s_i \in \Str^G_i = \Str^{\hat{G}}_i$,

\begin{equation*}
\E\big(u_i(G^1_i, \bs{s})\big)
= \E\Big(u_i\big(\hat{G}^1_i, (s_i, \bs{\cp}_{-i})\big)\Big)
\end{equation*}
\begin{equation*}
= \E\Big(u_{\Pl}\big(\hat{G}^1_i, (s_i, \bs{\cp}_{-i})\big)\Big) - \E\Big(u_0\big(\hat{G}^1_i, (s_i, \bs{\cp}_{-i})\big)\Big)
- \sum_{j \in \Ag \setminus \{i\}} \E\Big(u_j\big(\hat{G}^1_i, (s_i, \bs{\cp}_{-i})\big)\Big)
\end{equation*}
\begin{equation*}
\mathop{\le}^{\eqref{V-opt}\eqref{pcvgnew}\eqref{cppnew}}
V(\btheta) - V_i\Big(\hat{\theta}_0, \big(\pi(s_i), \bs{\pi^*}(\hat{\theta}_{-i})\big)\Big) - \sum_{j \in \Ag \setminus \{i\}} 0
\end{equation*}
\begin{equation*}
\mathop{=}^{\eqref{sgnval}}
V(\btheta) - V\big(\bs{\hat{\theta}}_0, \bs{\pi^*}(\hat{\theta}_{-i})\big) - v^{\pm}_i\big(\hat{\theta}_0, \bs{\pi}(\bs{s})\big),
\end{equation*}
and therefore, $\eqref{interest1eq} \le V(\btheta) - V\big(\bs{\hat{\theta}}_0, \bs{\pi^*}(\hat{\theta}_{-i})\big)$ which is independent from $s_i$.
The only inequality we used is \eqref{V-opt}, and Proposition~\ref{efficient2} shows that if $s_i = \cp_i$, then we have equality here.
\end{proof}

$\E\big(u_i(s_i)\big) \leq V_i\big(\fii(s_i)\big)$, therefore, the strategy form of $i \in \Ag$ with the highest value gains him the highest potential for his expected utility. Theorem~\ref{interest1} implies that the fair strategy form has the highest value. Moreover,
\begin{equation*}
\E\big(u_i(\cp_i + x)\big) = \big\{x \text{ if }V_i(\cp_i) > x;\text{ and 0 otherwise}\big\},
\end{equation*}
which is a quite good way of the exploitation of this potential. Of course, this only shows that the fair strategy with appropriate profit is a good heuristics for an agent, under competition.

\bigskip

The following arguments show that under practically reasonable approximations or assumptions, the agents have very weak or no incentives in using not fair strategies.

Let us fix an agent $i \in \Ag$ and his belief at $t_0$ in a Bayesian Project Management Model.
This belief can be identified with a probability distribution $\mu$ on $\G$, where $\theta_i$ is a constant.
Let us fix the strategies of the other players $\bs{s}_{-i}$.
Now $s_i$ is the only parameter, and the stochastic variables are the random choice of $G$ from $\mu$, the pure strategies chosen from the (mixed) strategies $\bs{s}$ and the chance events.
Let $\Pr_i$ and $\E_i$ denote the probability and expectation on these stochastic parameters.
With $\pi_i = \pi(s_i)$, let us define the following stochastic variables.
\begin{align*}
 \text{(value) }&& V^\pm &= v^\pm_i(\pi^*_i) &&\\
 \text{(difference) }&& D = D(\pi_i) &= V^\pm - v^\pm_i(\pi_i) &&\\
 && e = e(\pi_i) &= \E_i(D \where D < V^\pm) &&
\end{align*}

In practice, $V^\pm$ and $D$ are typically almost independent and both have ``nice" distributions, therefore, we may expect that $\Pr_i(e < V^\pm) = \Pr\big(\E_i(D \where D < V^\pm) < V^\pm\big) \ge \Pr_i(D < V^\pm)$.

\begin{Theorem}
If $\bs{s}_{-i}$ consists only of fair-looking strategies and $\Pr_i\big(\E_i(D \where D < V^\pm) < V^\pm\big) \ge \Pr_i(D < V^\pm)$ holds for all $\pi_i$, then a fair strategy of $i$ provides him the highest expected utility.
\end{Theorem}

\begin{proof}
Let $\bar{u}(s_i) = \E_i\big(u_i(s_i)\big)$ and let $x$ be the number by which $\bar{u}(\cp_i + x)$ is the largest possible. What we need to prove is that $\forall s_i\: \bar{u}(\cp_i + x) \ge \bar{u}(s_i)$.

For calculation purposes, let us allow for $i \in \Ag$ to submit his fair proposal with the signed value $v^\pm_i(\pi_i)$, namely, submitting $\pi_i + v^\pm_i(\pi^{\prime}_i)$, for an arbitrary $\pi^{\prime}_i$ chosen by $i$. This is an invalid action in $G$, because $i$ does not know the values of his possible proposals at this time, but for calculation, we allow him this proposal, and we denote the fair strategy but with this proposal by $\text{fair}_i(\pi^{\prime}_i)$.

$\pi_i$ is accepted if and only if $v^\pm_i(\pi_i) > 0$, or equivalently, $D < V^\pm$. By the equation
\begin{equation*}
\bar{u}(s_i) = \Pr_i(i \in \Acc) \cdot \E_i\big(u_i(s_i) \bigwhere i \in \Acc\big),
\end{equation*}
we get $\bar{u}(\cp_i + e) \mathop{=}\limits^{\eqref{fairpieq}} \Pr_i(e < V^\pm)e$, and $\bar{u}\big(\text{fair}_i(\pi_i)\big) = \Pr_i(D < V^\pm)e$, whence we can simply get that
\begin{equation*}
\bar{u}(\cp_i + x) - \bar{u}(s_i) =
\end{equation*}
\begin{equation*}
= \Big(\bar{u}(\cp_i + x) - \bar{u}(\cp_i + e)\Big) + \Big(\Pr_i(e < V^\pm) - \Pr_i(D < V^\pm)\Big)e + \Big(\bar{u}\big(\text{fair}_i(\pi_i)\big) -
\bar{u}(s_i)\Big).
\end{equation*}

\begin{itemize}
\item $\bar{u}(\cp_i + x) - \bar{u}(\cp_i + e) \geq 0$ by the definition of $x$.
\item If $e \leq 0$, then $\bar{u}(s_i) \leq 0 = \bar{u}(\cp_i) \leq \bar{u}(\cp_i + x)$, therefore, $s_i$ cannot be a better strategy. Assume that $e > 0$. In this case, because of the assumption in the theorem,
\begin{equation*}
\big(\Pr_i(e < V^\pm)-\Pr_i(D < V^\pm)\big)e \geq 0.
\end{equation*}
\item Theorem~\ref{optimality1} implies that $\bar{u}\big(\text{fair}_i(\pi_i)\big) - \bar{u}(s_i) \geq 0$.
\end{itemize}

To sum up, $\bar{u}(\cp_i + x) - \bar{u}(s_i) \geq 0$, which proves the theorem.
\end{proof}

\subsection{Negative results} \label{negative}

First, we show what kind of deviations from fair strategies we should expect. Consider a case when an agent $i \in \Ag$ will be asked to work on either a larger or a smaller task, and this choice depends only on the proposals of others. Then $i$ should demand more profit (payment beyond his costs) in the former case and less profit in the latter case, due to the different competition settings in the two cases. Say, if the larger task has the very same structure, but it has double valuations for everyone, then it is easy to check that the agents should demand exactly double profit in this case. And this is not a fair proposal.

The opinion of the author is that a rational agent would typically use a strategy very similar to a fair strategy, mainly with deviations like in the previous case. This strategy would still provide approximate efficiency. However, we describe a situation when an agent would use a highly non-fair-looking strategy, under a very extreme belief system.

Consider an agent $i$ working a subtask. Suppose that agent $i$ knows that his capability tree is clearly the best one for the task, and agent $j$ is clearly the second best. Suppose that $i$ knows very well the capability tree and belief of $j$, but $j$ believes that with probability close to 1, $i$ knows almost nothing about the capability tree of $j$. In this case, it can easily happen that $i$ can predict well the proposal of $j$. If $i$ does not know the capability trees of agents working on other tasks, then $i$ should submit the proposal of agent $j$ but claiming slightly smaller payment, in order to very slightly underbid $j$. This can provide $i$ with almost the maximum possible expected utility $V\big(\theta_0, (\pi^*_i, \bs{\pi}_{-i})\big) - V(\theta_0, \bs{\pi}_{-i})$.
Further techniques related to this question are shown in Appendices \ref{modification} and \ref{riskaverse}.

We note that the belief structure of the example above is exceptional in the sense that under common prior assumption, this must have very low probability, and even this unlikely event causes moderate inefficiency.

\section{First-price mechanism with commonly known capability trees} \label{commonknowsection}

Assume that the capability trees of all players are common knowledge between them, and the principal uses the conspiracy-fearing strategy $\cp_0$. We emphasize that the planner does not know the capability trees and the entire executions might be completely hidden.

In order to reduce the complexity of this section, somewhat imprecisely, we introduce a positive infinitesimal amount of money $\eps$, and we consider it at the decision about which agents to accept, but \textbf{we treat} $\bs{\eps = 0}$ otherwise. (This could be handled by limits of $\eps$-equilibria, where $\eps \rightarrow 0$.) Furthermore, for the sake of simplicity, we assume that there is no tie, namely, $\{\Acc\} = \argmax\limits_{S \subseteq N^+} V_S(\btheta_{S \cup \{0\}})$, or equivalently, if $\bs{\pi}\big(\theta_0, \bs{\pi^*}(\btheta_{-0})\big) = \bs{\pi^*}$, then $\Acc = \big\{i \in \N^+: v^+_i\big(\theta_0, \bs{\pi^*}(\btheta_{-0})\big) > 0\big\}$.

First, we show an example why we should expect multiple equilibria when the capability trees are common knowledge. Assume that there are 3 agents. Agent $A$ could complete one half of the project with a cost of 4, agent $B$ could complete the other half also with a cost of 4, and agent $C$ could complete the entire project with a cost of 10. Now, for any $x \in (4, 6)$, if agents $A$, $B$ and $C$ ask for $x$ and $10 - x - \eps$ and 10, respectively, then this is an ($\eps$-)equilibrium.

Consider the strategy profile $\bs{\cp + c} = (\cp_i + c_i)_{i \in N}$, where the vector $c_i$ satisfies
\begin{alignat}{2}
c_i &=
\begin{cases} \label{cidef}
   0 & \text{ if } i = 0 \text{ or } v^+_i(\bs{\pi^*}) = 0,
\\ v^+_i\big(\pi^*_i, (\bs{\pi^* + c})_{-i}\big) - \eps > 0 & \text{ if } i \ne 0 \text{ and } v^+_i(\bs{\pi^*}) > 0.
\end{cases}
\end{alignat}
Or equivalently, the choice of $\bs{c}$ satisfies that if $v^+_i(\bs{\pi^*}) > 0$, then $\big($ $v^+_i(\bs{\pi}) = \eps$ and $c_i > 0$ $\big)$, otherwise $c_i = 0$. This expresses that all players use a fair proposal, the best set of agents overbid their real costs by $\eps$ less than what would cause a tie, while others submit truthful proposals.

This way, $u_0 = V(\theta_0, \bs{\pi^* + c}) = V(\bs{\theta - c})$, and for $i \in N^+$, $u_i = c_i$, and Theorem~\ref{optimality1} implies that they maximize the expected total utility, therefore,
\begin{equation} \label{V-c}
V(\btheta) = V(\bs{\theta - c}) + \sum_{i \in N} c_i.
\end{equation}

The vector $\bs{c}$ is an $\eps$-Pareto-efficient core solution of the superadditive coalitional game for the agents defined by $V$. The existence of such a vector is a very simple fact and well-known in cooperative game theory. We can find one by the following algorithm. We start from $\bs{c} \equiv 0$, and in an arbitrary ordering, we keep increasing each value $c_i$, $i \in \Acc$ as long as $v^+_i(\bs{\pi^* - c}) > \eps$. The resulting vector will satisfy \eqref{cidef}. It is also easy to see that each $c_i \le v_i^+(\bs{\pi^*})$.\footnote{The payoff of each player $i$ by any core solution of a superadditive coalitional game is at most the marginal contribution of $i$. Otherwise the other players would get more payoff in total by excluding $i$ from the coalition.}

$\bs{\cp + c}$ maximizes the expected total utility and provides at most as much expected utility for the agents as under the second-price mechanism. Therefore, by exclusion, \emph{the principal gets at least as much utility here as under the second-price mechanism}.

Now we show that $\bs{\cp + c}$ is also a quasi-dominant equilibrium, which completes the proof of Theorem~\ref{price1common}.

\begin{Theorem}[Specified version of Theorem~\ref{price1common}] \label{price1commonspec}
If the capability trees $\btheta$ are commonly known, then $\bs{\cp + c}$ is a weak quasi-dominant equilibrium\index{quasi-dominant equilibrium!weak quasi-dominant eq.} under the first-price mechanism.
\end{Theorem}

\begin{proof}
We need to find functions $f_i$ which satisfy \eqref{fi5W} and \eqref{f-i5W}, which equations are the following in our case.
\begin{align}
\forall G \in \G,\  \forall i \in N,\  \forall \bs{s}_{-i} \in \bs{\Str}_{-i}\:&&
\E\Big(u_i\big(G, (\cp_i + c_i, \bs{s}_{-i})\big)\Big) &\ge f_i\big(G, \bs{\pi}_{-i}(G, \bs{s}_{-i}) \big) \label{fiC}
\end{align}
\\ $\forall G \in \G,\ \forall i \in N,\ \big( \forall \bs{s} \in \bs{\Str}\bigwhere
\bs{\pi}_{-i}(G, \bs{s}_{-i}) = (\bs{\pi^* + c})_{-i} \big)\:$
\begin{equation}
\E\big(u_N(G, \bs{s})\big) \le f_i\big(G, (\bs{\pi^* + c})_{-i}\big)
+ \sum_{j \in N \setminus \{i\}} f_j\Big(G, \big(\pi(G, s_i), (\bs{\pi^* + c})_{-i-j}\big)\Big) \label{f-iC}
\end{equation}

\noindent
We prove that the following functions $f_i$ satisfy \eqref{fiC} and \eqref{f-iC}. $I(\text{true}) = 1$ and $I(\text{false}) = 0$.

\begin{align} \label{fidefC}
\forall i \in N^+\:&&&&
f_i(G, \bs{\pi}_{-i}) &= I \big( i \in \Acc(\pi^*_i + c_i, \bs{\pi}_{-i}) \big) \cdot c_i &&
\end{align}
\begin{equation} \label{fCdefC}
f_0(G, \bs{\pi}) = V(\theta_0, \bs{\pi})
\end{equation}

\noindent
Proof of \eqref{fiC} for the agents.
\noindent $\forall G \in \G^1,\  \forall i \in N^+,\  \forall \bs{s}_{-i} \in \bs{\Str}_{-i}\:$
\begin{equation*}
\E\Big( u_i \big(G^1, (\cp_i + c_i, \bs{s}_{-i})\big) \Big)
\mathop{=}^{\eqref{fairpieq}} I \Big( i \in \Acc \big(\pi^*_i + c_i, \bs{\pi}_{-i}(G, \bs{s}_{-i}) \big) \Big) \cdot c_i
\mathop{=}^{\eqref{fidefC}} f_i\big(G, \bs{\pi}_{-i}(G, \bs{s}_{-i})\big)
\end{equation*}

\noindent
Proof of \eqref{fiC} for the principal. $\forall G \in \G^1,\  \forall \bs{s}_{-0} \in \bs{\Str}_{-0}\:$
\begin{equation*}
\E\Big( u_0 \big(G^1, (\cp_0, \bs{s}_{-0}) \big) \Big)
\mathop{\ge}^{\eqref{V(pi)-ineq}} V\big(\theta_0, \bs{\pi}(G, \bs{s}_{-0})\big)
\mathop{=}^{\eqref{fCdefC}} f_0 \big(G, \bs{\pi}(G, \bs{s}_{-0})\big)
\end{equation*}

\noindent
Proof of \eqref{f-iC} for the agents. Let $\Acc^* = \Acc\big(\pi(G, \bs{s}_i), (\bs{\pi^* + c})_{-i}\big)$.
\medskip

\noindent $\forall G \in \G^1,\ \forall i \in N^+,\ \big( \forall \bs{s} \in \bs{\Str} \bigwhere \bs{\pi}_{-i}(G, \bs{s}_{-i}) = (\bs{\theta + c})_{-i} \big)\:$
\begin{equation*}
f_i\big(G, (\bs{\pi^* + c})_{-i}\big)
+ \sum_{j \in N \setminus \{i\}} f_j\Big(G, \big(\pi(G, s_i), (\bs{\pi^* + c})_{-i-j}\big)\Big)
\end{equation*}
\begin{equation*}
\mathop{=}^{\eqref{fidefC}\eqref{fCdefC}\eqref{cidef}}
V \Big(\theta_0, \big(\pi(G, s_i), (\bs{\pi^* + c})_{-i}\big)\Big) + \sum_{j \in \Acc^* \cup \{i\}} c_j
\mathop{\ge}^{\eqref{V-monotone}}
V \big(\theta_0, (\bs{\pi^* + c})_{-i}\big) + \sum_{j \in \Acc^* \cup \{i\}} c_j
\end{equation*}
\begin{equation*}
\mathop{=}^{\eqref{cidef}}
V \big(\theta_0, (\bs{\pi^* + c})_{-i}\big) + v^+_i(\theta_0, \bs{\pi^* + c}) + \sum_{j \in \Acc^* \cup \{i\}} c_j
\mathop{=}^{\eqref{vpdef}}
V (\theta_0, \bs{\pi^* + c}) + \sum_{j \in \Acc^* \cup \{i\}} c_j
\end{equation*}
\begin{equation*}
= V_{W^*} \big(\theta_0, (\bs{\pi^* + c})_{W^*}\big) + \sum_{j \in \Acc^* \cup \{i\}} c_j
= V_{W^*} (\theta_0, \bs{\pi}^{\bs{*}}_{W^*}) - \sum_{j \in \Acc^*} c_j + \sum_{j \in \Acc^* \cup \{i\}} c_j
\ge V_{W^*} (\theta_0, \bs{\pi}^{\bs{*}}_{W^*})
\end{equation*}
\begin{equation} \label{line1981}
= V(\theta_0, \bs{\pi^*})
\mathop{=}^{\eqref{Veq}}
V(\btheta)
\mathop{\ge}^{\eqref{V-opt}}
 \E\big(u_N(G^1, \bs{s})\big)
\end{equation}

\noindent
Proof of \eqref{f-iC} for the principal. Let $\Acc^* = \Acc(\theta_0, \bs{\pi^* + c})$.
\medskip

\noindent
$\forall G \in \G,\ \big( \forall \bs{s} \in \bs{\Str}\bigwhere \bs{\pi}(G, \bs{s}_{-0}) = \bs{\pi^* + c} \big)\:$
\begin{equation*}
f_0\big(G, (\bs{\pi^* + c})\big)
+ \sum_{i \in N^+} f_i\big(G, (\theta_0, \bs{\pi^* + c})\big)
\mathop{=}^{\eqref{fCdefC}\eqref{fidefC}} V(\theta_0, \bs{\pi^* + c}) + \sum_{i \in \Acc^*} c_i
\mathop{=}^{\eqref{line1981}} \E\big(u_N(G^1, \bs{s})\big) \qedhere
\end{equation*}
\end{proof}



\section*{Acknowledgements}

I am thankful to \textbf{Péter Eső} for his suggestions for revision of the paper, and to \textbf{Miklós Péter Pintér} for the suggestions about the earlier version (MSc Thesis) of the paper. I am also very grateful for the help of
Mih\'aly B\'ar\'asz,
P\'eter Bencz\'ur,
Gergely Csap\'o,
Katalin Friedl,
L\'aszl\'o Lov\'asz,
\'Eva Tardos,
Alexander Teytelboym,
\'Agnes T\'oth,
L\'aszl\'o V\'egh
and many other people.

\bibliography{synchronization}

\newpage
\begin{center}
{\huge Appendix}
\end{center}
\bigskip
\appendix

\section{Negative results and counterexamples}

\subsection{Dominant strategy equilibrium is not possible} \label{nodominant}

Consider a project consisting of two tasks, carried out by two players: $A$ and $B$. First, $A$ completes the first task with two possible qualities: high or low. Then $B$ completes the second task, using two different technologies in the two cases.

Assume that $B$ can make some preparation for using either or both technologies before $A$ starts. Preparation to each case costs $1$, but it saves him $3$ cost in the corresponding case. Which preparation(s) $B$ makes remains hidden throughout the game, therefore, this decision has no effect on the transfer to him. The probability of the high quality depends on the non-observable effort level of $A$. Therefore, the optimal decision of $B$ depends on the hidden strategy of $A$. This proves the impossibility of a dominant strategy equilibrium.

\subsection{A counterexample about belief-independence} \label{belief}

Assume that there is a game with perfect information, and there is an equilibrium $\bs{s}$ with which each player uses only a part of his information. Consider now the same game with an imperfect information setup satisfying that all players still have the information which is required to follow the same strategy in $\bs{s}$. There is a well-known principle saying that under some assumptions, $\bs{s}$ must be an equilibrium in this new game. The following example shows that this principle is not valid in dynamic stochastic environments including our Project Management Model. This is the reason why we use different techniques.

There are two players: $A$ and $B$. At the beginning $A$ decides whether to participate in the game for a cost of $1$. If not, then the game ends with utilities $(0, 0)$. If yes, then $B$ can choose not playing, resulting in utilities $(-1, 0)$, or continuing. Then $B$ should make a decision between two options called red and blue, and then $A$ makes a guess for the decision. If the guess is right, then $A$ gets $5$, and $B$ gets $-1$, which are utilities $(4, -1)$ in total, but if wrong, then $B$ gets $5$ and $A$ gets $-1$, which are utilities $(-2, 5)$ in total.

In the game with perfect information, $B$ would not play, because $A$ would see his choice, and would make the right guess. Therefore $A$ is better off not participating in the game. To sum up, non-participation strategy for both players (with arbitrary decision and guess) is the (essentially) unique equilibrium. With this equilibrium, $A$ does not use the information on $B$'s choice of color. Therefore, the principle would say that this must be an equilibrium under imperfect information even if $A$ does not observe the decision of $B$.

Consider now the same game when $A$ cannot observe the decision of $B$. In this case, $B$ would be happy to participate, because this would provide him with utility $\frac{5-1}{2} = 2$ by making a uniform random choice. Therefore, $A$ should participate as well, and make a random guess, which provides her with utility $\frac{5-1}{2} - 1 = 1$. This contradicts with the principle.

\subsection{Collusion under the second-price mechanism} \label{problem2}

Consider a case when two agents can send proposals defined in such a way that their individual work is useless without each other. For example, agent 1 has a pink dummy spaceship in neo-Hawaiian style, and he offers to bring it; and agent 2 offers to complete the task provided that he gets a pink dummy spaceship in neo-Havaiian style, as a tool for his work. And, of course, nobody else has or requires such a toy. Assume that these proposals have positive values. If agent 2 asks for a payment decreased by a constant $x$, then the value of both proposals would increase by $x$. Consequently, they would get $2x$ more second-price compensation in total, which means $x$ more total utility for them. Using this trick, these players can get as much payment as they want.

\section{Simplified model} \label{simple-sec}

The primary purpose of this section is to help the comparison of our model with earlier models. The secondary goal is to show the reasons why the difficulties in the original model were necessary.

\medskip

We have a set $N = \{0, 1, 2, ..., n\}$ of players, where $0$ is the principal (personalizing the owner of a project) and $N^+ = \{1, 2, ..., n\}$ are the agents (who will potentially be hired). There is a finite number of periods $T = \{0, 1, 2, \dots t_{\max}\}$, each period consists of a finite number of steps. $\Theta$ denotes the finite set of possible types.

\medskip
\noindent
Period $0$ consists of the following steps.
\begin{itemize}
\item Nature chooses the type vector $\btheta_N^0 = (\theta_0^0, \theta_1^0, \theta_2^0, ..., \theta_n^0) \in \Theta^N$ from a fixed and publicly known joint distribution.
\item Each agent $i \in N^+$ privately observes his type $\theta_i^0 \in \Theta$.
\item The planner observes $\theta_0^0$.
\item Each agent $i \in N^+$ submits a report $\hat{\theta}_i^0$ about his type, or quits with utility $0$.
\item The planner chooses a (winner) subset of agents $\Acc \subset N^+$ from those who did not quit. $\Acc_0 = \Acc \cup \{0\}$.
\end{itemize}

\smallskip
\noindent
Each period $t \in \{1, 2, \dots t_{\max}\}$ consists of the following steps.
\begin{itemize}
    \item Each player $i \in \Acc$ makes a private decision $x_i^t \in X$.
    \item A public consequence vector $\bs{c}_N^t \in \C^N$ is announced. If $i \in \Acc$, then $c_i^t = \kappa(\theta_i^{t-1}, x_i^t)$ for a given function $\kappa \: \Theta \times X \rightarrow \C$. If $i \notin \Acc$, then $c_i^t = \emptyset$.
    \item Next, there are $|\Acc|$ more steps, indexed by the elements of $\Acc$, say, in increasing order. In the $i$th step, nature chooses $\theta_i^t \in \Theta$ from a probability distribution $\mu(\theta_i^{t-1}, x_i^t, \bs{c}_N^t)$ (where $\mu \: \Theta \times X \times \C^N \rightarrow \Delta(\Theta)$ is given) and player $i$ privately observes it, and sends a report $\hat{\theta}_i^t$ about it.
\end{itemize}

At the end, the transfers $y_i$ from the principal to agent $i$ are determined as the function of the entire history of consequences $\bs{c}_N^T$ and the history of communication.

\smallskip
The utility of each agent $i \in \Acc^+$ is $u_i = v(\theta_i^{t_{\max}}) + y_i$ (for a given $v \: \Theta \rightarrow \R$).

\smallskip
The utility of the principal is $u_0 = v(\theta_i^{t_{\max}}) - \sum\limits_{i \in \Acc^+} y_i$.



\subsection{Interpretation}

The following interpretation (or special case) was used in the Example (Section~\ref{example1}), this is the primary motivation of the model.
The type of each agent can be interpreted as his working capabilities starting from the current state of the working process, but including his costs which have already arisen. The valuation of the final state $v(\theta_i^{t_{\max}})$ is the negative of the total cost of agent $i \in \Acc^+$. The type of the principal can be identified with the history of the consequences of the agents. $v(\theta_0^{t_{\max}})$ shows the value for the principal of the entire result produced by the agents.
The principal may make no decisions and provide no consequences.

The changes of $\theta^i_t$ are interpreted as chance events during the working process of agent $i \in \Acc^+$.
The unusual and counterintuitive usage of them is a side effect of the discretization of the model.
In order to understand the right interpretation of it, we suggest understanding the main model, Section~\ref{model}.

We note that using consequences or public decisions (or both) are equivalent according to the following reductions. On one hand, a consequence $c$ of a player $i$ can be replaced by a public decision $c' \in A$ of player $i$ so that if $c' \ne c$, then $\theta^i_{t_{\max}}$ will satisfy $\theta^i_{t_{\max}} = - \infty$ (or a sufficiently low value guaranteeing that $i$ will never want to choose $c'$). On the other hand, having a public decision is equivalent to having a private decision with a consequence rule which reveals the decision.

\subsection{A simplified and very weak version of the results}

The truthful strategy of a player $i$ means that he always reports $\hat{\theta}_i^t = \theta_i^t$ and makes the socially optimal private decisions $x_i^t$ according to the efficient stochastic dynamic joint strategy corresponding to the reported states.

\smallskip

The pivot mechanism means the following. All agents are believed to be truthful and they receive from the principal the flow marginal contribution to the expected social welfare according to the reported states, deduced by the valuation of his own reported final state.

Formally, denote by $V(\btheta_S)$ the maximum possible total expected utility of the set of players $S$ if $\btheta_S$ is the vector of their types and they play the efficient joint strategy.
For each vector $\btheta_N^T$, each agent $i \in \Acc^+$ and each period $t \in T$, we use the notation $\bs{\theta^-}[i, t]$ and $\bs{\theta^+}[i, t]$ expressing the current states of the other players just before and after $i$ receives his new state in period $t$. Namely, $\bs{\theta^-}[i, 0] = \btheta_{N \setminus \{i\}}^0$, $\bs{\theta^+}[i, 0] = \btheta_N^0$, and for $t > 0$,
$\bs{\theta^-}[i, t] = \big( (\theta_j^t)_{j \in \Acc; j < i}, (\theta_j^{t-1})_{j \in \Acc; j \ge i} \big)$ and
$\bs{\theta^+}[i, t] = \big( (\theta_j^t)_{j \in \Acc; j \le i}, (\theta_j^{t-1})_{j \in \Acc; j > i} \big)$.
Then the payment from the principal to each agent $i$ is defined as follows.
\begin{equation} \label{simple-payment}
  t_i = \sum_{t \in T} \big( V(\hat{\theta}^+[j,t]) - V(\hat{\theta}^-[j,t]) \big) - v(\hat{\theta}_i^{t_{max}}).
\end{equation}

\begin{Theorem} \label{simple-thm}
  The pivot mechanism implements efficient outcome, namely, the truthful strategy profile is a perfect Bayesian equilibrium.
  The mechanism is individually rational for the agents: each agent can secure himself utility at least 0.
\end{Theorem}

Without proper definitions, we also mention the followings.
\begin{enumerate}
  \item The mechanism avoids free-riding.
  \item We do not need to assume any common prior, and Theorem~\ref{simple-thm} is true with a stronger equilibrium concept.
  \item If we modify \eqref{simple-payment} by excluding $t = 0$ from the summation, then we get to another mechanism called first-price mechanism. This is collusion-resistant in a sense, but in exchange, it is only approximately efficient and requires a strong competition. 
  \item We do not really need to assume that the capabilities of the principal is observed by the planner.
\end{enumerate}

\subsection{The disadvantages of using the simplified model}

\begin{enumerate}
  \item The simplified model defines the capabilities of the players dependent to each other, therefore, it is hard to speak about the capabilities of one player. This makes difficult to define free-riding as well as the level of competition (e.g.\ perfect competition). These also make the simplified model inconvenient for showing why the first-price mechanism works well in practice.
  \item In a continuous-time model, it is a more natural assumption that the chance events occur (new private states are received) one by one. Same-time chance events are just degenerate cases, and we can eliminate them by an arbitrary tie-breaking rule. In the discretized model, this natural issue becomes an unnatural assumption.
  \item Partially because of these two issues (\emph{1.} and \emph{2.}), the original (indirect) mechanism would be very difficult to define, and it would be even more difficult to show why it works much better in practice than the (direct) pivot mechanism.
  \item The assumption that the planner knows the initial type of the principal looks a strong assumption here. In the simplified model, it is more difficult to show why it is not the case.
  \item The discretization makes the information structure of the players a bit messy. We did not even use a very precise definition about what we mean that each chance event is observed first by the corresponding player. Separation of information and beliefs in the main model solves it nicely.
  \item The main model is a bit more general about the belief structures.
\end{enumerate}

\section{Comparison with existing models} \label{comparison}

 The simplified model in Appendix~\ref{simple-sec} may be helpful for the following comparisons.

\subsection{Comparison to Athey--Segal} \label{AtSeComp}

Similar questions in dynamic mechanism design were analyzed by Athey and Segal (2013) \cite{AtSe}.
Here we consider a comparison between their model with the ``balanced team mechanism" and our Project Management Model with our second-price mechanism.

In their model, the initial state $\theta_N^0 \in \Theta^N$ is assumed to be publicly known. In contrast, in our Project Management Model, all players but the principal has a private initial state chosen by nature, with no prior, meaning that our solution concept will be ex post with respect to the initial types. But we assume that each agent can participate only after he agrees with the principal, otherwise he exists with utility 0.

Their model uses a countable number of time periods. During each period, the players may receive private signals which they should report at the end of the period. Therefore, in that paper they had to use ex post incentive compatibility with respect to the same-period signals of the other players. E.g.\ they had to consider both of the possibilities that an agent $i$ misreports depending on the same-period chance event of $j$, and $j$ misreports depending on the same-period chance event of $i$. However, either of them would not need to be considered if reports were immediate. With this extra requirement, they implemented efficiency in a perfect Bayesian equilibrium. In contrast, we are using quasi-dominant strategy equilibrium, which is a stronger concept, and we will shortly see why it is an important difference.

We emphasize that the balanced mechanism of Athey and Segal also assume that at the end of each period, it cannot happen that an agent $j$ has already observed some information about the private state of $i$ which $i$ has not observed yet. Also, these periods should be very short in order to utilize all possibilities of making decisions dependent of all earlier chance events of all other players. Therefore, the difference is essentially that Athey and Segal assume this requirement at frequent time points, and we assume it at all time points, hereby avoiding the problems with same-period chance events. This difference does not extend to the (initial) types of the players, just to the changes of the states (chance events) during the game.

In their paper, first they show a mechanism for dependent capability trees (types) but with unbalanced transfers. This is a different problem from ours. Then they consider independent capability trees and balanced transfers, this is the more related to ours.

The unbalanced team mechanism is the following. The agents always report all their private information, and according to these reports, the mechanism makes the socially efficient public decision, makes recommendations for the private decisions, and everybody receives the monetary transfer equal to the sum of the valuations of the other agents. They show that truth-telling is a perfect Bayesian equilibrium.

Their ``balanced team mechanism" is the following. Whenever an agent $i$ reports a new private signal, the player gets the monetary transfer equal to the expected change on the total expected valuations of all other players during the game. This (signed) monetary transfer is paid by the other agents who share this \emph{equally}.

We show two examples about weaknesses with the proposed two perfect Bayesian equilibria in the two models.

Consider first the unbalanced mechanism, and the following setup with two agents. In each round, each agent receives a signal \$1000 or -\$1 independently, with probabilities $1/2$. Then they make a public decision ``yes'' or ``no''. If they choose ``no'', then both of their valuations are 0. But if they choose ``yes'', then their valuations equal the amount in the message.
Consider the case when an agent receives signal $-\$1$. If he reports $\$1000$ instead, then this costs him at most $\$2$, but this provides at least $\$999$ more utility to the other agent. Therefore, reporting $\$1000$ as long as the other agent also does so is another perfect Bayesian equilibrium in the infinite-horizon game.

This seems to be not only an abstract counterexample. Similar but less extreme phenomena seems to be very common, unless if the valuations are public information. Namely, whenever the report of an agent has a significant effect on his own reported valuation, then he may report higher valuation in order to build a long-term cooperation relationship with the others, even if his fake report causes some loss in social efficiency.

Consider now the balanced mechanism with the following setup with $101$ agents including Alice, Bob playing a repeated game. In each round, there is a good to be allocated to one agent. In every round, the true valuation of Alice and Bob of the good are around $\$1000$, and the valuations of all other agents are clearly lower. In some rounds, Alice knows in advance that her valuation is $\$1000$; the valuation of Bob is either $\$999$ or $\$1001$ with probability $1/2$ for each, and he gets to know which one just before the decision. If Bob receives the signal $\$1001$, but he reports $\$999$ instead, then Bob will get utility $\$500$ instead of $\$1001 - \$500 = \$501$, but Alice gets utility $\$1000 - \$500 / 100 = \$995$ instead of $\$500 / 100 = \$5$. Assume that Alice and Bob change role alternately. Then reporting $\$999$ as long as the other agent does so is another perfect Bayesian equilibrium which is better for Alice and Bob, but not socially efficient.

In contrast to their balanced mechanism, in our mechanism, we do not share the transfer equally, but we do the following. First, we choose an arbitrary ordering of the concurring chance events, excluding the reports of the initial states which we keep to be concurrent.
The marginal contribution by the reported initial state is compensated by the principal, but for reports of non-initial states of $i$, each agent $j \neq i$ pays to $i$ the change of $j$'s own expected total valuation.

A further important difference is that they assume publicly known initial states as a parameter of the mechanism.
In contrast, in our Project Management Model, all but one players have private capability trees with no prior.
(We think of the exceptional player as the owner of the project.)

Our mechanism also satisfies some important requirements: to be individually rational but to avoid free-riding.
We also present a version of our mechanism that is collusion-resistant, but which is less efficient under imperfect competition.

\subsection{Comparison to Bergemann--V\"alim\"aki} \label{BeVaComp}

Another related paper was presented by Bergemann and V\"alim\"aki (2010) \cite{BeVa}. They used an unbalanced dynamic mechanism for social allocations with independent private signals. Their setup is similar to the special case of our setup but without private decisions of the agents, and with discrete time periods. They introduce a mechanism different both from our mechanism and from the mechanism of Athey and Segal, and they show that truthfulness is a perfect Bayesian equilibrium under that mechanism. This equilibrium has similar weakness as in the result of Athey and Segal, as we show in the example below. In return, their mechanism satisfies an exit condition. This is a dynamic version of our conditions of individual rationality without avoiding free-riding, but they have this condition before each round  rather than just in the beginning. (This dynamic extension makes sense if the agents can play outside of the mechanism, which is not the case in our model.) The mechanism is roughly the following. In each of an infinite sequence of rounds, the agents are asked to report all of their private information, then we make the socially efficient public decision, and each agent $i$ pays the total expected decrement of the valuations of others caused by considering the preference of $i$ with the decision. Budget is not balanced.

Consider the following setup with at least three agents: Alice, Bob and others. In each round, there are two options: ``yes'' or ``no'', and the rounds are independent. Both Alice and Bob have a preference of about $\$1000$ for ``yes'', meaning that their valuations are $\$1000$ for ``yes'' and $\$0$ for ``no''. The other agents prefer ``no'', their total preference is about $\$2000$. The precise amounts will be private information, but their public \emph{a priori} distributions are highly concentrated. Suppose that everybody reports truthfully. If the total preference for ``no'' is the higher, then this will be the decision, and Alice and Bob will pay nothing. But if ``yes'' wins by a small amount $x$, then ``yes'' will be the decision and both agents pay their reported preference minus $x$, therefore, both of their utilities will be $x$. To sum up, the utilities of Alice and Bob are around 0 anyway.

Consider now what happens if Alice, instead of her truthful strategy, reports much more, say, $\$3000$ for ``yes''. If ``yes'' had won even with true reports, then there is no difference for Alice. If ``no'' had won by $x$, then this misreport changed the decision to ``yes'', but Alice had to pay her true preference plus $x$, therefore, she got $-x$ utility instead of 0. This is still a marginal difference. However, in both cases, Bob gets his more favorable decision and he does not have to pay anything, therefore, his utility equals his valuation, which is about $\$1000$. Therefore, Alice is able to have Bob getting a much higher utility, for at most marginal losses in her own utility. This works vice versa, and if both of them report a much higher preference, then ``yes'' will be the decision and they will not need to pay anything. Therefore, in the infinite repeated game, it is very likely that Alice and Bob will report higher preferences as long as the other one also does so.

This kind of problem cannot occur in all situations, but still, one should be very careful when applying this mechanism. In contrast, we use a much stronger equilibrium concept, therefore, our mechanism never has such problems.

\section{Observations for application and special cases} \label{obsapp}

\subsection{Risk-averse agents} \label{riskaverse}

Assume that an agent $i$ has a non-quasi-linear utility function $u_i(x_i, t_i)$, where $x_i$ is the execution of the capability tree of $i$. We assume that $\forall x\: \lim\limits_{t \rightarrow \infty} u_i(x, t) = \infty$. For example, if $u_i = w_i\big(v(x_i) + t_i\big)$ with a monotone increasing concave function $w_i$, then this expresses risk aversion. We define a proposal to be \textbf{reasonable} in the same way as the fair proposal with the only difference being that the agent demands a transfer $t_i$ satisfying
\begin{equation*}
u_i(\hat{\l}_i, t_i) \ge \sum\limits_{\chi \in \Chi_i} \delta^*(\chi)
\end{equation*}
By a reasonable proposal, in the case of acceptance, the expected utility of the agent is independent of the choices of the principal. If all proposals are reasonable, then the expected utility of the principal remains a function of her type and the proposals. If the agent is risk-neutral, then his reasonable proposal is fair. These are some reasons why reasonable proposals work ``quite well". We do not state that it is optimal in any sense, but a reasonable proposal may be better than a fair proposal in the risk-averse case.

We note that the evaluation of reasonable proposals can be much more difficult than of fair proposals, but for each agent $i$, if $h_i(x) = a_i - b_i\cdot e^{-\lambda_i x}$, then the analogous recursive evaluation algorithm of the proposals works as for the risk-neutral case.

\subsection{Modifications during the process} \label{modification}

In practice, the capability trees can be extremely difficult, therefore, submitting precise fair proposals cannot be expected. Hence players can only present a simplified approximation of their capability tree. Generally, such inaccuracies do not significantly worsen the optimality; nevertheless, this loss can be reduced far more with the following observation.

Assume that someone whose proposal has been accepted can refine his capability tree during the process. It would be beneficial to allow him to modify his proposal correspondingly. The question is: on what conditions?

The answer for us is to allow him to modify his capability tree if he pays the difference between the maximin utility of the principal with the original and the new proposals. Or, equivalently, the contract as a payment function automatically decreases by this difference. It is easy to see that, for an agent, whether and how to modify his proposal is the same question as which proposal to submit among the proposals with a given value. Consequently, Theorem~\ref{interest1} shows that the agent has incentives to change to his true fair proposal.

For a fair-like proposal $\pi(\htheta_i)$, this is equivalent to the following. For each possible modification, the agent extends $\htheta_i$ with a chance node with probabilities $1 - \eps$ and $\eps$ for continuing with the original and the modified tree, respectively, and we take the limit $\eps \rightarrow 0$. At such a chance node, in the limit $\eps \rightarrow 0$, the principal assigns 0 to the branch of not modifying; and to the modification, she assigns the difference between the values of the reported states of the project after and before.

It may happen that at the beginning, it is too costly for some agent to explore the many improbable branches of his decision tree, especially if he does not yet know whether his proposal will be accepted; but later, it would be worth exploring better the ones that became probable. These kinds of in-process modifications are what we would like to make possible. We show that each player has approximately the same incentives as the ``total incentives'' of the players in better scheduling of these small modifications.

The expected utility of an agent with an accepted fair proposal is fixed, and for a nearly fair agent, the little modifications of the other proposals have negligible effect. As the modifications of each agent have no influence on the utility of the principal and only this negligible influence on the expected utility of other agents, the change in the expected total utility is roughly the same as the change in the expected utility of this agent. This confirms the above statement.

A very similar argument shows that the principal can also do such in-process modifications, and her incentives in doing this is about as much as the total incentive of all players.

\subsection{Necessity of being informed about one's own process} \label{infchance}

We assumed that none of the chosen players knew anything better about a chance event of any other chosen agent.
We show an example that does not satisfy this assumption and the mechanism is inefficient. Consider two agents $i$ and $j$ that will definitely be accepted. Assume that $i$ believes the probability of an unfavorable event $\chi$ in his work to be $P(\chi = e_0) = 50\%$, but $j$ knows that the probability is $P(\chi = e_0) = 60\%$, $j$ knows the estimation of $i$, and $j$ also knows that at a particular decision node of $j$, he will be asked to make a move corresponding to the outcome of this chance event of $i$. If the proposal of $i$ is fair, namely, $\pi_i = \cp_i + x_i$, then $j$ can claim that $\chi$ would affect him, and he can increase the asked payment in his proposal for $\chi = x_0$ by an amount of money and decrease it for $\chi \ne x_0$ by the same amount. This way, the value of the proposal of $j$ remains the same, and he bets $1:1$ with $i$ on an event of 60\% probability.

However, if the assumption (in the first sentence) \emph{almost} holds, then we can apply some techniques which limits these problems.
If an agent $i$ has only a small risk that his information might not dominate the belief of someone else about $i$'s probabilities, then in order to limit potential losses, he could rightfully say that a larger bet can only be increased on worse conditions.
Submitting a reasonable proposal with risk-averse utility function makes something similar, which is another reason to use this.

Another technique is that the agent reports a slightly lower flexibility. In the example in the first paragraph, this means that after $i$ gets to know that one of the events could provide him a larger payment, then he uses his unreported capability to increase the probability of this event.

If the chance event is contractible, then all these problems can be avoided. In this case, we can even ``backdate'' the chance event as long as the player can prove that he had no influence on its probabilities between its backdated and actual dates. Under this time interval, all weights $w$ to all chance events of all agents should be functions of $\chi$.

\subsection{Controlling and controlled players} \label{control}

For an example, consider a task of building a unit of railroad. An agent $i$ can make this task for a cost of $100$, but with $1\%$ probability of failure which would cause a huge loss of $10,000$. Another agent $j$ could inspect and, in the case of failure, correct the work of $i$ under the following conditions. The inspection costs $1$. If the task was made correctly, then he does nothing else. If not, then he detects and corrects the failure with probability $99\%$ for a further cost of $100$. But with probability $1\%$ he does not detect the failure, and therefore does nothing. If both of them use truthful strategy and they are the accepted agents for the task, then the mechanism works in the following way.

In the end, $i$ gets $101.99$, but pays $199$ (totally he pays $97.01$) compensation if he fails. $j$ gets $1$ if he correctly finds the task to be correct, and, beyond this, he gets $200$ if the task was made badly but he corrects it, but he pays $9800$ if he misses correcting it.

It can be checked that the expected utility of each agent is his profit independently of the behavior of the others, and that the utility of the principal is fixed.

\subsection{Agents with limited liability} \label{limited}

This subsection is only a suggestion for the cases when we have some agents with limited liability, and we cannot provide any clear mathematical statement about its level of efficiency.

Our mechanism requires each agent to be able to pay as much money as the maximum possible damage he could have caused. However, in many cases, there may be some agents who cannot satisfy this requirement. Despite this, accepting such an agent $i$ may still be a good decision, if $i$ is reliable to some degree.

To solve this problem, $i$ should find someone who has enough funds, and who takes the responsibility, for example for an appropriate fee. If the agent is reliable to some degree, then he should be able to find such insurer player $j$. (It can even be the owner of the project.) This method may also be useful when $i$ has enough funds, but is very risk-averse.

Here, $i$ and $j$ work similarly as the controlled and controlling parties in Appendix~\ref{control}. The differences are that $j$ does not work here, but he has some knowledge about the capability trees (mainly about the capability tree of $i$). Furthermore, the role of $j$ here can be combined with his role in Appendix~\ref{control}.

\subsection{Simplified capability trees}

In many situations, this mechanism cannot be applied directly because of its complex administrative requirements. For example, consider a supermarket employing many cashiers.
Some of the cashiers can go to work whenever they are asked to, while others need to get to know their schedule well in advance.
And everyone can be ill, etc. Clearly, it would be unrealistic to ask them to precisely define their future as a capability tree.

However, it might be useful to construct a weaker but simpler class of proposals which require very few and simple communications, but which could still express the preferences and requests quite well. 

We show an example about how the author believes this could work in practice. A cashier who reports a level of permanent availability gets a base salary $X$, but he may receive messages like ``We need you tomorrow form 7:00 to 15:00 for an increased salary $Y$, but you get $Z$ deduction if you do not come''. $X$, $Y$ and $Z$ are chosen in a fair way, depending on the level of availability the cashier reported. And of course, everyone can report modifications in their availability (e.g.\ via an online system) with the fair conditions defined by the mechanism.

\bigskip

Another example is about contracting for gas or electricity. In practice, some companies get cheaper electricity if they accept that in the case of problems in the current supply, they may be switched off. But if the companies give a rough description of their incentives about it, and the electric supplier also makes a dynamic stochastic description of the possible problems, then we will be able to make socially more efficient decisions.


\subsection{No parallel executions}

The messages the principal sends depend only on her not strictly earlier chance events and the previous messages she received.
Thus, if an agent $i \in \Ag$ is sure that the principal receives no message from anyone else and has no chance event in the time interval $[a, b]$, then, without decreasing the value of his proposal, $i$ can ask the principal to send him already at $a$ what messages she would send during $[a, b]$ depending on the messages she would have received from $i$. In addition, if $i$ is sure that the principal will not send any messages to anyone else during $[a, b]$, then, without changing the value of his proposal, $i$ can simplify his offer so that he sends all his messages only at $b$f that he would have originally sent during $[a, b]$.

Consequently, consider a project consisting of two tasks, where the second task can only be started after the first one has been accomplished. We put this into our model in the following way. The consequence provided by each agent working on the first task (called first agent) consists of his completion time $C_1$. The consequence provided by of each agent working on the second task consists of his starting time $S_2$ and the time $C_2$ he completes; and his capability tree starts with doing nothing until an optional time point $S_2$, and then he can start his work. The valuation function of the principal is of the form $\tilde{u}(C_2)$ for a decreasing function $\tilde{u}\: \R^{\text{(time)}} \rightarrow \R$ if $C_1\geq S_2$, and $-\infty$ otherwise ($\R^{\text{(time)}}$ means the space of time points). The valuation of each agent is simply the minus of his costs. In this case, the principal always communicates only with the agent who is working at the time. Therefore, using the above observation, we can make simplified proposals of the following form, with the same values as of the fair proposals.

If the principal makes $ \tilde{u} $ public at the beginning, then the penalty for the chosen second agent is the loss from the delayed completion of the project, therefore, each second agent should demand $\tilde{u}(C_2) - g_2(S_2)$ money, if he can start his work at $S_2$ and complete it at $C_2$. The penalty for the chosen first agent is $- g_2(C_1)$, and each first agent declares how much money he demands for the first task depending on the penalty function. Then the principal chooses the pair by which she gains the highest utility.

Formally, the form of the simplified proposal for the first agents is a function $g_1\: (\R^{\text{(time)}} \rightarrow \R) \rightarrow \R$, and for the second agents this is a function $g_2\: \R^{\text{(time)}} \rightarrow \R$. If all proposals are so, then the principal chooses a pair for which $g_1(g_2)$ is the greatest. Then she tells the penalty function $g_2$ to the chosen first agent at the beginning of his capability tree, and, after his completion, she pays him $g_2(C_1) - g_1(g_2)$. Then the chosen second agent can start his work at $C_1$ and after his completion, he gets $\tilde{u}(C_2) - g_2(C_1)$. This way, the principal gets utility $\tilde{u}(C_2) - \big(g_2(C_1) - g_1(g_2)\big) - \big(\tilde{u}(C_2) - g_2(C_1)\big) = g_1(g_2)$.

In the simplified fair proposals, $g_1$ and $g_2$ are chosen in such a way that makes their expected utility independent of the arguments ($h$ and $C_2$, resp.), if the agents use their best strategies afterwards.

If a first agent has no choice in his capability tree, that is, his completion time $C_1$ is simply a probabilistic variable, then he should choose $g_1(h) = \E\big(h(C_1)\big) - x$, where $x$ is his costs plus his profit.

\subsection{Another example}

We sketch another example which shows much more about the power of our results.

Assume that we have $k$ mines of different goods. We need $g_i$ number of goods from mine $i$, and our valuation $v(t) = C - \alpha \cdot t$ linearly decreases by the finishing time $t$ when we get all (the last one) of the goods. There are miners applying for the mining jobs, and each mine has a fixed capacity of miners (maybe 1 or $\infty$). Goods are discrete units and mining is a Poisson-process: the probability distribution of future success of finding goods depends only on the abilities of the miner about that mine and on his efforts, but independent from the previous efforts and successes. Efforts are measured by cost-per-time rates.

The abilities and the effort levels of the miners are hidden, all what we can observe are the goods they find. Moreover, we assume only that we will observe the total amount of goods found by each miner, but the miners can report earlier or later finding times.

Our goal is to choose and allocate the miners and choose their effort levels optimally, namely, to maximize the expected valuation of the result minus the expected total costs of the miners.

Our mechanisms applied to this problem are the following. At the beginning, the miners report their abilities. Then we calculate the best plan assuming that the reports were truthful, and we allocate the miners accordingly. We ask the miners for the effort level in the plan (but we cannot observe whether they follow it) and to report whenever a good is found. Each time a good is reported to be found the miner gets an amount of money for the success. But there is a deduction rate for the time spent.

All the suggested effort levels, the prices for finding each kind of good and the deduction rates are recalculated whenever any of the miners report that he finds a good. These can be calculated as follows. With a simple backward induction, we calculate the expected total utility $v(\bs{g'})$ for each demand vector $\bs{g'} = (g^\prime_1, g^\prime_2, ..., g^\prime_k)$ assuming that the abilities of the miners are the same as what they reported, and we have full control on their actions. Then the price to be paid to the miner for finding a good of type $i$ is $v(r_i - 1, \bs{r}_i) - v(\bs{r})$, where $r_i$ is the remaining demand of the good of type $i$. The penalty rate for the time spent is $\lambda_i(f)\big(v(r_i - 1, \bs{r}_{-i}) - v(\bs{r})\big) - f$, where $f$ is the suggested effort level (cost-per-time rate) and $\lambda_i(f)$ is the reported intensity of finding goods at effort level $f$. Beyond these payments, under the second-price revelation mechanism, the miners get second-price compensations for their initial proposals. Or under the first-price revelation mechanism, the fair offer includes a demand for a constant additional payment.

We note that if we had tried to solve the problem directly, without using any of the results in this paper, then it would have seemed to be a reasonable idea to also consider the tool of one-time compensations to the agents whenever the price for finding a good or the deduction rate changes. But our results show that efficiency can be implemented (in quasi-dominant equilibrium) without using this tool. Without our result, it would be difficult to explain why we do not need this tool.

\section{A general framework} \label{conjecturesection}

We present a framework for the analysis of first and second-price and other mechanisms on a larger class of problems. Therefore, we redefine the notions and notations, but these remain analogous to which we used so far.

When we apply it to the Project Management Model, the action of a player should be interpreted as his strategy given his own type. We may exclude some strategies from the ``action sets'' which are not rational \emph{after} the first move.

Let $N = \{0, 1, 2, ..., n\}$, where 0 represents the principal and $N^+ = \{1, 2, ..., n\}$ is the set of agents. We have a prior distribution $\mu$ on $\Theta^N$, a set $A$ of feasible offers (or actions), $\varnothing \in A$, and two functions (valuation) $v\: A^N \rightarrow \R$ and (utility) $u\: \Theta^N \times A^N \rightarrow \R^N$. We assume that
\begin{align*}
\forall \btheta \in \Theta^N,\ \forall i \in N^+,\ \forall \bs{a}_{-i} \in A^{N \setminus \{i\}}\:&&
u_i\big(\btheta, (\bs{a}_{-i}, \varnothing)\big) = 0,
&&\phantom{\forall \btheta \in \Theta^N,\ \forall i \in N,\ \forall \bs{a}_{-i} \in A^{N \setminus \{i\}}\:}
\end{align*}
which will express that rejected players get utility 0. There is a fixed function (decision) $d: A^N \rightarrow A^N$ satisfying that
\begin{align*}
\forall \bs{a} \in A^N\:&&
d(\bs{a}) &\in \argmax_{\bs{b} \in \{a_0\} \times \{\varnothing, a_1\} \times \{\varnothing, a_2\} \times ... \times \{\varnothing, a_n\}} v(\bs{b}).
&&\phantom{\forall \bs{a} \in A^N\:}
\end{align*}

The strategy set of player $i$ is the set of functions $S_i = \{\Theta \rightarrow A\}$.
Assume that for each $i \in N$, there exists a (truthful) strategy $\cp_i \in S_i$ satisfying the following conditions.
\begin{align*}
\forall \btheta \in \Theta^N,\ \forall i \in N^+,\ \forall \bs{a}_{-i} \in A^{N \setminus \{i\}}\: &&
u_i\Big(\btheta, \big(\bs{a}_{-i}, \cp_i(\theta_i)\big)\Big) &= 0
&&\phantom{\forall \btheta \in \Theta^N,\ \forall i \in N^+,\ \forall \bs{a}_{-i} \in A^{N \setminus \{i\}}\:}
\end{align*}
\begin{align*}
\forall \bs{a}_{N^+} \in A^{N^+}\:
&&&&&&&&&&&&
u_0\Big(\btheta, \big(\bs{a}_{N^+}, \cp_0(\theta_0)\big)\Big) &= v\Big(d\big(\bs{a}_{N^+}, \cp_0(\theta_0)\big)\Big).
&&&&&&&&&&&&\phantom{\forall \bs{a}_{N^+} \in A^{N^+}\:}
\end{align*}
\begin{align*}
\forall \btheta \in \Theta^N,\ \forall \bs{a} \in A^N\:
&&&&&&&&&&&&&&&&&&
\sum_{i \in N} u_i \big( \btheta, d(\bs{a}) \big) &\le v \Big( d \big( \bs{\cp}(\btheta) \big) \Big)
&&&&&&&&&&&&&&&&&&
\phantom{\forall \btheta \in \Theta^N,\ \forall \bs{a} \in}
\end{align*}

Assume that for each $r \in A$ and $x \in \R$, there exists an offer denoted by $r + x \in A$ satisfying $\forall \btheta \in \Theta^N$, $\forall i \in N^+$, $\forall \bs{a}_{-i} \in A^{N \setminus \{i\}}\:$
\begin{equation*}
v(\bs{a}_{-i}, r + x) = v(\bs{a}_{-i}, r) - x,
\end{equation*}
\begin{equation*}
u_i\big(\btheta, (\bs{a}_{-i}, r + x) \big) = u_i\big(\btheta, (\bs{a}_{-i}, r) \big) + x.
\end{equation*}

These were the main assumptions. Optionally we might have further assumptions, we will come back to them, but let us see now how to use the framework.

Let us define the marginal contribution of an offer $a_i$ of a player $i \in N$ by
\begin{equation*}
v^+_i(\theta_0, \bs{a}) = v \big( \theta_0, d(\theta_0, \bs{a}) \big) - v \big( \theta_0, d(\theta_0, \bs{a}_{-i}, \varnothing) \big).
\end{equation*}


A strategy profile $\bs{s} = (s_1, s_2, ..., s_n) \in \bs{S} = S_1 \times S_2 \times ... \times S_n$ is an equilibrium under the first-price mechanism if
\begin{align*}
\forall i \in N^+,\ \forall s_i' \in S_i\:&&
\E_{\mu} \bigg( u_i\Big(\btheta, d\big(\bs{s}_{-i}(\btheta_{-i}), s_i'(\theta_i) \big) \Big) \bigg) &\le \E_{\mu} \bigg( u_i \Big(
\btheta, d \big( \bs{s}(\btheta) \big) \Big) \bigg).
&&\phantom{\forall i \in N,\ \forall s_i' \in S_i\:}
\end{align*}

A strategy profile $\bs{s} \in \bs{S}$ is an equilibrium under the second-price mechanism if \\ $\forall i \in N^+,\ \forall s_i' \in S_i\:$
\begin{equation*}
\E_{\mu} \bigg( u_i\Big(\btheta, d\big(\bs{s}_{-i}(\btheta_{-i}), s_i'(\theta_i) \big)\Big) +
v^+_i\big(\bs{s}_{-i}(\btheta_{-i}), s_i'(\theta_i)\big) \bigg) \le \E_{\mu} \bigg( u_i \Big( \btheta, d \big( \bs{s}(\btheta) \big) \Big) + v^+_i(\bs{s}\big(\btheta)\big) \bigg),
\end{equation*}
\begin{align*}
\text{and }\forall s_0' \in S_0\:&&&&&&
\E_{\mu} \bigg( u_0\Big(\btheta, d\big(\bs{s}_{N^+}(\btheta_{N^+}), s_0'(\theta_0) \big)\Big) - \sum_{i \in N^+}
v^+_i\big(\bs{s}_{N^+}(\btheta_{N^+}), s_0'(\theta_0)\big) \bigg)&&
\end{align*}
\begin{equation*}
\le \E_{\mu} \bigg( u_0 \Big( \btheta, d \big( \bs{s}(\btheta) \big) \Big)  - \sum_{i \in N^+}
v^+_i\big(\bs{s}(\btheta)\big) \bigg).
\end{equation*}

We can generalize Theorems \ref{price1perfect} and \ref{price1commonspec} under this general framework, because the proofs used only properties and assumptions what we stated above. Therefore, this framework can be useful for further analysis about the level of efficiency of the first-price mechanism. Furthermore, we can define the generalized version of the conjecture mentioned in Section~\ref{results2}.

\begin{Conjecture} \label{conjecture}
For each common prior $\mu$, there exists an equilibrium $\bs{s} \in \bs{S}$ under the second-price mechanism satisfying
\begin{equation*}
\E_{\mu} \bigg( u_0 \Big( \btheta, d \big( \bs{\cp}(\btheta) \big) \Big)  - \sum_{i \in N^+}
v^+_i\big(\bs{\cp}(\btheta)\big) \bigg)
\le \E_{\mu} \bigg( u_0 \Big( \btheta, d \big( \bs{s}(\btheta) \big) \Big)  - \sum_{i \in N^+}
v^+_i\big(\bs{s}(\btheta)\big) \bigg).
\end{equation*}
\end{Conjecture}



Now we define further potential assumptions which might be useful and which were satisfied in the Project Management Model.

\smallskip

We may assume that whether a player wins with an offer or with a constant higher offer has no effect on the other players:
\begin{align*}
\forall j \in N \setminus \{i\}\:&&
u_j\big(\btheta, (\bs{a}_{-i}, a_i + x) \big) &= u_j(\btheta, \bs{a}).
&&\phantom{\forall j \in N \setminus \{i\}\:}
\end{align*}

\smallskip

We may assume that the set of offers is convex, namely, $\forall p, q \in A$ and $\lambda \in (0,1)$, there exists a strategy denoted by $\lambda p + (1 - \lambda) q \in A$ satisfying that $\forall \btheta \in \Theta^N$, $\forall i \in N$, $\forall \bs{a}_{-i} \in A^{N \setminus \{i\}}\:$
\begin{equation*}
v\Big(\theta_0, \big(\bs{a}_{-i}, \lambda p + (1 - \lambda) q \big) \Big) = \lambda \cdot v\big( \theta_0, (\bs{a}_{-i}, p) \big) + (1 - \lambda) \cdot
v\big(\theta_0, (\bs{a}_{-i}, q) \big),
\end{equation*}
\begin{equation*}
u_i\Big(\btheta, \big(\bs{a}_{-i}, \lambda p + (1 - \lambda) q \big) \Big) = \lambda \cdot u_i \big( \btheta, (\bs{a}_{-i}, p) \big) + (1 - \lambda) \cdot u_i \big(\btheta, (\bs{a}_{-i}, q) \big).
\end{equation*}

\smallskip

We may assume that each player can submit multiple offers, namely, $\forall p, q \in A$, there exists an offer denoted by $p \wedge q \in A$ satisfying that $\forall
\btheta \in \Theta^N$, $\forall i \in N$, $\forall \bs{a}_{-i} \in A^{N \setminus \{i\}}\:$
\begin{equation*}
v\big(\theta_0, (\bs{a}_{-i}, p \wedge q) \big) = \max_{r \in \{p, q\}} \Big( v\big(\theta_0, (\bs{a}_{-i}, r) \big) \Big),
\end{equation*}
and with the very same $r$,
\begin{equation*}
u_i \big( \btheta, (\bs{a}_{-i}, p \wedge q) \big) = u_i \big( \btheta, (\bs{a}_{-i}, r) \big).
\end{equation*}

\printindex

\end{document}